\documentclass[epj,nopacs]{svjour}
\pdfoutput=1
\usepackage{amsmath,amssymb,amsfonts,graphicx,cite,multirow,color,subfig}
\usepackage{soul}
\usepackage[utf8]{inputenc}
\usepackage[retainorgcmds]{IEEEtrantools}
\usepackage{ulem}
\usepackage{indentfirst}
\usepackage{xspace}

\graphicspath{{graphics/}}
\makeatletter
\ifx\input@path\@undefined
\def\input@path{{graphics/}}
\else
\g@addto@macro\input@path{{graphics/}}
\fi
\makeatother
\newcommand{\HerwigPP}{\textsc{Herwig}\texttt{++}\xspace}
\newcommand{\Herwig}{\textsc{Herwig}\xspace}
\newcommand{\Matchbox}{\textsc{Matchbox}\xspace}
\newcommand{\powhegbox}{\textsc{PowhegBox}\xspace}
\newcommand{\Rivet}{\textsc{Rivet}\xspace}

\newcommand{\hfact}{\textsf{hfact}\xspace}
\newcommand{\resummation}{\textsf{resummation}\xspace}

\newcommand{\FIz}{z_i}
\newcommand{\FIx}{x_{ij,b}}
\newcommand{\FIinv}{s_{ij,b}}
\newcommand{\FIinvprime}{s_{ij,b}^\prime}

\newcommand{\IFu}{u_j}
\newcommand{\IFx}{x_{jk,a}}
\newcommand{\IFinv}{s_{aj,k}}
\newcommand{\IFinvprime}{{s_{aj,k}^\prime}}

\newcommand{\FFz}{z_i}
\newcommand{\FFy}{y_{ij,k}}
\newcommand{\FFinv}{s_{ij,k}}
\newcommand{\FFzplus}{z_{i,+}}
\newcommand{\FFzminus}{z_{i,-}}
\newcommand{\FFzpm}{z_{i,\pm}}

\newcommand\clH{{\mathcal H}}
\newcommand\clS{{\mathcal S}}

\newcommand{\hardProcScale}{\mu_\mathrm{H}}
\newcommand{\vetoScale}{Q_\perp}
\newcommand{\showerScale}{\mu_\mathrm{S}}

\newcommand{\ie}{{\it i.e.}\xspace}

\newcommand{\eg}{{\it e.g.}\xspace}

\preprint{CERN-TH-2018-219\\HERWIG-2018-03\\IPPP/18/91\\LU-TP 18-33\\MCNET-18-28\\UWTHPH-2018-22}

\title{Parton Shower and Matching Uncertainties in\\Top Quark
  Pair Production with \Herwig\,7}

\author{Kyle Cormier\inst{1}, Simon Pl\"atzer\inst{2}, Christian
  Reuschle\inst{3,4}, Peter Richardson\inst{5,6}, Stephen
  Webster\inst{6}}

\institute{Department of Physics, University of Toronto, Toronto, Canada \and
Particle Physics, Faculty of Physics, University of Vienna, Vienna, Austria\and
Physics Department, Florida State University, Tallahassee, FL, 32306-4350, U.S.A.\and
Theoretical Particle Physics, Department of Astronomy and Theoretical Physics, Lund University, SE-223 62 Lund, Sweden\and
Theoretical Physics Department, CERN, 1211 Geneva 23, Switzerland\and
IPPP, Department of Physics, Durham University, Durham, UK}

\date{\today}

\abstract{We evaluate the theoretical uncertainties in
  next-to-leading order plus parton shower predictions
  for top quark pair production and decay in hadronic collisions.
  Our work is
  carried out using the \Herwig\,7 event generator and presents an
  in-depth study of variations in matching schemes with two
  systematically different shower algorithms, the traditional
  angular-ordered and alternative dipole shower. We also present all of
  the required extensions of the \Herwig\ dipole shower algorithm to
  properly take into account quark mass effects, as well as its
  ability to perform top quark decays. The predictions are compared at
  parton level as well as to LHC data, including in the boosted regime.
  We find that the regions where predictions with a non-top-quark-specific tune
  differ drastically from data are plagued by large uncertainties
  which are consistent between our two shower and matching algorithms.}

\begin{document}

\authorrunning{K. Cormier et al.}
\titlerunning{Parton Showers and Matching for Top Quark Pair Production}
\maketitle


\setlength{\parskip}{1ex}

\section{Introduction}
\label{sec:Introduction}

Top quark pair production is an extremely important process at the Large
Hadron Collider~(LHC) due to its significant cross section.  As the top quark
decays before it can hadronize, top quark pair production provides us with a
unique opportunity to study Quantum Chromodynamics~(QCD) radiation and
corrections involving massive particles. This includes measurements of the top
quark mass, which is important to constrain the higher-order corrections in
the electroweak sector of the Standard Model. Top quark pair production at
hadron colliders has become a `standard candle' due to the accurate
calculation~\cite{Czakon:2013goa} and measurement of the total cross
section. However, the large production rate also allows the measurement of an
ever increasing range of kinematic quantities. This means that different
kinematic reconstruction strategies for the top quark, and its mass in
particular, including in the boosted regime, can be evaluated.  It also means
that we can study QCD in detail by comparing with less inclusive calculations
and Monte Carlo event generators.  The large production rate also means that
top quark pair production, particularly with the presence of extra jets from
QCD radiation, is often the main background to searches for physics Beyond the
Standard Model. A number of measurements are available both extracting the top
quark mass \cite{Khachatryan:2015hba,Aaboud:2016igd}, as well as a number of
kinematic properties, \eg \cite{Aad:2015eia}.

Monte Carlo event generators
\cite{Gleisberg:2008ta,Bahr:2008pv,Sjostrand:2014zea} used for predictions of
top quark pair production have seen several improvements, which mainly
concentrated on combining next-to-leading order QCD corrections with
subsequent parton shower algorithms
\cite{Frixione:2002ik,Frixione:2007vw,Platzer:2011bc,Hoeche:2011fd}, and the
production of additional jets using multi-jet merging algorithms, {\it e.g.}
those employed in the \Herwig~7 event generator
\cite{Platzer:2012bs,Bellm:2017ktr}. Some of the matching algorithms have
addressed off-shell effects in the calculation of the hard process
\cite{Jezo:2016ujg}, though none of the event generators yet features shower
algorithms which properly take into account the effect of the finite width of
the top quark and its interplay with the parton shower infrared cutoff.

As compared to indirect approaches based on total cross section measurements,
these state-of-the-art simulations provide a very sophisticated description of
kinematic properties and thus allow to extract the top quark mass from
kinematic properties with an unprecedented precision through template
fits. These fits determine the top quark mass parameter used by the event
generator simulation. The question in what scheme this mass parameter needs to
be interpreted, and what uncertainties need to be taken into account, is still
subject to ongoing research
\cite{Fleming:2007qr,Beneke:2016cbu,Nason:2017cxd}, and for coherent parton
shower evolution in $e^+e^-$ collisions first analytic and numeric insights
have been gained on the role of the mass parameter, including measurements of
the top quark mass from reconstruction of its decay products
\cite{Hoang:2018zrp}. Some aspects of the hadronization effects in such
observables have recently been evaluated \cite{Corcella:2017rpt}, however a
comprehensive analysis of variations in parton shower evolution, and the
impact of different paradigms to include higher order corrections has not yet
been performed.

The present work therefore, in comparison to what one would typically
consider state of the art, makes a deliberate step back and is centred
around a thorough investigation of how reliable predictions by
established paradigms are across phase space. This question has not
yet been answered by an in-depth comparison of similar, yet
algorithmically very different, predictions and their associated
variations which can be established to constitute a set of
uncertainties when meeting well-defined constraints
\cite{Bellm:2016rhh}. We do this particularly in the light of event
generator predictions which are highly specialized in their parameter
choices and thus might generate a wrong impression of how well
theoretical understanding is under control, and the associated
question of what improvements, specifically concerning multi-jet
merging, are required. Also it seems likely that these simulations
will remain the main tool used by the LHC experiments to study top
quark physics for the foreseeable future.

We also use this study to introduce some improvements to both
radiation from heavy quarks and the handling of their decays in the
\Herwig\ dipole shower module. These changes enable us to perform this
study between different matching and shower algorithms in a consistent
way, using the same hadronization and underlying event models and with
control over shower starting scales and resummation in the hard
emission region.

\section{Outline of this Work}
\label{sec:Simulation}

In this study we use the most recent version, 7.1.4, of
the \Herwig\ event generator
to make use of the various improvements to the simulation of heavy
quarks in production, shower emissions and decays. The modelling of
this physics will be discussed in detail in the following
sections. In the version considered \Herwig\ sets up the next-to-leading order~(NLO) QCD
corrections to the top quark pair production process using the
automated facilities outlined in \cite{Platzer:2011bc}, using external
libraries \cite{Alwall:2014hca,Cascioli:2011va} to evaluate the required amplitudes on
a phase-space point by phase-space point basis. The production of top
quark pairs has been validated against MCFM
\cite{MCFM,Campbell:1999ah,Campbell:2011bn,DanielThesis}. NLO corrections to the decays are included
within a NLO matched parton shower simulation, while we neglect
non-factorizable corrections which are beyond the leading contribution
in the narrow-width approximation.

Matching the production process to the parton shower is possible
within both the subtractive (MC@NLO-type\,\cite{Frixione:2002ik}) and
the multiplicative (Powheg-type\,\cite{Nason:2004rx}) matching
paradigms, using the matching subtractions obtained by the
\Matchbox\ module along with the QCD corrections required.
The matching of the decay to the parton shower is available within 
the multiplicative paradigm in both the QTilde\footnote{An old-style matrix element correction is
  used by default in the angular-ordered shower, which is formally equivalent to the Powheg method.}
(angular-ordered) \cite{Gieseke:2003rz} and Dipole (Catani-Seymour
\cite{Catani:1996vz,Catani:2002hc} dipole-type) \cite{Platzer:2009jq} shower modules.

The details of the simulation inputs including the scale choices, parton distribution functions (PDFs), strong coupling running and analyses used to obtain results are included in the subsequent
sections.  The default electroweak parameters of \Herwig\,7.1 are used in all runs and for the decay corrections the top mass is used as the renormalization scale.

The remainder of this work is organized as follows. In
Section~\ref{sec:ProductionRadiation} we consider 
QCD radiation from the top quark pair production process.
In Section~\ref{sec:DecayRadiation} the parton shower simulation
of the decay stage is discussed in detail.
We then proceed with an in-depth discussion of the
NLO matching in Section~\ref{sec:Matching}.  We use the framework to
assess phenomenologically relevant uncertainties in the matched NLO+PS
predictions in Section~\ref{sec:UncertaintyBenchmarks} and conclude
with a detailed analysis of our predictions compared to available data
from the LHC in Sections~\ref{sec:Boosted}~and~\ref{sec:data-benchmarks}.
Finally we present our conclusions.

\section{Radiation in Production of Heavy Quarks}
\label{sec:ProductionRadiation}

\subsection{Generalities}

For both parton shower algorithms used in the \Herwig\ event generator,
a colour flow is assigned to the hard process on the basis of
the tree-level colour sub-amplitudes sq.
This is a consequence of evaluating the colour
correlations relevant to the soft radiation pattern in the limit of a
large number of colour charges, $N_c\to\infty$. The chosen colour flow
is used to set the initial conditions in both parton shower modules,
in particular identifying which `dipole'-type systems radiate
coherently. Radiation in both parton showers is also subject to a veto on
hard emissions, as set by the hard shower scale,
to be discussed in more detail in Sec.~\ref{sec:UncertaintyBenchmarks}.

Since a comprehensive treatment of non-factorizable QCD effects which
connect the production process and the decay beyond the
narrow-width-approximation is not available both parton shower algorithms
evolve the production process down to the infrared cutoff which, in
the current version, is a cutoff on the relative transverse momentum
of the emissions. Once the cutoff has been reached by the evolution of
the hard process, the decay of the top quark(s) is performed, and
further showering of the decay system is simulated as discussed
in Sec.~\ref{sec:DecayRadiation}.

\subsection{Angular-Ordered Shower}
\label{sec:AOprod}
The improved angular-ordered shower used by default in \Herwig\ is described
in detail in Refs.\,\cite{Gieseke:2003rz,Bahr:2008pv}. Here we will only
summarize the important details relevant for heavy quark production together
with recent improvements not described in
Refs.\,\cite{Gieseke:2003rz,Bahr:2008pv}.
The momenta of the partons produced in the parton shower are decomposed 
in terms of the 4-momentum of the parton initiating the jet, $p$ ($p^2=m^2$,
the {\em on-shell} parton mass-squared), a light-like reference vector, 
$n$, in the direction of
the colour partner of the parton initiating the jet 
and the momentum transverse to the direction of $p$ and $n$. The four momentum of 
any parton produced in the evolution of the jet can be decomposed as
\begin{equation}
q_i = \alpha_i p + \beta_i n + q_{\perp i},
\end{equation}
where $\alpha_i$ and $\beta_i$ are coefficients and $q_{\perp i}$ is the transverse four momentum 
of the parton ($q_{\perp i}\cdot p = q_{\perp i}\cdot n =0$).
If we consider the branching of a final-state parton $i$ to two partons $j$
and $k$, {\it i.e.} $i\to j k$, the evolution variable is
\begin{equation}
\tilde{q}^2_i = \frac{q^2_i-m^2_i}{z_i(1-z_i)}, 
\end{equation}
where $q^2_i$ is the square of the virtual mass developed by the parton $i$ in
the branching, $m_i$ is the physical mass of parton $i$, and $z_i$ is the
momentum fraction of the parton $j$ defined such that
\begin{equation}
  \alpha_j = z_i \alpha_i, \ \ \ \ \ \ \ \ \ \ \ \ \ \ \ \ \ \ \ \alpha_k = (1-z_i)\alpha_i.
\label{eq:z_defn}  
\end{equation}
The transverse momenta of the partons produced in the branching are
\begin{equation}
q_{\perp j} =    z_i q_{\perp i} + k_{\perp i},\ \ \ \ \ \ \ \ \ 
q_{\perp k} = (1-z_i)q_{\perp i} - k_{\perp i},
\label{eq:fs_pt}
\end{equation}
where $k_{\perp i}$ is the transverse momentum generated in the branching.
In this case the virtuality of the parton $i$ is
\begin{equation}
q_i^2 = \frac{p^2_{Ti}}{z(1-z)} + \frac{m_j^2}z+\frac{m_k^2}{1-z},
\label{eqn:mass}
\end{equation}
where $p_{Ti}$ is the magnitude of the transverse momentum produced
in the branching defined such that \mbox{$k_{\perp i}^2=-p^2_{Ti}$}.

In this case the probability for a single branching to occur is
\begin{equation}
{\rm d} \mathcal{P} = \frac{{\rm d}\tilde{q}^2_i}{\tilde{q}^2_i}\frac{\alpha_S}{2\pi}
\frac{{\rm d} \phi_i}{2\pi} {\rm d} z_i P_{i\to jk}(z,\tilde{q}),
\label{eqn:prob}
\end{equation}
where $P_{i\to jk}(z,\tilde{q})$ is the quasi-collinear splitting function
and $\phi_i$ is the azimuthal angle of the transverse momentum $k_{\perp i}$
generated in the splitting.

As described in Ref.\,\cite{Gieseke:2003rz} this choice of evolution variable,
including the mass of the radiating parton, together with the use of the
quasi-collinear splitting functions gives a better treatment of radiation from
the parton in the small-angle region.  In this region we expect a suppression
of soft radiation for angles $\theta \lesssim m/E$, where $\theta$ is the
angle of emission, $m$ and $E$ the mass and energy of the radiating parton,
respectively.  The choices used in \Herwig\,7 give the expected smooth
turn-off of soft radiation rather than the `{\it dead-cone}'\footnote{{\it i.e.}
  radiation was forbidden for $\theta < m/E$}~\cite{Marchesini:1989yk} used in
{\textsf HERWIG 6}~\cite{Corcella:2000bw}.

The angular-ordered shower is simulated as a series of individual emissions,
and only the shower variables~($\tilde{q},z,\phi$) are calculated for each emission.
Once the evolution has terminated, {\it i.e.} there is no phase space available for
further emissions, the external particles are taken to be on-shell
and the physical momenta reconstructed.

If we set $\alpha_i=1$ for final-state progenitors\footnote{the partons from the
  hard process which initiate the parton shower.}  and $\alpha_i=x$, the
light-cone momentum fraction, for initial-state progenitors then using
Eq.~\eqref{eq:z_defn} and the momentum conservation relation
$\alpha_{i}=\alpha_{j}+\alpha_{k}$, all the $\alpha$ values can be iteratively
calculated, starting from the hard process and working outward to the external
legs. For final-state radiation the transverse momenta can be calculated in
the same way using Eq.~\eqref{eq:fs_pt}, whereas for initial-state radiation
the transverse momentum is calculated iteratively assuming that the parton
extracted from the proton as a result of the backward evolution has zero
transverse momentum.\footnote{or a non-perturbative `{\it intrinsic}'
  transverse momentum.}  The $\beta$ variables for the external partons can
then be calculated using the on-shell condition and those for radiating
partons using momentum conservation, {\it i.e.} $\beta_{i}=\beta_{j}+\beta_{k}$.
The latter step may be iterated until the progenitor is reached giving all
the $\beta$ coefficients.

As a result of the shower evolution all the progenitor partons, $I$, produced
in the hard process gain a virtual mass, {\it i.e.} the progenitor partons, which
initiated the jets, are no longer on mass shell, $q_{I}^{2}\neq m_{I}^{2}$.
We need to restore momentum conservation in a way that disturbs the internal
structure of the jet as little as possible. The easiest way to achieve this is
by boosting each jet along its axis so that their momenta are rescaled,
{\it i.e.} for every jet a Lorentz boost is applied such that
\begin{equation}
  q_I=\left(\mathbf{q}_I;\sqrt{\mathbf{q}_I^{2}+q_I^{2}}\right)\stackrel{{\rm boost}}{\longrightarrow}q_I^{\prime}=\left(k_I\mathbf{p}_I;\sqrt{k^{2}\mathbf{p}_I^{2}+q_I^{2}}\right),
\end{equation}
where $k_I$ is the rescaling factor.  The rescaling factors, and the choice of
frame in which to apply the boosts, are determined by the choice of which
kinematic variables we wish to preserve in the rescaling process. In
Ref.\,\cite{Gieseke:2003rz} an approach was suggested based on the colour
connections between the partons initiating the jets:
\begin{itemize}
\item for colour-connected final-state partons the reconstruction was performed in
  the centre-of-mass frame of the partons and the momenta rescaled such that the centre-of-mass
  energy was conserved, {\it i.e.}
\begin{equation}
    \sum_{I=1}^{n}\sqrt{k^{2}\mathbf{p}_{I}^{2}+q_{J}^{2}}=\sqrt{s},
\end{equation}
where $\sqrt{s}$ is the centre-of-mass energy and the same rescaling factor
$k$ is used for all the jets;
\item for colour-connected initial-state partons the reconstruction is
  performed in the hadronic centre-of-mass frame and the partonic
  centre-of-mass energy is preserved. In order to fully specify the
  kinematics an additional constraint is required which in
  Ref.\,\cite{Gieseke:2003rz} was chosen such that the rapidity of the
  partonic collision was preserved;
\item for partons with a colour connection between the initial and final
  state, such as Deep Inelastic scattering~(DIS), the system is
  reconstructed in the Breit frame of the partons such that the virtuality of
  the system is preserved.
\end{itemize}
As the majority of hadronic collisions cannot be decomposed into separate
colour-singlet systems in early versions of \HerwigPP\ hadronic collisions
were all reconstructed by first using the procedure for colour-connected
initial-state partons and then that for final-state partons.
This was changed such that if possible
the hard process was decomposed into separate colour-singlet
systems\footnote{in the large number of colours, $N_C$ limit.}, for example in
$q\bar q \to t \bar t$, then the separate colour-singlet systems were reconstructed as described
above.

In \Herwig\,7 we have adopted an approach which attempts to use as much information as possible
on the colour structure of the hard process when performing the reconstruction. In order to
achieve this we now consider all the partons in the hard process and commence the
reconstruction with the parton which had the hardest, {\it i.e.} largest $p_T$, emission in the parton
shower. The system formed by this parton and its colour partner is then reconstructed, with
either a full reconstruction of the jet produced by the colour partner, the default, or optionally
just using the partner to absorb the recoil leaving it on its partonic mass shell
and not performing the reconstruction of the full jet. This procedure is repeated for the parton with
the hardest shower emission which has not been reconstructed until all the kinematics of
all the jets have been reconstructed. Together with an additional option of preserving the momentum
fraction of the softer incoming parton in the hard process, for systems with colour connections
between initial-state partons, this means that for a single emission the kinematics
reduce to those of the Catani-Seymour~\cite{Catani:1996vz,Catani:2002hc} dipoles making matching in the MC@NLO approach simpler.

\subsection{Dipole Shower}

\begin{figure}
  \centering
  \subfloat[Final-initial dipole.] {%
    \includegraphics[width=0.4\textwidth]{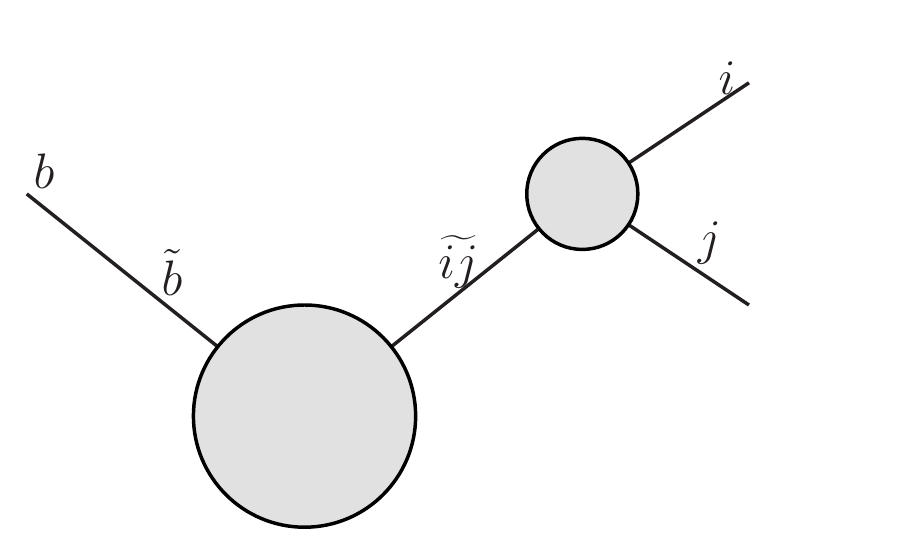}
    \label{fig:FIDip}
  }
  \\
  \centering
  \subfloat[Initial-final dipole.] {%
    \includegraphics[width=0.4\textwidth]{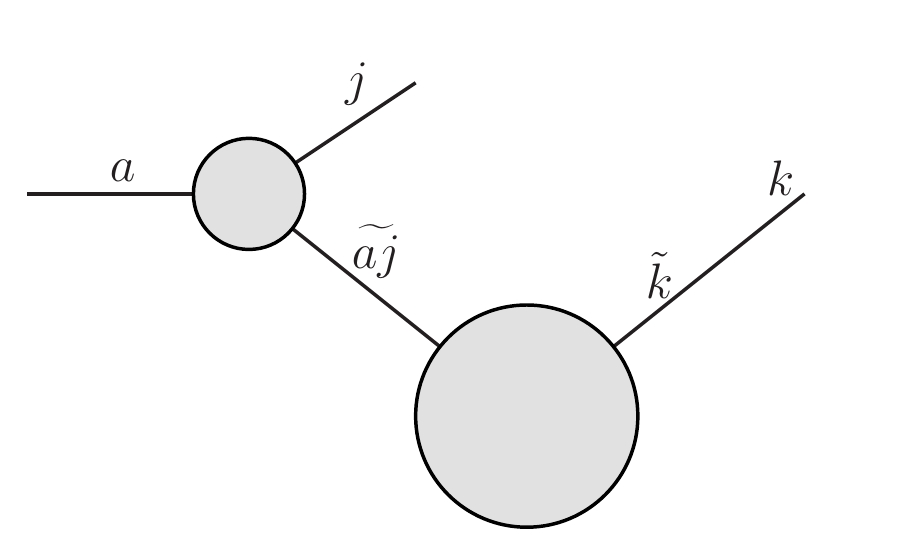}
    \label{fig:IFDip}
  }
  \\
  \centering
  \subfloat[Final-final dipole.] {%
    \includegraphics[width=0.4\textwidth]{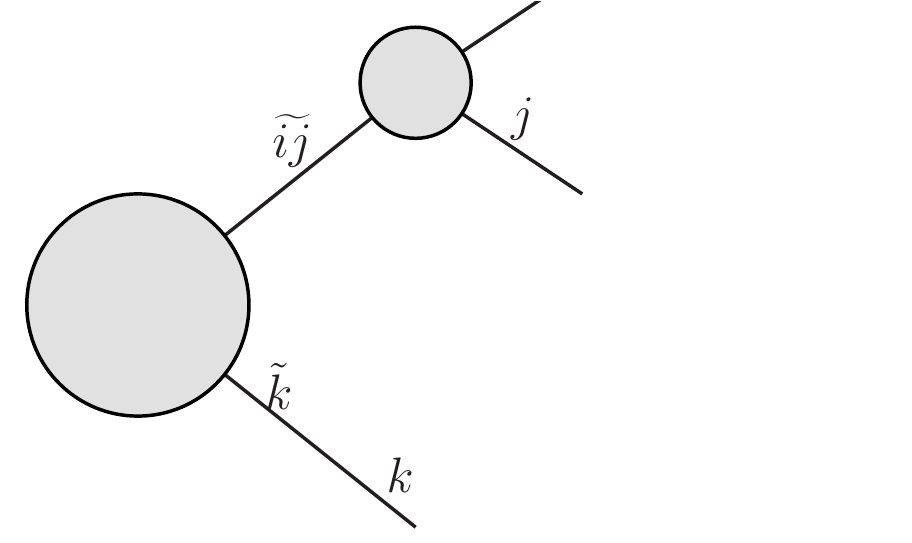}
    \label{fig:FFDip}
  }
  \caption{Diagrams of the massive dipoles.}
  \label{fig:Dipoles}
\end{figure}

The dipole shower algorithm evolves singlet systems of colour connected
dipoles, referred to chains \cite{Platzer:2011bc}, based on the colour
flow information assigned to the hard process. For the massless case the
details of the dipole shower algorithm in \Herwig\ have been discussed in
Refs.\,\cite{Platzer:2009jq,Platzer:2011bc}. In this paper we focus on the
generalization of the algorithm to radiation from heavy quarks, and radiation
in the decays of coloured objects, to be covered in detail in
Sec.~\ref{sec:DecayRadiation}. While the heavy quark treatment in
Ref.\,\cite{diplmartin} has previously been based on Ref.\,\cite{Schumann:2007mg},
an improved description is presented here
which is in one-to-one correspondence to the
massless case, and in particular adopts the transverse momentum relevant in
the quasi-collinear limit \cite{Catani:2002hc}, with a smooth massless
limit. Throughout this work we use the terminology `massive dipole' to refer
to a dipole that includes at least one massive parton and/or splits to produce at least one massive parton.

Splittings involving massive incoming partons are not currently implemented in the
\Herwig\ dipole shower. This means that there are three possible dipole configurations
involving massive partons. These are shown diagrammatically in
Fig.~\ref{fig:Dipoles}. Massive final-final (FF) dipoles, with final-state
emitter and spectator, Fig.~\ref{fig:FFDip}, must include at least one massive
outgoing parton before or after the splitting. Massive final-initial (FI)
dipoles, Fig.~\ref{fig:FIDip}, consist of a massless incoming spectator and an
outgoing emitter. At least one of the outgoing partons before or after the
splitting must be massive. Massive initial-final (IF) dipoles,
Fig.~\ref{fig:IFDip}, consist of a massless incoming emitter and massive
final-state spectator.

Due to its large mass, parton shower emissions from top quarks are highly suppressed. This
means that emissions from massive FI dipoles do not make a significant
contribution in the parton shower. Similarly emissions from FF dipoles with a top
quark emitter are highly suppressed, however emissions from FF dipoles with a
massless emitter and top quark spectator are not suppressed in this
way. Therefore both massive FF and IF dipole splittings make a significant
contribution in the parton shower.

A detailed understanding of these radiation processes with full mass effects
is therefore mandatory, and the main goal of this work is to formulate the
relevant kinematic parametrization and evolution quantities in a similar way
to the massless case, with emphasis on a covariant
formulation and an evolution variable which reflects the transverse momentum
relevant to the enhancements present for collinear radiation. We present the
kinematics used for splittings of all massive dipoles in the following
sections. The kernels used to describe the splittings are those given in
Ref.~\cite{Catani:2002hc}.

For each dipole the kernels and kinematics used to describe a splitting are
parametrized by two splitting variables and an azimuthal angle. In \Herwig\,7.1
we use spin-averaged dipole splitting kernels, therefore we randomly generate
the azimuthal angle for each splitting according to a uniform
distribution. The splitting variables used to parametrize the splitting for
each dipole are those used in Ref.~\cite{Catani:2002hc} and are given for each
dipole in the following sections.

In the dipole shower in \Herwig\ we actually generate the transverse momentum,
$p_\perp$, and the light-cone momentum fraction, $z$, as used in the
standard quasi-collinear Sudakov parametrization of the momenta following a
splitting. This is the parametrization used in the angular-ordered
shower~\cite{Gieseke:2003rz}, see Sec.\,\ref{sec:AOprod}.
We choose a light-like vector $n$ to define the
collinear direction and for a splitting from a final-state emitter with
momentum $\tilde{p}_{ij}$ we write the momentum, $q_j$, of the emitted parton
as
\begin{equation}
	q_j = (1-z)\tilde{p}_{ij} + \frac{m_j^2 - (1-z)^2 m_{ij}^2 +
          p_\perp^2}{2 \tilde{p}_{ij}\cdot n (1-z)} n - k_\perp ,
	\label{eq:qjforward}
\end{equation}
where $m$ is the mass of the emitted parton and $m_{ij}$ is the mass of the
emitter. The space-like vector $k_\perp$ satisfies $k_\perp
\cdot \tilde{p}_{ij} = k_\perp \cdot n = 0$ and $k_\perp^2 = -
p_\perp^2$ .

Similarly for a splitting from a massless incoming parton we write the momentum
of the emitted parton as
\begin{equation}
   q_j = (1-z)q_a + \frac{ p_\perp^2}{2 q_a\cdot n (1-z)} n - k_\perp ,
   \label{eq:qjbackward}
\end{equation}
where $q_a$ is the momentum of the parton incoming from the proton following
the splitting and $k_\perp \cdot q_a = k_\perp \cdot n = 0$.

In view of these parametrizations, which are the ones relevant in the
(quasi-)collinear limit, we choose to set up kinematic mappings for a
dipole splitting including momentum conservation in a way that we express the
resulting kinematics in terms of these physical variables, $p_\perp$ and $z$,
rather than the ones which most conveniently allow the separation and integration over the phase space.
This has been done in the massless case, and the mappings below generalize this to the
massive case with a smooth massless limit.

\subsubsection{Final-Final Dipoles}
\label{sec:DSFFKin}

We consider the splitting process $\tilde{p}_{ij},\tilde{p}_k\to q_i,q_j,q_k$
where all momenta before and after the splitting are on-shell,
$\tilde{p}_{ij}^2 = m_{ij}^2$, $\tilde{p}_{k}^2 = q_k^2 = m_k^2$, $q_{i,j}^2 =
m_{i,j}^2$ and satisfy momentum conservation for the dipole
considered, {\it i.e.} $Q=\tilde{p}_{ij}+\tilde{p}_{k}=q_i+q_j+q_k$ with $s=Q^2$.
Splittings from FF dipoles are conveniently parametrized by the splitting
variables $\FFz$ and $\FFy$ which are defined in terms of
the physical momenta as
\begin{subequations}
\begin{IEEEeqnarray}{C}
   \FFz = \frac{q_i \cdot q_k}{ (q_i + q_j) \cdot q_k }\,,
   \label{eq:FFKin:z}
   \\
   \FFy = \frac{q_i \cdot q_j}{q_i \cdot q_j + q_i \cdot q_k + q_j \cdot q_k}\,.
   \label{eq:FFKin:y}
\end{IEEEeqnarray}
\end{subequations}
A fully consistent mapping
from $\tilde{p}_{ij},\tilde{p}_k \to q_{i,j,k}$ written in terms of $\FFz$ and $\FFy$ is presented in Appendix~\ref{app:FFKin:altForm}, however we do not consider it further here.
This is because, while this mapping and the corresponding mapping from $q_{i,j,k}\to \tilde{p}_{ij},\tilde{p}_k$ defined in
Ref.~\cite{Catani:2002hc} are formulated for arbitrary particle masses, the identification of the physical degrees of freedom relevant in the quasi-collinear limit \cite{Catani:2000ef} is not directly obvious. 

For a massless spectator the relevant direction can be directly identified,
however for a massive spectator we first need to map both of the massive
dipole momenta prior to emission into light-like momenta $n_{ij}$ and $n_k$,
which in general have $n_{ij}+n_k\neq Q$. We therefore define
\begin{equation}
  (n_{ij}+n_k)^2= 2 n_{ij}\cdot n_k \equiv \FFinv \ .
\end{equation}
We can write these light-like vectors in terms of the emitter and spectator momenta as
\begin{subequations}
\begin{IEEEeqnarray}{rCl}
      n_{ij} & = & \frac{\FFinv^2}{\FFinv^2- m_{ij}^2 m_k^2} \left( \tilde{p}_{ij} - \frac{m_{ij}^2}{\FFinv}\tilde{p}_k \right),
      \\
      n_k & = & \frac{\FFinv^2}{\FFinv^2- m_{ij}^2 m_k^2} \left( \tilde{p}_{k} - \frac{m_{k}^2}{\FFinv}\tilde{p}_{ij} \right),
\end{IEEEeqnarray}
\end{subequations}
which gives
\begin{multline}
   \FFinv = 2n_{ij}\cdot n_k =\\
 \frac{1}{2} \left(s-m_{ij}^2-m_k^2 + \sqrt{(s-m_{ij}^2-m_k^2)^2-4 m_{ij}^2 m_k^2}\right).
\end{multline}
The scaled emitter and spectator momenta can be\linebreak parametrized as
\begin{IEEEeqnarray}{rCl}
	q_{ij} & = & x_{ij} n_{ij} + \frac{m_{ij}^2}{x_{ij}\FFinv} n_k\,,
	\\\nonumber
    q_{k} & = & x_k n_{k} + \frac{m_{k}^2}{x_k \FFinv} n_{ij}\,.
\end{IEEEeqnarray}
The emitter and spectator momenta relevant in the quasi-collinear limit for
the definition of $z$ and $p_\perp$ are expressed as
\begin{IEEEeqnarray}{rCl}
      q_i & = & z q_{ij} + \frac{m_i^2-{z}^2 m_{ij}^2 - k_\perp^2 }{x_{ij} \FFinv z} n_k + k_\perp\,,
      \label{eq:FFKin:qi}
      \\\nonumber
      q_j & = & (1-z)\ q_{ij} + \frac{m_j^2-(1-z)^2 m_{ij}^2 - k_\perp^2 }{x_{ij} \FFinv (1-z)} n_k - k_\perp\,.
      \label{eq:FFKin:qj}
\end{IEEEeqnarray}
Notice that the limit $m_k\to 0$ smoothly reproduces the parametrization where
one works with a light-like collinear direction along the spectator.
Comparison to Eq.~\eqref{eq:qjforward} allows us to identify the physical
branching variables $p_\perp$ and $z$, which relate to the propagator involved
in the splitting as
\begin{IEEEeqnarray}{rCl}
\frac{1}{z(1-z) }\left( p_\perp^2 + z m_j^2 + (1-z) m_i^2 - z(1-z) m_{ij}^2  \right) &=& \nonumber  \\
   \left[ (q_i+q_j)^2 -m_{ij}^2 \right].
\end{IEEEeqnarray}

The remaining details of this formulation, including expressions for the scaling variables
$x_{ij}$ and $x_k$ and expressions for $\FFz$ and $\FFy$ in terms of the variables $p_\perp$ and $z$,
are provided in Appendix~\ref{app:FFKin}.
A formulation similar to that presented here is described in Ref.~\cite{Hoeche:2009xc},
however it differs in the definition of the momenta of the splitting products and the variables used.

The probability for a single branching to occur from a final-final dipole is
\begin{equation}
  \mathrm{d}\mathcal{P}_{\mathrm{branching}} =
  \frac{1}{(q_i + q_j)^2 - m_{ij}^2} \langle V_{i j, k} \left(\FFz,\FFy\right) \rangle
  \mathrm{d} q_j \ ,
  \label{eq:FFKin:branchProb}
\end{equation}
where $\langle V_{i j, k} \left(\FFz,\FFy\right) \rangle$ is the spin-averaged
dipole splitting kernel used to 
describe the branching of a final-state emitter into partons $i$ and $j$ with final-state
spectator, $k$. The single-particle emission phase space, discussed in more detail
in Appendix~\ref{app:FFKin}, is denoted by $\mathrm{d}q_j$.

Finally, we show that this formulation of the splitting momenta is consistent
with the definitions of the kernels and requirements
in Ref.~\cite{Catani:2002hc}. Following the splitting there are three momenta that
must be determined, $(q_i,q_j,q_k)$, with no considerations this system
contains twelve degrees-of-freedom. Given that we know the identity and
therefore the mass of each parton, we can immediately remove three
degrees-of-freedom. We are now left with nine degrees-of-freedom, namely the
energy, $E_n$, polar angle, $\theta_n$, and azimuthal angle, $\phi_n$, for
each parton $n$, {\it i.e.}
\begin{IEEEeqnarray}{rCl}
	q_i & : & \left\{ E_i, \theta_i, \phi_i \right\},
	\nonumber
	\\
	q_j & : & \left\{ E_j, \theta_j, \phi_j \right\},
	\\
	q_k & : & \left\{ E_k, \theta_k, \phi_k \right\} \ .
	\nonumber
\end{IEEEeqnarray}
We choose to work in the rest frame of the dipole with $\tilde{p}_{ij}$ along
the positive z-axis. Implicitly $\tilde{p}_{k}$ must lie along the negative
z-axis and the mapping from $q_{i,j,k}\to \tilde{p}_{ij},\tilde{p}_k$ defined in
Ref.~\cite{Catani:2002hc} requires that, in this frame, the spectator only absorbs
longitudinal momentum in the splitting. Therefore $\theta_k = \phi_k = 0$
which eliminates two degrees-of-freedom. Furthermore we require that the
momentum $Q$ is conserved in the splitting which eliminates a further four
degrees of freedom. Finally we generate the azimuthal angle of the splitting
$\phi = \phi_i = -\phi_j$, where the second equality follows from momentum
conservation, according to a uniform distribution. We are now left with two
degrees-of-freedom.

It is important to note that the above constraints on the degrees-of-freedom
follow from the requirement of momentum conservation in the splitting and the
requirements in Ref.~\cite{Catani:2002hc}. We have also chosen to
simplify the picture by working in a convenient frame which additionally
defines the meaning of the azimuthal angle $\phi$. Therefore, given $\phi$, the
momenta following the splitting must be fully constrained by two independent
variables. Hence for a given $\FFz$ and $\FFy$ the momenta are fully
constrained. Therefore regardless of the variables we generate and the explicit
covariant expressions that we use, so long as $\FFz$ and $\FFy$ can be uniquely
expressed in terms of the generated variables, the splitting momenta are
uniquely defined. Importantly, we can use the splitting kernels and phase-space
limits given in Ref.~\cite{Catani:2002hc} with our covariant formulation of the
splitting kinematics.

\subsubsection{Final-Initial Dipoles}
\label{sec:FIKin}

As the spectator in a FI dipole is necessarily massless, one can use
the standard quasi-collinear parametrization of the kinematics to
describe splittings from massive FI dipoles. In order to be consistent
with the formulation used to describe splittings from IF dipoles,
Section~\ref{sec:IFKin}, we instead choose to provide a parametrization in
terms of the dipole splitting variables.
The four-momenta of the spectator and emitter prior to the splitting are $\tilde{p}_b$ and
$\tilde{p}_{ij}$, respectively.
The four-momenta of the spectator, emitter and emission following
the splitting are $q_b$,
$q_i$ and $q_j$, respectively. The mass of the emitter prior to
the splitting and the masses of the emitter and emitted partons
following the splitting are $m_{ij}$, $m_i$ and $m_j$, respectively.

Splittings from FI dipoles are parametrized by the splitting variables $\FIz$
and $\FIx$ which are defined in terms of the physical momenta as
\begin{IEEEeqnarray}{C}
  \FIz = \frac{q_i \cdot q_b}{(q_i + q_j)\cdot q_b},\,
  \\
\!\!\!  \FIx = \frac{(q_i +q_j)\cdot q_b - q_i \cdot q_j + \frac{1}{2} \left( m_{ij}^2 - m_i^2 - m_j^2 \right)} {(q_i + q_j)\cdot q_b}.
\end{IEEEeqnarray}
As the spectator is incoming and therefore massless, $\FIz$ is identical to the generated variable $z$. We define the conserved momentum transfer
\begin{equation}
  Q = \tilde{p}_{ij} - \tilde{p}_{b} = q_i + q_j - q_b\,,
\end{equation}
and for convenience the invariant
\begin{equation}
  \FIinv = 2 \tilde{p}_{ij} \cdot \tilde{p}_b\,.
\end{equation}

The momenta prior to the splitting are written in terms of the momenta following the splitting as
\begin{IEEEeqnarray}{rCl}
  \tilde{p}_b & = & \FIx \, q_b \ ,
  \\
  \tilde{p}_{ij} & = & q_i + q_j - (1-\FIx)q_b \ .
\end{IEEEeqnarray}
These expressions are satisfied by writing the momenta following the splitting
as
\begin{subequations}
\begin{IEEEeqnarray}{rCl}
  q_i & = & \FIz\tilde{p}_{ij} +k_\perp + \left[ (1-\FIz)\left(\frac{1-\FIx}{\FIx}\right)
    \right.\nonumber\\
    &&\ \ \ \ \left.   + \frac{1}{\FIinv} \left( m_i^2 - m_j^2 + (1-2\FIz)m_{ij}^2 \right) \right] \tilde{p}_b\,,
  \\
  q_j & = & (1-\FIz)\tilde{p}_{ij} - k_\perp + \left[ \FIz\left(\frac{1-\FIx}{\FIx}\right)
    \right.\nonumber\\
    &&\ \left.  + \frac{1}{\FIinv} \left( -m_i^2 + m_j^2 - (1-2\FIz)m_{ij}^2 \right) \right]\tilde{p}_b\,,
  \\
  q_b & = & \frac{1}{\FIx} \tilde{p}_b\,.
\end{IEEEeqnarray}
\end{subequations}
We obtain an expression for the splitting variable $\FIx$ in terms of
the generated variables $p_\perp$ and $z$ by comparison with
Eq.~\eqref{eq:qjforward}, giving
\begin{equation}
  \FIx = \left[ 1 + \frac{p_\perp^2 + (1-z) m_i^2 + z m_j^2 - z(1-z) m_{ij}^2}{ \FIinv z(1-z)} \right]^{-1}.
\end{equation}

The probability for a single branching to occur from a FI dipole is given by
\begin{IEEEeqnarray}{rCl}
  \mathrm{d}\mathcal{P}_{\mathrm{branching}} &=& \frac{1}{(q_i+q_j)^2 - m_{ij}^2}
  \frac{1}{\FIx} \frac{f_b(x_s/\FIx)}{f_b(x_s)}\nonumber\\
&&  \times \langle V_{i j}^b \left( \FIz, \FIx \right) \rangle
  \mathrm{d} q_j\,,
  \label{eq:FIKin:branchProb}
\end{IEEEeqnarray}
where $\langle V_{i j}^b \left( \FIz, \FIx \right) \rangle$  is the spin-averaged
dipole splitting kernel used to 
describe the branching of a final-state emitter into the
partons $i$ and $j$ with an initial-state spectator,~$b$.
The parton density function of the incoming spectator is $f_b(x)$
and $x_s$ is the proton momentum fraction carried by the spectator prior
to the splitting, and
$\mathrm{d}q_j$ denotes the single-particle emission phase space.
A detailed description of the emission phase space is given in Appendix~\ref{app:FIKin}.

\subsubsection{Initial-Final Dipoles}
\label{sec:IFKin}

The momenta of the incoming emitter and outgoing spectator prior to the splitting are $\tilde{p}_{aj}$ and $\tilde{p}_k$, respectively.
The new emitter following the splitting is defined to be the parton incoming from the proton while the emitted particle
is the emitted final-state parton.
The momenta of the emitter, emitted particle and spectator following the splitting are $q_a$, $q_j$ and $q_k$, respectively.
The mass of the spectator is $m_k$.

Splittings from IF dipoles are parametrized by the splitting variables $\IFx$ and $\IFu$ which
are defined in terms of the physical momenta as
\begin{IEEEeqnarray}{rCl}
	\IFx &=& \frac{q_a \cdot q_j + q_a \cdot q_k - q_j \cdot q_k}{(q_j + q_k) \cdot q_a}\,,
	\\
	\IFu & = & \frac{q_a \cdot q_j} {(q_j + q_k) \cdot q_a}\,.
\end{IEEEeqnarray}
We define the conserved momentum transfer
\begin{equation}
	Q =\tilde{p}_k -\tilde{p}_{aj} = q_j + q_k - q_a\,,
\end{equation}
and the invariant
\begin{equation}
	\IFinv = 2 \tilde{p}_{aj} \cdot \tilde{p}_k\,.
\end{equation}

The momenta prior to the splitting are written in terms of the momenta following the splitting as
\begin{subequations}
\begin{IEEEeqnarray}{rCl}
    \tilde{p}_{aj} & = & \IFx \, q_a\,,
    \\
    \tilde{p}_k & = & q_j + q_k - (1-\IFx)q_a\,.
\end{IEEEeqnarray}
\end{subequations}
These expressions are satisfied by writing the momenta following the splitting
as
\begin{subequations}
\begin{IEEEeqnarray}{rCl}
	q_a & = &  \frac{1}{\IFx} \tilde{p}_{aj}\,,
	\\
	q_j & = & \left[ \left( \frac{1-\IFx}{\IFx} \right) (1-\IFu) - \IFu \frac{2 m_k^2}{\IFinv} \right] \tilde{p}_{aj}\nonumber\\
        &&{}+ \IFu \tilde{p}_k - k_\perp\,,
	\label{eq:IFKin:qj}
	\\
	q_k & = & \left[ \left( \frac{1-\IFx}{\IFx} \right)\IFu + \IFu \frac{2 m_k^2}{\IFinv} \right] \tilde{p}_{aj}\nonumber\\
        && {}+ (1-\IFu) \tilde{p}_k + k_\perp\,.
\end{IEEEeqnarray}
\end{subequations}
We need to write the splitting variables in terms of the variables generated in the parton shower, $p_\perp$ and $z$.
We set \mbox{$n = \tilde{p}_k - (m_k^2/\IFinv) \tilde{p}_{aj}$} in Eq.~\eqref{eq:qjbackward} and
equate this to Eq.~\eqref{eq:IFKin:qj} giving
\begin{subequations}
\begin{IEEEeqnarray}{rCl}
	\IFx & = & \frac{\IFinv}{2r(\IFinv-m_k^2)}(1-z+r) 
	\\
	&& \times \left[ 1 - \sqrt{1 - \frac{4r(\IFinv-m_k^2)}{\IFinv}\frac{z(1-z)}{(1-z+r)^2}} \right]\,,
	\nonumber
	\\
	\IFu & = & \IFx \left( \frac{r}{1-z} \right)\,,
\end{IEEEeqnarray}
\end{subequations}
where we have defined $r = p_\perp^2 / \IFinv$. These expressions again relate
the backward-evolution, dipole picture recoil to the quantities involved in
the physical forward branching process, Eq.~\ref{eq:qjbackward}.

The probability for a single branching to occur from an IF dipole is
\begin{IEEEeqnarray}{rCl}
  \mathrm{d}\mathcal{P}_{\mathrm{branching}} &=&
  \frac{1}{2 q_j \cdot q_a} \frac{1}{\IFx}   \frac{f_a(x_e/\IFx)}{\tilde{f}_{aj}(x_e)}\nonumber\\
  &&
  {}\times \langle V^{a j}_k \left( \IFu, \IFx \right) \rangle
  \mathrm{d} q_j\,,
  \label{eq:IFKin:branchProb}
\end{IEEEeqnarray}
where $\langle V^{a j}_k \left( \IFu, \IFx \right) \rangle$  is the spin-averaged dipole splitting
kernel used to describe the branching of an initial-state emitter $\widetilde{aj}$ into an
initial-state parton $a$ and a final-state parton $j$ with a final-state spectator $k$.
The parton density function of the incoming partons $\widetilde{aj}$ and $a$ are
$\tilde{f}_{aj}(x)$ and $f_a(x)$, respectively.
The proton-momentum fraction carried by the
parton $\widetilde{aj}$ is $x_e$ and $\mathrm{d}q_j$ denotes the single-particle emission phase space.
A detailed description of the emission phase space is given in Appendix~\ref{app:IFKin}.

\section{Radiation in the Decays of Heavy Quarks}
\label{sec:DecayRadiation}

In both  \Herwig\ parton showers the production and decay processes are showered
independently, following a factorized approach. In the case of top quark pair
production, the hard process, {\it e.g.} $pp \to t\bar{t}$, is first evolved
down to the IR cutoff $\mu_\mathrm{IR} \approx 1 \text{GeV}$, as described in
Section~\ref{sec:ProductionRadiation}. This involves radiation from both
the initial- and final-state partons, including the top quarks.
When simulating predictions with unstable top quarks, these then
undergo a perturbative decay, and further shower evolution from the
decaying system, and possible further decay products, {\it e.g.} those originating
form a hadronic $W$ decay. The hard scale relevant for emissions from the
decaying top quark is the mass of the top quark, and the evolution will
preserve its four-momentum including the virtuality.

Matchbox is currently limited to generating hard processes with
on-shell outgoing particles, because in the factorized approach a smearing of
the mass with some input distribution consistently is only possible at leading
order~(LO), and poses major challenges at next-to-leading order unless one resorts
to a complete off-shell calculation, which can in principle be handled by the
framework. While the angular-ordered shower can handle off-shell coloured particles, the
dipole shower can currently only deal with on-shell coloured particles, such
that we do not consider a reconstructed resonance hierarchy from a full
calculation as an input to the showers. This also implies that in the hard
process the top width is set to zero, as we could otherwise not treat it as an
on-shell particle at the level of the hard process.

In \Herwig\ by default top quarks, $t$, are decayed according
to the 3-body matrix element to a bottom quark, $b$, and
two fermions, $f$ and $\bar{f^\prime}$, via  an intermediate W-boson
in order that off-shell effects are included for the
W-boson.
The decay system is then showered as described in
Section~\ref{sec:AODecayShower} for the angular-ordered shower and
Section~\ref{sec:DSDecayShower} for the dipole shower, which presents a new
development which we cover in detail.

In both cases we first shower the
top-bottom-W-boson, $tbW$, system followed by the
W-boson-fermion-antifermion, $Wf\bar{f^\prime}$, system.
In the shower the $tbW$ and $Wf\bar{f^\prime}$ systems are considered
to be colour isolated from each other and the rest of the
process. In this sense each decay system is showered independently
from the rest of the process. 
This pattern
of evolving `down' decay trees, {\it i.e.} from the hard process towards the
final-state particles, is true for all decays in \Herwig\,7.

In both parton showers we have the option of performing
the first emission from the decay system according to the
real-emission matrix element using the
builtin Powheg decay correction~\cite{Richardson:2013nfo} for all SM decay processes,
including both the top quark and $W$-boson decay.
In practice this is sufficient as the NLO virtual corrections only effect
the calculation of the width and not the physical distributions.
This is switched on by default in the dipole shower whereas the angular-ordered
shower uses a matrix-element correction by default.
While the angular-ordered shower also includes QED radiation
this is not currently available in the dipole shower.
However, in the case of SM decays involving no coloured particles,
for example a leptonic W-boson decay,
QED radiation is generated using the SOPHTY implementation
in \Herwig~\cite{Hamilton:2006xz} which is used by default in the dipole shower.

\subsection{Angular Ordered Shower}
\label{sec:AODecayShower}

\begin{figure}
  \centering
  \includegraphics[width=0.5\textwidth]{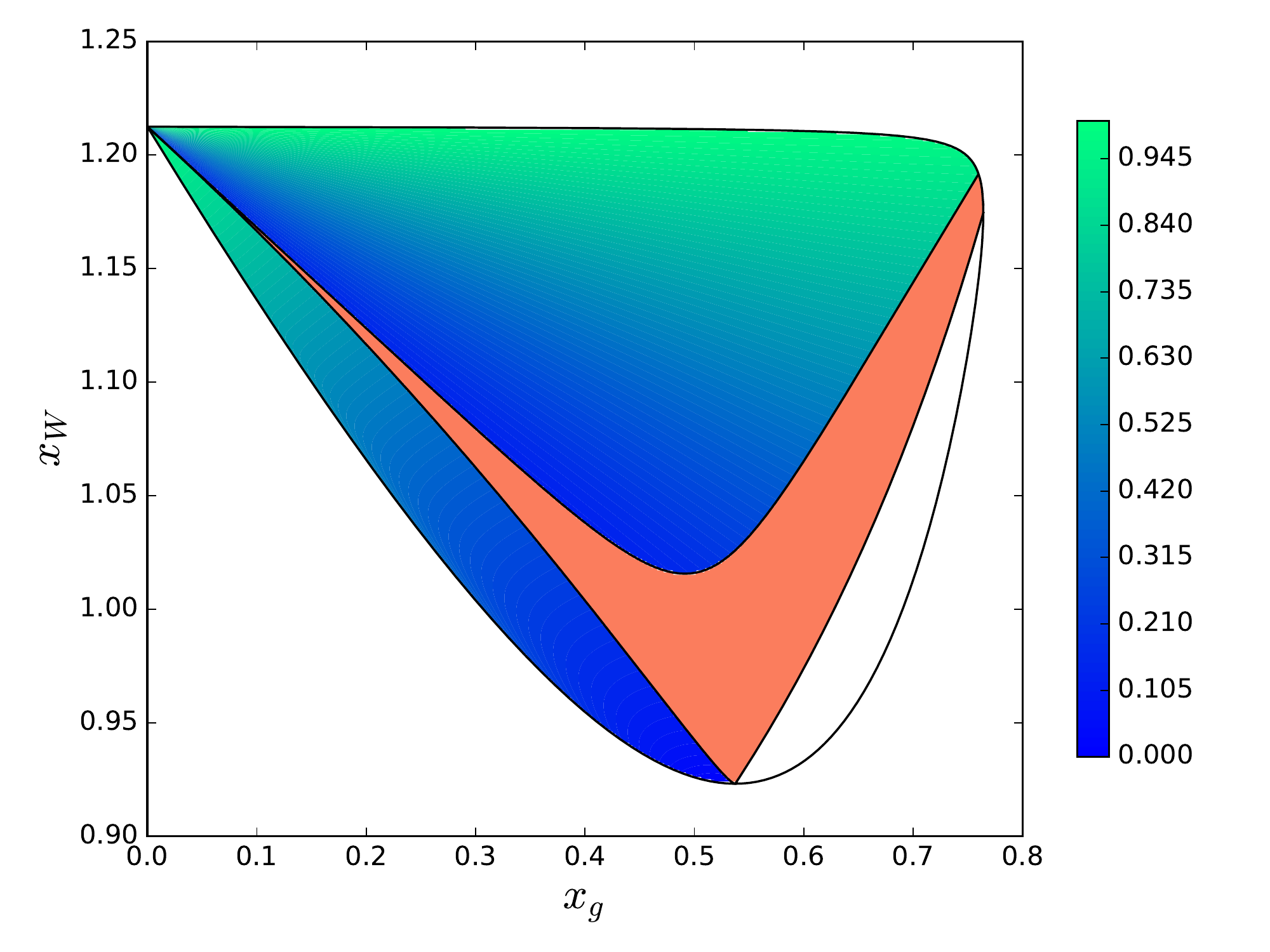}
  \caption{Dalitz plot for $t\to b W^+ g$ where the gluon is
    emitted by the angular-ordered parton shower.
    The energy fractions of the gluon and $W^+$ boson
    are $x_g=2E_g/m_{\rm top}$ and $x_W=2E_W/m_{\rm top}$, respectively.
    In the regions of
    allowed emission in the angular-ordered parton shower the plot shows the
    ratio of the leading-order matrix element result over the parton-shower approximation.
    The red region, the `dead-zone', is not filled by the parton shower while the empty region for large $x_g$
    could be filled by the parton shower, in practice the the shower provides
    a poor approximation in this region and it and the `dead-zone' are filled using a hard matrix-element
    correction.}
  \label{fig:AOdalitz}
\end{figure}

The improved angular-ordered shower used in \Herwig\ proceeds in
much the same way for decays as for hard processes.
The main difference is the handling of radiation with
a coloured decaying particle connected to one of the decay products,
{\it e.g.} $t\to bW^+$.
In order to cover the full soft phase-space region we must have radiation
from both the decaying particle and the decay
product~\cite{Gieseke:2003rz}.\footnote{The original angular-ordered
shower in {\textsf HERWIG 6}~\cite{Corcella:1998rs} did not have radiation from the
decaying particle and therefore did not cover the full soft phase-space region.}
This can be seen in Fig.\,\ref{fig:AOdalitz} where in
order to cover the full phase-space region for soft emission, {\it i.e.} $x_g\to0$,
we need radiation in both the upper region, from $b\to b g$ branchings, and the lower
region from $t\to t g$ branchings. As can be seen in Fig.\,\ref{fig:AOdalitz}
the shower approximation overestimates the leading-order real emission
matrix element over all the filled phase space and the two results agree in
the soft $x_g\to0$ and collinear limit where $x_W$ tens to its maximum value.
The angular-ordered shower has a `dead-zone' where there is no
emission from the parton shower, and a region at large $x_g$ which could in
principle be filled by the parton shower. In this region
the parton shower significantly underestimates the real emission matrix element
and therefore as this region contains to soft or collinear enhancements
we choose not to generate parton shower emissions in it.
As described in detail in Refs.~\cite{Gieseke:2003rz,Bahr:2008pv}
the recoil from any shower emissions in this case is absorbed such that any recoil
perpendicular to the direction of the $W$ boson in the top rest frame is
absorbed by the bottom quark, while the remaining component parallel to the $W$ boson
direction is absorbed by the $W$ boson.

As with \Herwig6~\cite{Corcella:1998rs} in \Herwig7
we apply both a hard matrix-element correction, 
to fill the `dead-zone' and unfilled shower region as well as a soft matrix
element to correct the hardest-so-far emission in the parton shower regions,
this is described in more detail in Ref.\,\cite{Hamilton:2006ms}.
This is the default option, however there is
also an option to apply a 
Powheg correction to the decay~\cite{Richardson:2013nfo}
including the truncated shower.

\subsection{Dipole Shower}
\label{sec:DSDecayShower}

In a top quark decay a dipole is formed by the decaying top quark
and the outgoing bottom quark. During showering the incoming
top quark can also form dipoles with other partons outgoing from the
decay. In the
current implementation of the dipole shower in \Herwig\ we include
emissions from final-initial decay (FI-decay) dipoles only and do not
include initial-final decay (IF-decay) dipoles. In other words we
explicitly consider emissions from outgoing emitter partons only and do
not explicitly include emissions from the incoming top quark. This
choice is justified in Section~\ref{sec:DSDecayKernels}.

The simulation of top quark decays is the primary motivation behind
the new developments outlined in this section, therefore we follow the
example of top quark decays throughout.
These developments have been implemented such that they
are applicable to general decays, including BSM processes. In
particular the new technical developments in the implementation of the
dipole shower, Section~\ref{sec:DSDecayImp}, and the kinematics for
splittings from decay dipoles, Section~\ref{sec:DSDecayKin}, are
independent of the identity of the particles involved.

\subsubsection{Implementation}
\label{sec:DSDecayImp}

In each decay system the colour chains
and dipoles are constructed and updated following each splitting using
exactly the same procedure as for the showering of hard production
processes~\cite{Platzer:2011bc}. The shower starting scale for each
decay system is chosen to be the mass of the incoming decayed
particle.

In the case of a top quark decay, with the default POWHEG correction,
we attempt to produce the first emission from the $tbW$ system using
the exponentiated real-emission matrix element.
Following this corrected real\linebreak emission we
shower the system starting from a scale equal to the transverse
momentum of the emission. In the rare case that there is no POWHEG emission
above the IR cutoff, we do not shower the system.

The $Wf\bar{f^\prime}$ system is a FF dipole, therefore we require no
new kinematics or kernels to shower the system. On-the-other-hand the
top quark decay introduces new complications. The momentum of the top
quark is set, prior to its decay, in the production process and we
must not change its momentum following its decay. Therefore in dipoles
with the top quark as spectator we cannot use the top quark to absorb
the recoil from the splitting. Instead we choose to apply a boost to
the rest of the outgoing particles in the decay system to absorb the
recoil. This is discussed in more detail in
Section~\ref{sec:DSDecayKin}.

The $tbW$ system is showered until no further emission above the IR cutoff can
be generated. This is followed by a `reshuffling' of the momenta of
the particles outgoing from the decay in order to put all partons on
their constituent mass-shell as required for hadronization. In the
case of a decay system we must ensure that the sum of the outgoing
momenta is equal to the four-momentum of the decayed particle. It is
this constraint that we enforce in the reshuffling procedure.

In the case where there are two or more outgoing partons, we simply
rescale the masses and 3-momenta of each parton such that all partons
are put on their constituent mass shell. In the rare case of no
emission from a $tbW$ system we must put the bottom quark on its
constituent mass-shell but without reshuffling the momenta amongst
other partons. In this case we conserve the momentum of the system by
using the W-boson to absorb the momentum change of the bottom quark
while keeping the virtuality of the W-boson unchanged.

Splittings from decay dipoles and the reshuffling procedure can modify
the momentum of the W-boson from the value set in the 3-body decay of
the top-quark. Therefore following the showering of the $tbW$ system
and the subsequent reshuffling, we must apply a boost to the decay
products of the W-boson to ensure that momentum is conserved in the
W-boson decay. This boost is applied prior to showering the
$Wf\bar{f^\prime}$ system. In longer decay trees, following the
showering of each decay, we work down the decay tree updating the
momenta of decay products as appropriate.

\subsubsection{Kinematics}

\label{sec:DSDecayKin}

As a colour partner of the emitter we refer to the incoming top quark
as the `spectator', however we wish to preserve the 4-momentum of the
top quark as its momentum has been set, before its decay, in the
showering of the production process. Therefore the top quark is not
used to absorb the recoil in splittings. Instead the recoil is
absorbed by all outgoing particles from the top decay system, except
for the emitter and the new emission. 

\begin{figure}
  \centering
  \includegraphics[width=0.4\textwidth]{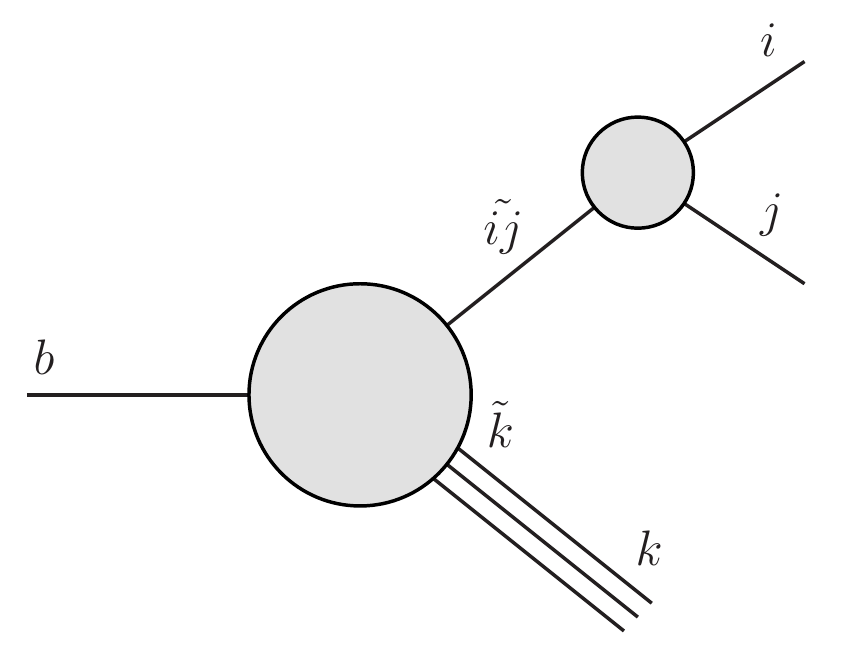}
  \caption{The final-initial decay dipole.}
  \label{fig:FI-Decay-Dip}
\end{figure}

Fig.~\ref{fig:FI-Decay-Dip} shows a diagram of a decay dipole. The momenta of
the incoming decayed parton and the outgoing emitter prior to the
splitting are $q_b$ and $\tilde{p}_{ij}$, respectively. The total
momentum of all other outgoing particles in the decay system is
$\tilde{p}_k$. Following the splitting the momenta of the
new outgoing emitter and emission are $q_i$ and $q_j$,
respectively and the total momentum of all other outgoing particles in
the decay system is $q_k$. It is implicit from our
definition of the recoil system as all particles outgoing from the
decay except the emitter that the incoming parton momentum $q_b$ is
the conserved dipole momentum
\begin{IEEEeqnarray}{rCl}
	Q = q_b  = \tilde{p}_{ij} + \tilde{p}_k 
	 =  q_i + q_j + q_k\,.
\end{IEEEeqnarray}
The splitting kinematics then exactly follow those for a splitting from a
massive final-final dipole given in Section~\ref{sec:DSFFKin}. The
only difference is that for splittings from a decay dipole the recoil,
$\tilde{p}_k \rightarrow q_k$, is absorbed through the application of
an appropriate Lorentz transformation to the recoil system rather than by a single
spectator parton.

\subsubsection{Decay Kernels}
\label{sec:DSDecayKernels}

As stated in Section~\ref{sec:DSDecayShower} we do not include explicit
splittings from IF-decay dipoles. This is because the kernel for a gluon emission
from the incoming top quark contains a negative term proportional to the top
quark mass-squared which gives rise to a kernel that is almost always negative.
We have therefore chosen to include the IF-decay
splitting kernels in the FI-decay splitting kernels which are usually large enough
to remain positive following the inclusion of the negative
mass-squared term. With these considerations there are two possible dipoles and
three possible splittings we must consider: the $t-q$ dipole where the
final state quark emits a gluon and the $t-g$ dipole where the final state gluon
can split into either a $q\bar{q}$-pair or a pair of gluons.

Following the discussion in Section~\ref{sec:DSDecayKin} the notation used to
express the kernels follows that used for splittings from FF dipoles given
in Section~\ref{sec:DSFFKin}. We denote the mass of the incoming decayed parton
as $m_b$.

There is only one possible splitting from the $t-q$ FI-decay dipole, $q \to qg$,
therefore we must include the entire contribution from the corresponding
$t \to tg$ splitting in this kernel. We have used some discretion with regard to
which finite pieces are included in the kernels. The kernel, $V_{q\to qg}$, used
to describe splittings from a $t-q$ FI-decay dipole in \Herwig\,7.1 is given in
Eq~\eqref{eq:DSqtoqgKernel}. Note that following the conventions
of Ref.~\cite{Catani:2002hc} there is a propagator factor of $1/q_i\cdot q_j$ taken
out of the kernel. This is the origin of the factor $\FFy/(1-\FFz(1-\FFy))$ in
front of the $t\rightarrow tg$ piece of the kernel and correctly reproduces the
eikonal formula that would otherwise be obtained by summing over all possible
splittings and configurations for each dipole.

In order to be consistent with the kernels used for splittings from
other massive dipoles, we follow the convention from 
Ref.~\cite{Catani:2002hc} of multiplying certain terms in the kernels by a 
finite ratio of relative velocities. The explicit forms of these terms are
given in Eq.~\eqref{eq:DSvijk}.

The $t-g$ dipole is more complicated because there are two possible
splittings, $g\rightarrow gg$ and $g\rightarrow q\bar{q}$.  The
splitting kernels $V_{g\to gg}$ and $V_{g\to q\bar{q}}$ used to
describe the $g \to gg$ and $g \to q\bar{q}$ splittings in
\Herwig\,7.1 are given in Eq.~\eqref{eq:DSgtoggKernel} and
Eq.~\eqref{eq:DSgtoqqKernel}, respectively.

The limits on $\FFz$, $\FFzpm$, and the
relative velocity term $v_{ij,i}$ required to express these kernels are given 
explicitly in Eq.~\eqref{eq:DSFFzpm} and Eq.~\eqref{eq:DSviji} respectively. We have 
followed the convention
of Ref.~\cite{Catani:2002hc} and used a parameter $\kappa$ to distribute finite
pieces between the two kernels. In \Herwig\,7.1, $\kappa$ is set to zero in all
dipole shower splitting kernels.

We include divergences arising from the IR limits of both $q_i$ and $q_j$
in $V_{g\to gg}$ such that $V_{g\to gg }$ is symmetric with regard to $q_i$ and
$q_j$. This is because this splitting produces indistinguishable
final-state gluons and it is consistent with the other $g \to gg$ kernels used
in the parton shower.\footnote{The $g\to gg$
  splitting can be adjusted to contain only one soft singularity as a
  means of selecting from the two possible colour flows in that
  splitting, if a different option has not been pursued, see also the
  discussions in \cite{Bellm:2018wwz,Platzer:2018pmd}.}

\begin{subequations}
\begin{IEEEeqnarray}{rCr}
  \lefteqn{V_{q->qg}= 8 \pi \alpha_S \mathrm{C}_\mathrm{F}} 
    \label{eq:DSqtoqgKernel}  \\
  &&
    \left \{ \vphantom{ \left[ \frac{\tilde{v}_{ij,k}}{v_{ij,k}} \right]} \right.
    \frac{2 \left( 2 m_i^2 + 2\FFy\bar{s} + \bar{s} \right)}{(1+\FFy)\bar{s} - \FFz(1-\FFy)\bar{s}}
    - \frac{\tilde{v}_{ij,k}}{v_{ij,k}} \left( (1+\FFz) + \frac{2 m_i^2}{y \bar{s}} \right)
    \nonumber
    \\
    &&
     +
    \frac{\FFy}{1-\FFz(1-\FFy)}
    \left[ \frac{2 \left( 2 m_i^2 + 2\FFy\bar{s} + \bar{s} \right)}
      {(1+\FFy)\bar{s} - \FFz(1-\FFy)\bar{s}}
      \right.\nonumber\\ &&\left.\left.
      - \frac{\tilde{v}_{ij,k}}{v_{ij,k}}
      \left( 2 + \frac{2 m_b^2}{\left(1-\FFz(1-\FFy)\right) \bar{s}} \right)
      \right]
    \right \},
    \nonumber\\
    \lefteqn{V_{g->gg} = 16 \pi \alpha_S \mathrm{C}_\mathrm{A} } \label{eq:DSgtoggKernel}\\
      && \left \{ \vphantom{\frac{1+2\FFy}{(1+\FFy)-\FFz(1-\FFy)}} 
   \frac{1+2\FFy}{(1+\FFy)-\FFz(1-\FFy)}  \right.\nonumber\\&&+ \frac{1+2\FFy}{(1+\FFy)-(1-\FFz)(1-\FFy)}\nonumber\\&&\left.
    + \frac{1}{v_{ij,k}} \left[ \FFz(1-\FFz) - (1-\kappa)\FFzplus\FFzminus -2 \right]
    \right \}
    \nonumber \\
  \lefteqn{   +8 \pi \alpha_S \mathrm{C}_\mathrm{F}} \nonumber\\
&&   \left \{ \vphantom{ \left[ \frac{\tilde{v}_{ij,k}}{v_{ij,k}} \right]} \right. 
     \frac{\FFy}{1-\FFz(1-\FFy)}
     \left[ \frac{2(1+2\FFy)}{(1+\FFy)-\FFz(1-\FFy)}
       \right. \nonumber\\&&
       \left.     - \frac{\tilde{v}_{ij,k}}{v_{ij,k}} \left( 2 + \frac{ 2 m_b^2}{(1-\FFz(1-\FFy))\bar{s}} \right) \right]
     \nonumber\\&&
    +
   \frac{\FFy}{1-(1-\FFz)(1-\FFy)}
   \left[ \frac{2(1+2\FFy)}{(1+\FFy)-(1-\FFz)(1-\FFy)}\right.\nonumber\\&&\left.\left.
     - \frac{\tilde{v}_{ij,k}}{v_{ij,k}}
     \left( 2 + \frac{ 2 m_b^2}{(1-(1-\FFz)(1-\FFy))\bar{s}} \right) \right]
   \right \},\nonumber\\
   \lefteqn{V_{g->q \bar{q}} = 8 \pi \alpha_S \mathrm{T}_\mathrm{R}}  \label{eq:DSgtoqqKernel} \\
   && \left \{
   1 - 2\left( \FFz(1-\FFz) - (1-\kappa)\FFzplus\FFzminus -
   \frac{\kappa m_i^2}{2m_i^2 + \bar{s}\FFy} \right) \right \} \nonumber
\end{IEEEeqnarray}
\end{subequations}

\begin{subequations}
\begin{IEEEeqnarray}{Crl}
\tilde{v}_{ij,k} &=& 
 \frac{\sqrt{ \lambda \left(s, m_{ij}^2, m_k^2 \right)} }{s-m_{ij}^2-m_k^2},\label{eq:DSvijk}\\
v_{ij,k} &=&\nonumber \\&& \!\!\!\!\!\!\!\!\!\!\!\!\!\!\!\!\! \!\!\!\!\!\!\!\!\!\!\!
     \frac{ \sqrt{ \left[ 2 m_k^2 
     + \left( s - m_i^2 - m_j^2 - m_k^2 \right) \left( 1 - \FFy \right) \right]^2
     - 4 m_k^2 }}
     { \left( s - m_i^2 - m_j^2 - m_k^2 \right) \left( 1 - \FFy \right)},
      \nonumber \\
    \FFzpm \left( \FFy \right) &=& 
    \frac{ 2 m_i^2 + \left( s - m_i^2 - m_j^2 - m_k^2 \right) \FFy }
    { 2 \left[ m_i^2 + m_j^2 + \left( s - m_i^2 - m_j^2 - m_k^2 \right) \FFy
    \right]} 
    \nonumber\\ &&
    \left( 1 \pm v_{ij,i} v_{ij,k} \right),\label{eq:DSFFzpm}
    \\ 
         v_{ij,i} &=& 
        \frac{ \sqrt{ \left( s - m_i^2 - m_j^2 - m_k^2 \right)^2 \FFy^2 
        - 4 m_i^2 m_j^2 }}
        {\left( s - m_i^2 - m_j^2 - m_k^2 \right) \FFy + 2 m_i^2 }.\nonumber\\
        \label{eq:DSviji}
  \end{IEEEeqnarray}
\end{subequations}

Finally we include a symmetry factor of $\frac12$, which is not written
explicitly here, in front of the $g\rightarrow gg$ pieces of $V_{g\to gg}$.
With the inclusion of this symmetry factor the factors in front of the eikonal
parts from the $g \rightarrow gg$ and $q \rightarrow qg$ pieces are consistent
in the large-$\text{N}_\text{C}$ limit and we reproduce the correct eikonal
expression.

\subsubsection{Validation}

\begin{figure}
  \centering
  \includegraphics[width=0.5\textwidth]{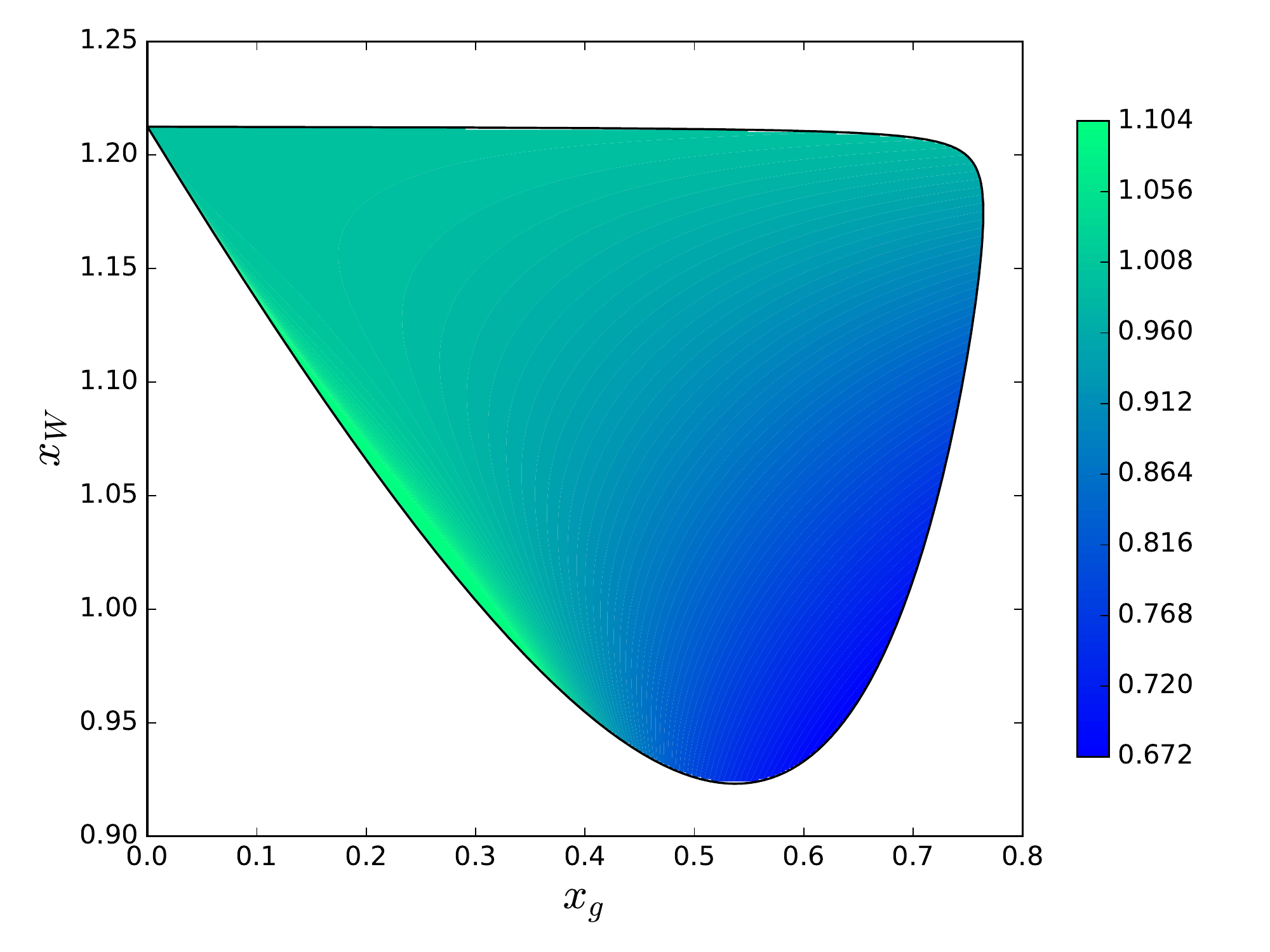}
  \caption{Dalitz plot for $t\to b W^+ g$ where the gluon is
    showing the ratio of the leading-order matrix element result over the dipole-shower approximation.
    The energy fractions of the gluon and $W^+$ boson
    are $x_g=2E_g/m_{\rm top}$ and $x_W=2E_W/m_{\rm top}$, respectively.}
  \label{fig:Dipoledalitz}
\end{figure}
We present results to validate the new decay kernels and kinematics in the
dipole shower. We consider observables which depend
primarily on the first, hardest, emission from the decay system and we compare
results obtained with and without the real emission decay correction. This comparison
directly evaluates how well $V_{q \to qg}$, Eq.~\eqref{eq:DSqtoqgKernel},
reproduces the full real emission correction.
As can be seen in Fig.\,\ref{fig:Dipoledalitz} the kernel overestimates
the leading-order matrix element over most of the phase space, apart from a small
region  near the lower phase-space boundary for $0.1<x_g<0.4$.

Our procedure for the following tests exactly follows that used in
Refs.~\cite{Orr:1996pe,Corcella:1998rs,Hamilton:2006ms}.
We generate $e^+ e^- \to t \bar{t}$ events at LO at a collision energy
of 360 GeV.  This collision energy is chosen to be close to the
threshold energy for the process, {\it i.e.} $2 m_\mathrm{top}$, in order to
reduce radiation from the production process. We work at parton level
and include only dileptonic processes. All final-state quarks and
gluons are clustered into three jets using the $k_\perp$
algorithm~\cite{Catani:1991hj} implemented in
FastJet~\cite{Cacciari:2011ma} and we exclude events containing a jet
with transverse energy less than 10 GeV.  We additionally exclude
events in which the minimum jet separation is less than $\Delta R =
0.7$ where $\Delta R^2 = \Delta\eta^2 + \Delta \phi^2$, where $\eta$
and $\phi$ denote pseudorapidity and azimuthal angle respectively.

We present results for two observables; the separation ${\Delta
  R}_\mathrm{min}$ of the closest pair of jets in the event and the
jet measure $y_3$, defined as the value of the jet resolution
parameter at which the three jet event would be identified as a two
jet event. This is given by
\begin{equation}
	y_3 = \frac{2}{s} {\min}_{ij} \left( \min \left( E_i^2 , E_j^2 \right) (1-\cos\theta_{ij})  \right),
\end{equation}
where $s$ is the centre-of-mass energy squared of the collision, $E_i$ and
$E_j$ are the energy of jets $i$ and $j$ respectively and
$\theta_{ij}$ is the angular separation of jets $i$ and $j$.

\begin{figure}
  \centering
   \includegraphics[width=0.4\textwidth]{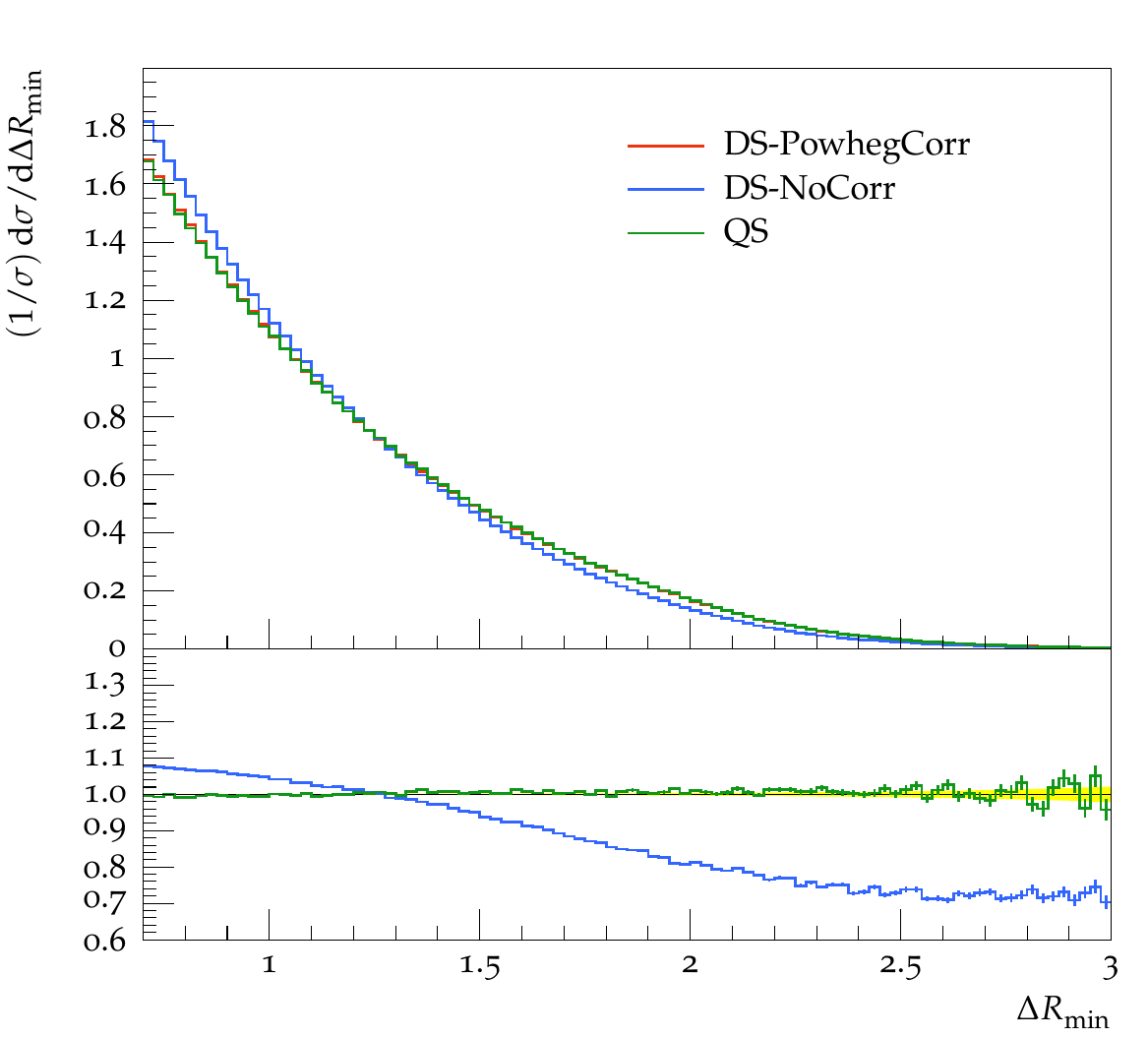}
  \\
  \includegraphics[width=0.4\textwidth]{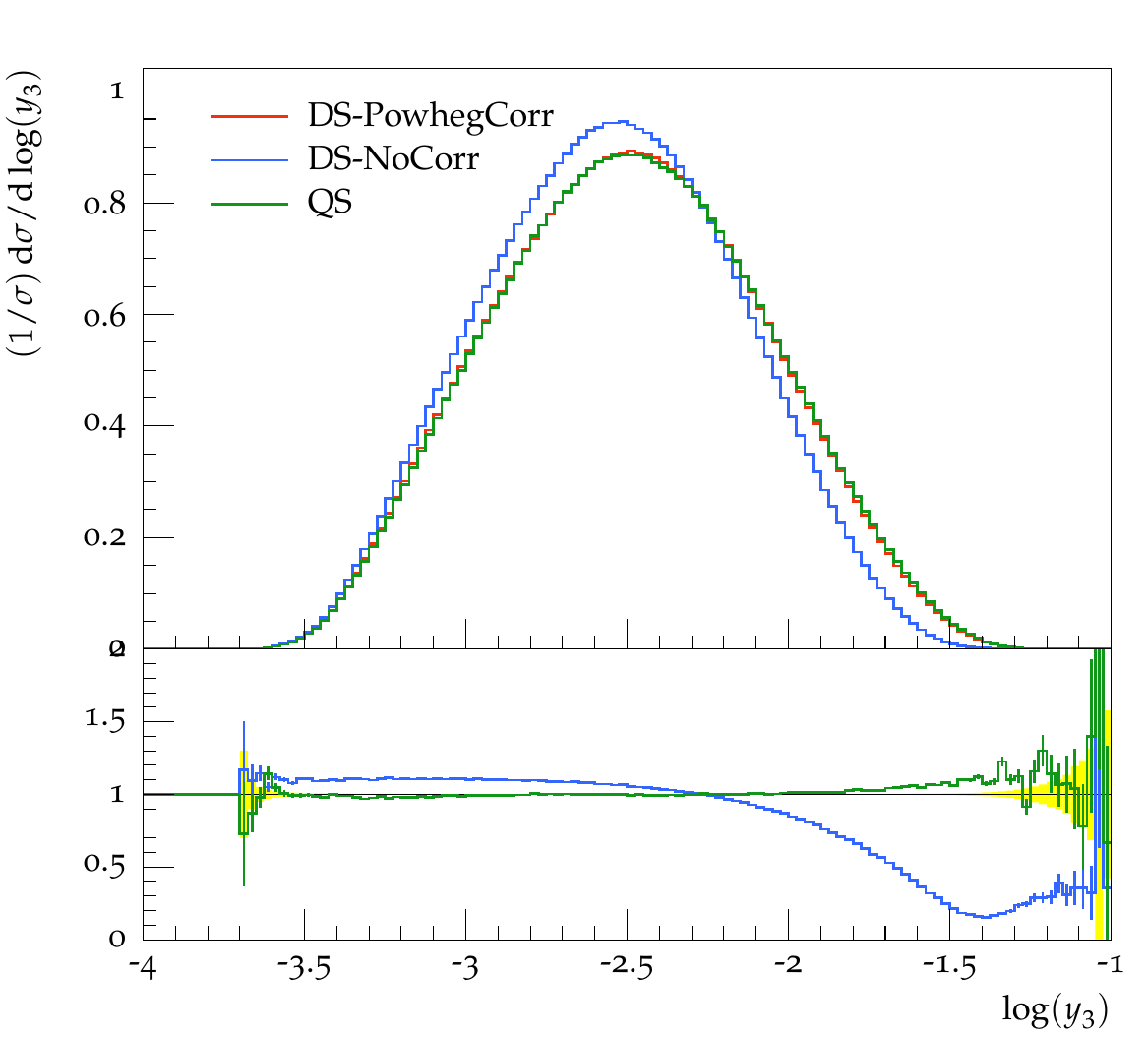}
  \caption{The distribution of (upper) the minimum jet separation and (lower) the
    jet measure $y_3$ in 3-jet $e^+ e^- \to t \bar{t}$ events.
    The distributions are shown for events showered using the dipole shower
    with~(DS-PowhegCorr) and without~(DS-NoCorr) the real emission decay correction.
    In addition we show the
    distributions obtained using the angular-ordered shower~(QS) with the full
    matrix-element decay correction.}
  \label{fig:dipole-shower-decay-tests}
\end{figure}

Fig.~\ref{fig:dipole-shower-decay-tests} shows the distribution of the minimum jet
separation for events showered with and without the real emission decay
correction. In general a harder first emission will produce a greater
separation of the two closest jets.  Therefore, as we expect, the
shower with the real emission decay correction produces more events with a
larger minimum jet separation. We see that the results with and
without the real emission decay correction agree well $(\sim10\%)$ at small jet
separations. Furthermore even at large minimum jet separations, where
we do not expect the splitting kernel alone to give a good description
of the emission, the results agree to within roughly $30\%$.

Fig.~\ref{fig:dipole-shower-decay-tests} also shows the distribution of $y_3$ for events showered
with and without the real emission decay correction. Again, a harder first
emission will in general lead to a larger separation of the two
closest jets and thus such 2-jet events can be resolved into 3-jet
events at a larger value of $y_3$. As we would expect there is a skew
towards larger values of $y_3$ for the results with the real emission-corrected
decay versus the results without the correction. We see that the
results at low $y_3$, corresponding to a softer first emission, are
well described by the shower without the real emission decay correction. The log
scale used for $y_3$ in Fig.~\ref{fig:dipole-shower-decay-tests} emphasises the limitations
of the splitting kernel in describing hard emissions. This is evident
from the increasing disagreement between the results with and without
the real emission decay correction at larger values of $y_3$.

These results show that the kernel $V_{q\to qg}$ behaves well in the IR region
as we require. It also performs reasonably well in the case of harder
emissions but its limitations are apparent in the distribution of $y_3$ in
Fig.~\ref{fig:dipole-shower-decay-tests}. There is a major
limitation to these tests in that they only directly probe the $q \to qg$
splitting kernel.
The effects of subsequent emissions are small and it is difficult to create a
test to probe $g \to gg$ and $g \to q \bar{q}$ emissions from decay dipoles
directly.

As a further comparison we have also included the results from showering with
the angular-ordered shower with the appropriate full matrix element correction
to the decay in both figures. In all except the lowest bins we see a good
agreement between the dipole shower with the real emission decay correction and the angular-ordered
shower. This verifies that the corrections in the two showers produce
the same behaviour, as we would expect. The disagreement in the lower bins is
not a concern as there are numerous differences between the showers and we do
not expect agreement to be exact in all regions of phase space.

\section{NLO Matching and Scale Choices}
\label{sec:Matching}

A major improvement to the simulation of top quark production and decay in the
\Herwig~7 event generator is the inclusion of NLO QCD corrections consistently
combined with the subsequent parton shower evolution. NLO matching paradigms
are typically less ambiguous than their merging counterparts and entirely
driven by solving a matching condition such that the combination of a NLO
cross section with a parton-shower evolution reproduces the NLO cross section
exactly, plus higher-order terms. In the following we will elaborate on the
basic matching algorithms available in \Herwig\,7 and their implementation,
and will consider in detail the sources of uncertainty involved in matched
predictions.

\subsection{Hard Process Setup and NLO Subtraction}

The partonic cross section for the hard process at leading order can be
written as
\begin{equation}
\sigma_{\text{LO}}[u] = \int {\rm d}\sigma_B(\phi_n) {\rm d}f\ u(\phi_n)\;,
\end{equation}
where ${\rm d}\sigma_B$ is the Born cross section, ${\rm d}f$ denotes the
partonic luminosity (parton distribution functions), and $u(\phi_n)$ 
represents a generic observable defined on
the Born phase-space point $\phi_n=\{p_a,p_b\to p_1,...,p_n\}$. The \Herwig\,7
\Matchbox\ module~\cite{Platzer:2011bc} identifies the possible 
subprocesses contributing to the
cross section, and sets up a multi-channel phase-space generator to map the
phase-space measure ${\rm d}\phi_n$, which includes the momentum conserving
$\delta$-function as well as mass-shell constraints.

For a NLO calculation, which we carry out in the dipole subtraction
formalism based on Catani-Seymour dipole subtraction~\cite{Catani:1996vz,Catani:2002hc},
real emission processes including an additional jet are then identified
in the same way as for the leading-order cross section,
and the NLO cross section is calculated as
\begin{eqnarray}
\label{eq:signlo}
\sigma_{\text{NLO}}[u] &=& \sigma_{\text{LO}}[u] +
\sigma_{V+A+C}[u]+\sigma_{R-A}[u]\ ,
\end{eqnarray}
with
\begin{eqnarray}
\sigma_{R-A}[u]
=\int &\Big[& {\rm d}\sigma_R(\phi_{n+1})u(\phi_{n+1})
  \\\nonumber &-&  \sum_i {\rm
    d}\sigma_A^{(i)}(\phi_{n+1} )u(\Phi^{(i)}_{n}(\phi_{n+1})) \,\Big]{\rm d}f \ .
\end{eqnarray}
The first two terms in Eq.~\eqref{eq:signlo} contain the leading-order cross
section, as well as the finite combination \linebreak
\mbox{$\sigma_{V+A+C}=\sigma_{V+I+P+K}$} of virtual corrections, analytically
integrated subtraction terms, as well as collinear counterterms, which are all
defined over the Born phase-space point $\phi_n$ and handled accordingly.  We
have further introduced the dipole subtraction terms ${\rm
  d}\sigma_A(\phi_{n+1})^{(i)}$ and the real-emission contributions ${\rm
  d}\sigma_R(\phi_{n+1})$ which are all functions of the real-emission
phase-space point $\phi_{n+1}$, and the index $i$ runs over the possible
dipole configurations, each of which is associated with a particular kinematic
mapping $\Phi^{(i)}_{n}(\phi_{n+1})$ onto the so-called `tilde' or underlying
Born kinematics. The phase-space mappings trigger phase-space convolutions
which can be cast into phase-space factorizations upon introducing suitably
adapted parton distribution functions
\begin{IEEEeqnarray}{C}
  \label{eq:psfactorization}
\left. {\rm d}\phi_{n+1}{\rm d}f
\right|_{\phi_{n+1}=\Phi^{(i)}_{n+1}(\phi_n,r)} = {\cal J}^{(i)}(\phi_n,r){\rm
  d}\phi_{n}{\rm d}f^{(i)}{\rm d}r\ ,
  \IEEEeqnarraynumspace
\end{IEEEeqnarray}
where $\Phi^{(i)}_{n+1}(\phi_n,r)$ is the inverse mapping to the mapping
$\Phi^{(i)}_{n}(\phi_{n+1})$, and $r$ here refers to the collection of
variables required to describe the additional emission, {\it i.e.} a scale of
the emission, a momentum fraction, and an azimuthal variable. We can also
associate the respective definitions as functions of the real emission
variables, $R^{(i)}(\phi_{n+1})$, such that
\begin{equation}
  \Phi^{(i)}_{n+1}(\Phi^{(i)}_n(\phi_{n+1}),R^{(i)}(\phi_{n+1})) = \phi_{n+1} \ .
\end{equation}
\Matchbox\ uses diagrammatic information to deduce\linebreak which subtraction terms
need to be included, and automatically sets up a cross section in the
form above.

\subsection{Parton-Shower Action and Matching}

The parton-shower action can conveniently be described as
\begin{equation}
  \sigma[u]\to \sigma[{\rm PS}_{\mu_{\text{IR}}}[u]],
\end{equation}
where the parton-shower operator up to the first emission is
\begin{multline}
\label{eq:psop}
  {\rm PS}_{\mu_{\text{IR}}}[u](\phi_n) = \prod_i\Delta^{(i)}(\phi_n,\mu_{\text{IR}})u(\phi_n) +\\
  \sum_i {\rm d}P^{(i)}(\phi_n,r)\kappa(Q^{(i)}(\phi_{n}),p_\perp(r))\theta(q(r)-\mu_{\text{IR}})\ \times\\
  \prod_j\Delta^{(j)}(\phi_n,\mu_{\text{IR}})u(\Phi^{(i)}_{n+1}(\phi_n,r)) \ .
\end{multline}
Here $q(r)$ is the evolution variable which we have singled out only
in the phase-space limits on the evolution, starting at a hard scale
$Q^{(i)}(\phi_{n})$ and ending at the infrared cutoff $\mu_{\text{IR}}$. The
differential splitting probability is the combination of the
respective phase-space factors and a ratio of parton luminosities, and
the Sudakov form factor starting at the hard configuration is
\begin{multline}
\label{eq:sudakov}
  -\ln \Delta^{(i)}(\phi_n,\mu_{\text{IR}}) =\\ \int{\rm
    d}P^{(i)}(\phi_n,r)\kappa(Q^{(i)}(\phi_{n}),p_\perp(r))\theta(q(r)-\mu_{\text{IR}}) \ .
\end{multline}
Notice that the constraint on the hard scale is in general not a sharp cutoff,
but might be imposed in different ways, see~\cite{Bellm:2016rhh} and the discussion
below in Sections~\ref{sec:profilescalechoices} 
and~\ref{sec:hardvetoscales}.  We have, not accidentally, chosen the same kinematic
mapping as has been used for the dipole subtraction terms. Indeed, the
kinematic reconstruction algorithm, and not least the kinematics used in
the dipole shower and the Powheg correction to be discussed below, resemble,
for one emission, exactly the dipole subtraction kinematics, such that we do
not need to consider any additional Jacobian factors.

At this point we can expand the shower action to first order in $\alpha_S$ and
subtract this contribution from the NLO cross section to set up the matched
cross section. To this extent it is worth noting that we can recast both, the
integrand of the Sudakov exponent as well as the emission rate multiplied by
the Born cross section into another approximate cross section using the
inverse of the kinematic mapping,
\begin{multline}
\label{eq:dsigps}
  {\rm d}\sigma^{(i)}_{\text{PS}}(\phi_{n+1}){\rm d}f \equiv\left[ {\rm d}\sigma_B{\rm
    d}f^{(i)} {\rm d}P^{(i)}(\phi_n,r)\right.\ \times \\ \left.\kappa(Q^{(i)}(\phi_{n}),p_\perp(r))\right]_{\phi_{n} =
    \Phi_{n}^{(i)}(\phi_{n+1}),r = R^{(i)}(\phi_{n+1})} \ .
\end{multline}
We have explicitly left out the infrared cutoff in this expression for 
reasons which will soon become clear. The NLO matching subtraction term 
is then
\begin{multline}
  \sigma^{\text{PS}}_{R-A}[u]=\sum_i  \int {\rm d}\sigma^{(i)}_{\text{PS}}(\phi_{n+1}){\rm d}f\ \times
  \\ \theta(q^{(i)}(\phi_{n+1})-\mu_{\text{IR}})\left(u(\phi_{n+1})-u(\Phi_n^{(i)}(\phi_{n+1}))\right)\ ,
\end{multline}
with the shorthand $q^{(i)}(\phi_{n+1})=q(R^{(i)}(\phi_{n+1}))$.
The NLO matched cross section is 
\begin{align}
\label{eq:sigmatched}
\sigma^{\text{matched}}_{\text{NLO}}[u]
=\sigma_{\text{NLO}}[u]
-\sigma^{\text{PS}}_{R-A}[u]\ ,
\end{align}
such that $\sigma^{\text{matched}}_{\text{NLO}}[{\rm
PS}_{\mu_{\text{IR}}}[u]]=\sigma_{\text{NLO}}[u]+\text{h.o.}$
This can be
conveniently combined with the dipole book keeping already employed for the
fixed-order NLO calculation to yield two contributions to the NLO matched cross
section:
\begin{align}
\label{eq:sigmatched2}
\sigma^{\text{matched}}_{\text{NLO}}[u]
= \sigma_{S}[u]+\sigma_{H}[u]\ ,
\end{align}
with
\begin{multline}
\label{eq:sigs}
  \sigma_S[u] = \sigma_{\text{LO}}[u] + \sigma_{V+I+P+K}[u] \\ 
  + \sum_i \int \left(
  {\rm d}\sigma^{(i)}_{\text{PS}}(\phi_{n+1})\theta(q^{(i)}(\phi_{n+1})-\mu_{\text{IR}})
  \right.\\ -\left.
  {\rm d}\sigma^{(i)}_{A}(\phi_{n+1})\right){\rm d}f\ u(\Phi_n^{(i)}(\phi_{n+1}))\ ,
\end{multline}
which constitutes Born-type configurations, also referred 
to as $\clS$ events, as well as
\begin{multline}
\label{eq:sigh}
  \sigma_H[u] = \int \Big(
  {\rm d}\sigma_R(\phi_{n+1}) \ -\\
  \sum_i {\rm d}\sigma^{(i)}_{\text{PS}}(\phi_{n+1})
  \theta(q^{(i)}(\phi_{n+1})-\mu_{\text{IR}})\Big){\rm d}f\ u(\phi_{n+1})\ ,
\end{multline}
to provide real-emission type configurations, also referred
to as $\clH$ events. We stress that these
contributions cannot yet be used to generate events with finite weights owing
to the presence of the infrared cutoff, which allows for configurations with
divergent weights, even if the parton-shower approximated cross section would
be able to reproduce the full singularity structure of the real
emission. Instead, an additional auxiliary cross section
\begin{multline}
  \sigma_{X}[u] = \sum_i  \int {\rm d}\sigma^{(i)}_{X}(\phi_{n+1}){\rm d}f\ \times
  \\ \theta(\mu_{\text{IR}}-q^{(i)}(\phi_{n+1}))\left(u(\Phi_n^{(i)}(\phi_{n+1}))-u(\phi_{n+1})\right),
\end{multline}
can be added to the matched cross section to eventually yield modified
versions of $\sigma_S$ and $\sigma_H$, which can be employed to generate
events. In practice, we use the dipole subtraction terms themselves to
facilitate this, {\it i.e.} \linebreak \mbox{${\rm d}\sigma_X={\rm d}\sigma_A$}.
Note that, for infrared-safe observables $u$, $\sigma_{X}$ only adds
power corrections below the infrared cutoff.

\subsection{Matching Variants}

Both the angular-ordered and the dipole showers fit into the framework
outlined above, which constitutes the subtractive, or MC@NLO-type, 
matching in \Herwig~7, and the sole task is to determine the shower matching
subtraction ${\rm d}\sigma^{\text{PS}}_{R-A}$, which we have implemented in a
process-independent way in the \Matchbox\ module. These subtractions are
indeed very similar to the dipole subtraction terms, but averaged over
azimuthal orientation and for colour correlators evaluated in the large-$N_c$
limit. With the recent development of spin-correlation algorithms in both
shower modules~\cite{Richardson:2018pvo}, spin correlations can be
restored in these subtractions, and full colour correlations can be justified
when using colour matrix-element corrections~\cite{Platzer:2012np,Platzer:2018pmd}, 
at least for the dipole shower algorithm.

Another choice is a multiplicative, or Powheg-type, matching for which we
employ a hardest emission generator, which performs a shower emission
using a modified splitting function, or matrix-element correction, determined
from the real-emission and Born matrix elements as
\begin{multline}
\label{eq:powsplit}
  P^{(i)}(\phi_n,r) \rightarrow \\
  \frac{w^{(i)}(\Phi^{(i)}_{n+1}(\phi_n,r))}{\sum_j
    w^{(j)}(\Phi^{(j)}_{n+1}(\phi_n,r))} \frac{|{\cal
      M}_R(\Phi^{(i)}_{n+1}(\phi_n,r)|^2}{|{\cal M}_B(\phi_n)|^2}\ ,
\end{multline}
for which no complications arise as the full divergent behaviour is
reproduced by construction. An additional truncated, vetoed shower
needs to be included if the hardest emission generated this way is not
the first one to occur. In practice, for the $w^{(i)}$ we use
dipole-type partitioned Eikonal factors to perform the weighting into
the different singular channels $i$ and use the \texttt{ExSample}
library~\cite{Platzer:2011dr} to generate emissions according to the
Sudakov form factor obtained from the matrix-element correction
defined above.

\subsection{Profile Scale Choices}
\label{sec:profilescalechoices}

The parton shower hard scale needs to be limited from above in order to avoid
the summation of an unphysical tower of logarithms in the Sudakov exponent.
To this extent, we have not chosen a fixed starting scale, but a profile scale
function $\kappa(Q^{(i)}(\phi_n),p_\perp(r))$. This function
encodes the possibility that not all of the emission phase space should
be available to the parton shower. From here on we will generically denote
$Q^{(i)}(\phi_n)=\vetoScale$, {\it i.e.} we choose the (upper) hard
scale $Q^{(i)}(\phi_n)$
manifest as a scale $\vetoScale$ which defines an upper limit on the
transverse momentum available to shower emissions.
    
Several possible parametrizations of the profile scale choices were
investigated for leading-order plus parton-shower predictions~\cite{Bellm:2016rhh}. 
We first introduce a hard veto scale $\vetoScale$, which
defines an upper limit on the transverse momentum available to shower
emissions.  By default this is chosen to be the hard process scale, $\hardProcScale$,
which in turn is typically set to the factorization and
renormalization scale, but may also be chosen independently in \Herwig\,7.  The
profile scale choice $\kappa \left(\vetoScale, p_\perp \right)$ is a function
of $\vetoScale$ and the transverse momentum $p_\perp$ of the splitting. 
For convenience, we define the quantity $x$ as the ratio of these scales
\begin{align}
\label{eq:profilescalex}
x=\frac{p_\perp}{\vetoScale}\ .
\end{align}

The default profile scale choice in \Herwig\,7 is the \resummation\ profile
\begin{equation}
\kappa\left(\vetoScale, p_\perp \right)= 
\begin{cases}
  1 &,\ x \leq 1-2\rho\ , \\
  1 - \frac{\left(1-2\rho-x\right)^2}{2\rho^2} &,\ x \in (1-2\rho,\,1-\rho]\ , \\
  \frac{\left(1-x\right)^2}{2\rho^2} &,\ x \in (1-\rho,\,1]\ , \\
  0 &,\ x > 1\ ,
\end{cases}
\end{equation}
where $\rho$ is a parameter which is set in \Herwig\,7.1.4 to $\rho=0.3$. The
\resummation\ profile is defined to be zero above the veto scale,
such that the shower does not populate this region in which it is expected to
perform poorly. Conversely it is equal to one at low scales, where the shower
is expected to perform well.

We compare the \resummation\ profile to the \hfact\ profile, which is the
damping factor used in \powhegbox~\cite{Alioli:2010xd}. The \hfact\ profile
is defined as
\begin{align}
\kappa \left(\vetoScale, p_\perp \right)= 
\frac{1}{1+x^2}\ .
\end{align}
While this function tends to one in the hard emission region, it does not
enforce a cutoff on the shower emission scale as in the \resummation\ profile.
Similarly, the \hfact\ profile tends to zero in the infrared limit but, unlike the
\resummation\ profile, never actually equals zero.

In this study we restrict ourselves to a simple investigation of the effects
of the profile scale choice on the simulation of $t\bar{t}$ production using
MC@NLO-type matching. To do
this we compare results obtained using the two profile scale choices defined
above (see Section~\ref{sec:MCatNLOProfileScaleChoiceResults}).  For
a detailed discussion of the exact properties of the various profile scale
choices available in \Herwig\,7 we refer the reader
to Ref.~\cite{Bellm:2016rhh}\,%
\footnote{
 As pointed out in Ref.~\cite{Bellm:2016rhh} the choice of the 
 profile scale, i.e. how to approach the boundary of hard emissions, 
 is non-trivial and highly relevant in the context of NLO plus 
 parton-shower matching. The choice of the profile scale is 
 essentially constrained by consistency conditions on central 
 predictions (i.e. it should not modify the input distributions of the 
 hard process) and uncertainties (i.e. large uncertainties are 
 expected in unreliable regions or regions where hadronisation effects 
 are dominant, as well as stable results are expected in the Sudakov 
 region). 
 It was found in Ref.~\cite{Bellm:2016rhh} that the \hfact profile does not 
 admit results compatible with these criteria. Instead, using the 
 \resummation profile it was found that the angular-ordered and dipole 
 showers are compatible with each other, both in central predictions 
 and uncertainties (despite their very different nature).
 In addition to studying some of these effects here for top-quark pair
 production, we would like to point out that choosing a profile scale
 reminiscent of the \resummation profile rather than the \hfact 
 profile might also shed some more light on the effects observed in
 Higgs-boson pair production in Ref.~\cite{Heinrich:2017kxx}.
}.

\subsection{Hard Veto Scale Choices in MC@NLO-type Matching}
\label{sec:hardvetoscales}

Both shower modules require an upper limit on the transverse momentum of emissions, which
is set by a hard veto scale (see previous section). 
This hard veto scale coincides with the starting
scale for the $p_\perp$-ordered dipole shower, and is explicitly
implemented as an additional veto for the angular-ordered
shower. By default in \Herwig\,7, in leading-order events, {\it i.e.}
Born-type events, we use $\vetoScale=\hardProcScale$.

For NLO matched predictions, the generated $\clS$ and
$\clH$ events (see previous section) separately undergo showering. While $\clS$
events constitute Born-type events and are treated as such, several complications arise for $\clH$ events.

In MC@NLO-type matching there is no requirement of exact cancellation between the real-emission matrix
element and the subtraction term in any
region of phase space, as it is possible for the subtracted
real-emission cross section
still to contain power corrections in the regions where the real
emission is soft or collinear.  Correspondingly we expect to see
a fraction of $\clH$ events
with a soft and/or collinear emission.
In the case of such an $\clH$ event it is unnatural to choose the hard veto
scale to be of the order of the small transverse momentum of
the real emission. Consider for example our case of $t\bar{t}$ 
production, and say we have an $\clH$ event in which the real emission has a 
transverse momentum of $\sim2$ GeV. Given the high energy scales involved in 
$t\bar{t}$ production, it would be unreasonable to veto all shower emissions with
transverse momentum greater than that of the real emission.
Instead we need to choose a hard veto scale which is
more representative of the scales involved in the process.

In general, as with most scale choices there is no `correct' choice and we
have some freedom in choosing the hard veto scale. By default in \Herwig\,7
we choose $\vetoScale = \hardProcScale$, for which we typically choose
$\hardProcScale=\mu_\mathrm{F}=\mu_\mathrm{R}$ with $\mu_\mathrm{F}$
and $\mu_\mathrm{R}$ denoting the factorization and renormalization scale
respectively. The hard veto scale and the scale of the hard process may also be
chosen independently.
Overall, given our previous discussion, we desire to choose
$\vetoScale$ to be representative of the scales of the objects outgoing from the hard
process. In the case of a hard real emission,
a hard veto scale that reflects the scale of the real emission should be used.
Conversely in the case of a relatively low-scale real emission, a larger scale
should be chosen.

Assume for now that we use $\vetoScale = \hardProcScale$ and consider an $\clH$ event.
Common choices for $\hardProcScale$ involve
the transverse masses of the top quark and antiquark, often in a linear or 
quadratic sum. In the case of a very low-$p_\perp$
real emission, the
transverse masses of the top quarks will be largely unaffected by the emission.
Therefore we would shower such an event from a scale similar to that had there
been no emission. Conversely a high-$p_\perp$ real emission on-average
increases the sum of the transverse masses of the top quarks, and the presence
of the hard real emission is reflected in the hard veto scale.
There are choices for $\hardProcScale$ that, while significantly affected by the scale of
the real emission, are relatively large over a wide range of real emission
scales. If $\hardProcScale$ is large enough, the actual maximum scale for showering will
be the maximum physically allowed scale, determined from the splitting
kinematics, for the first shower emission. In this case, while $\hardProcScale$ may be directly
affected by the scale of the real emission, the scale of the real
emission will have only a small impact on the subsequent showering.

In the case described above one should consider using an alternative
choice for $\vetoScale$. We have introduced such a scale, which we denote as
$\mu_\mathrm{a}$, in \Herwig\,7.1 for use in $t\bar{t}$ production
\begin{equation}
  \mu_\mathrm{a} = 
  \sqrt{ \frac{1}{n_\mathrm{out}} \sum_i m_{\perp,i}^2 } \ ,
  \label{eq:ShowerScale}
\end{equation}
where $n_\mathrm{out}$ is the number of particles outgoing from the
hard process prior to showering and the sum is over these outgoing particles.
This is simply the quadratic mean of the transverse masses of the outgoing 
particles in the lab frame. 
In an $\clH$ event with a hard real emission,
the scale $\mu_\mathrm{a}$ is sensitive to the scale of this real emission.
In the case of an $\clH$ event with a low-$p_\perp$ real emission, 
$\mu_\mathrm{a}$ is much larger than the scale of the real emission and better reflects the scales in the process. We note that this scale is not smooth in the limit of a
soft/collinear emission, {\it i.e.} the 
transition from $\clH$ to $\clS$ events. In the case of an $\clH$ event with a
low-$p_\perp$ real emission this returns a scale smaller than that in
an $\clS$ event by a factor $\sqrt{2/3}$ in the soft/collinear limit.
We expect the effects of this discontinuity on results to be very small.

In the following (see Section~\ref{sec:mcatnlo-veto-scale-prod-level} and Section~\ref{sec:mcatnlo-veto-scale-full-process})
we investigate some of the impacts of the choice of the
hard veto scale on the prediction of observables using MC@NLO-type matching, and how the effects change depending on the choice for the hard process scale.
To do this we compare, for each of three different choices for $\hardProcScale$, results obtained using $\vetoScale = \hardProcScale$ and $\vetoScale = \mu_\mathrm{a}$.
The three choices for $\hardProcScale$ that we compare
are
\begin{subequations}
\begin{IEEEeqnarray}{rCl}
  \mu_1 &=& \frac{m_{\perp,t} +
m_{\perp,\bar{t}}}{2} ,
  \\
  \mu_2 &=& \frac{m_{\perp,t} + m_{\perp,\bar{t}}}{4} ,
  \\
  \mu_3 &=& m_{t\bar{t}} ,
\end{IEEEeqnarray}
\label{eqn:scalechoices}
\end{subequations}
where $m_{t\bar{t}}$ is the invariant mass of the $t\bar{t}$-pair\,%
\footnote{
 We refer the reader to Ref.~\cite{Czakon:2016dgf} for a detailed
 discussion on dynamical scale choices in top-quark pair production.
}.

As always in discussions of scale choices
there is no right or wrong choice.  The aim of this discussion is to highlight
that when we use MC@NLO-type matching we have to make a choice for the hard veto scale.
We will show that, depending on the choice for
$\hardProcScale$, different choices for $\vetoScale$ can have differing and significant
effects on our predictions for observables.

\section{Uncertainty Benchmarks}
\label{sec:UncertaintyBenchmarks}

In order to estimate the uncertainty for the event generator predictions we
pursue both, variations of the scales involved in the hard
production process
as well as  the scales involved in the subsequent parton showering
(see Section~\ref{sec:scale-variations-prod-level}). We also
consider the impact of different profile scale choices (see
Section~\ref{sec:MCatNLOProfileScaleChoiceResults}), and
of different choices for the hard veto scale, depending on the
scale of the hard process (see Section~\ref{sec:mcatnlo-veto-scale-prod-level}).
We consider $t\bar{t}$ pair production in proton-proton (pp)
collisions at a centre-of-mass energy of $13\ {\rm GeV}$ using parton-level
predictions for stable top quarks.

All parton-level simulations use the `benchmark' settings of Ref.~\cite{Bellm:2016rhh}.
Except for the variations of interest in each section, we use identical input settings for the parton showers and matching schemes in every run. Only QCD radiation is included in the simulations
and the same infrared cutoff of $\mu_\mathrm{IR}=1$ GeV
(implemented as minimum transverse momentum cutoff on shower
emissions) is used in both showers.
We use a mass parameter of $m_t=174.2$ GeV in the hard process as well as in the subsequent showering algorithms and all other quarks are considered to be massless.

The factorization and renormalization scales
are set to the same value $\mu_\mathrm{R} = \mu_\mathrm{F} \equiv
\hardProcScale$, where our default for the
central hard process scale choice 
(as used in Sections~\ref{sec:scale-variations-prod-level} 
and~\ref{sec:MCatNLOProfileScaleChoiceResults}) is
\begin{equation}
  \hardProcScale = 
  \frac{m_{\perp,t} + m_{\perp,\bar{t}}}{4} \,,
\label{eq:uncertaintyscalechoice}
\end{equation}
{\it i.e.} half of the average transverse masses of the top and
anti-top quarks, unless stated otherwise. This scale choice is motivated by the results of
Ref.~\cite{Czakon:2016dgf}.
We use the default choice, $\vetoScale = \hardProcScale$, for the hard veto scale in all runs
apart from those in which this is the scale of interest. Similarly, the \resummation profile scale is used in all runs unless otherwise stated.

We use the \texttt{MMHT2014nlo68cl} parton distribution functions (PDFs) along with a 
two-loop running of $\alpha_S$ with $\alpha_S(M_Z)=0.12$
both in the parton shower and the hard process\,%
\footnote{This
  refers to an input value which is not used in conjunction with a CMW
  correction and is only used for the parton-level benchmark settings
  considered here. Typically a tuned value will include the CMW
  correction numerically. Also, note that in \Herwig\,7 we
  perform the running of $\alpha_S$ ourselves rather than using the
  running determined by the PDF set.}.
All runs use a four-flavour scheme.
All cross sections are rescaled to the
NNLO cross section of 815.96\,pb\,%
\footnote{This 
  is the reference cross section calculated by the CMS and ATLAS 
  collaborations.}, 
calculated using Top++2.0\,\cite{Czakon:2011xx} assuming a top mass of
173.2\,GeV and including soft-gluon resummation to next-to-next-to-leading-log order, 
as are the variations we consider and the envelopes resulting from these variations.

We use a purpose-built analysis written in
\Rivet~\cite{Buckley:2010ar} to analyse the simulated events.  Our
analysis considers objects with pseudo-rapidity $|\eta| < 5$, with
transverse momentum ordered jets obtained from the
anti-$k_\perp$ jet algorithm~\cite{Cacciari:2008gp,Cacciari:2011ma}
with a jet radius of $R=0.4$.

\subsection{Scale Variations}
\label{sec:scale-variations-prod-level}

In this section we discuss the parton shower and matching scheme uncertainties
that arise from scale variations. We present results for chosen observables that
probe various aspects of the simulation. We first compare results generated with
LO matrix elements plus parton shower simulations, using
both the angular-ordered (PS) and dipole showers
(DS).
We use LO plus parton-shower results primarily to compare and contrast the two showers in
addition to discussing the uncertainties on the predictions.
This is followed by a discussion of results produced by NLO matrix elements matched to a parton shower, 
{\it i.e.} NLO matched simulations.
In this discussion, in addition to considering the uncertainties, we focus on
the differences between the results obtained using the MC@NLO and Powheg matching
schemes.

Following the approach used in Ref.~\cite{Bellm:2016rhh}, 
we estimate the uncertainty on the
predictions by considering the variations of three scales:
\begin{itemize}
\item the factorization and renormalization scale in
  the hard process, \ie the hard process scale
$\hardProcScale=\mu_\mathrm{R}=\mu_\mathrm{F}$;
\item the boundary on the
  hardness of emissions in the shower, \ie the hard veto
  scale $\vetoScale$;
\item the argument of $\alpha_S$ and the PDFs in the parton shower,
  \ie the shower scale $\showerScale$\,%
\footnote{
 In this study we are concerned only with variations of the 
 arguments of $\alpha_S$ and the PDFs in the parton showers, 
 therefore, even though they can differ, we use the common terminology 
 `shower scale' for these scales.
 In the angular-ordered shower the argument of the 
 strong coupling is related to the transverse 
 momentum of the emitted parton and differs for 
 initial- and final-state evolution, while the 
 argument of the PDFs is simply the ordering 
 variable for initial-state 
 evolution\,\cite{Gieseke:2003rz}.
 In the dipole shower the transverse momentum of 
 the emitted parton is used for both scales.
}.
\end{itemize}
We apply multiplicative factors of 0.5, 1 and 2 to each of
the corresponding central scales such
that the full set of variations consists of 27 different scale combinations.
The complete uncertainty envelope corresponding to this set of variations is
shown in each plot. 
In addition, for each result, we include ratio plots that breakdown the
uncertainties according to the individual scale variations.
For each of the three scales considered we separately plot the envelope produced
by the upward and downward variations of that scale about the central result,
{\it i.e.} only two variations are included for each envelope in addition to the central
result.

\begin{figure*}
  \centering  \includegraphics[width=0.4\textwidth]{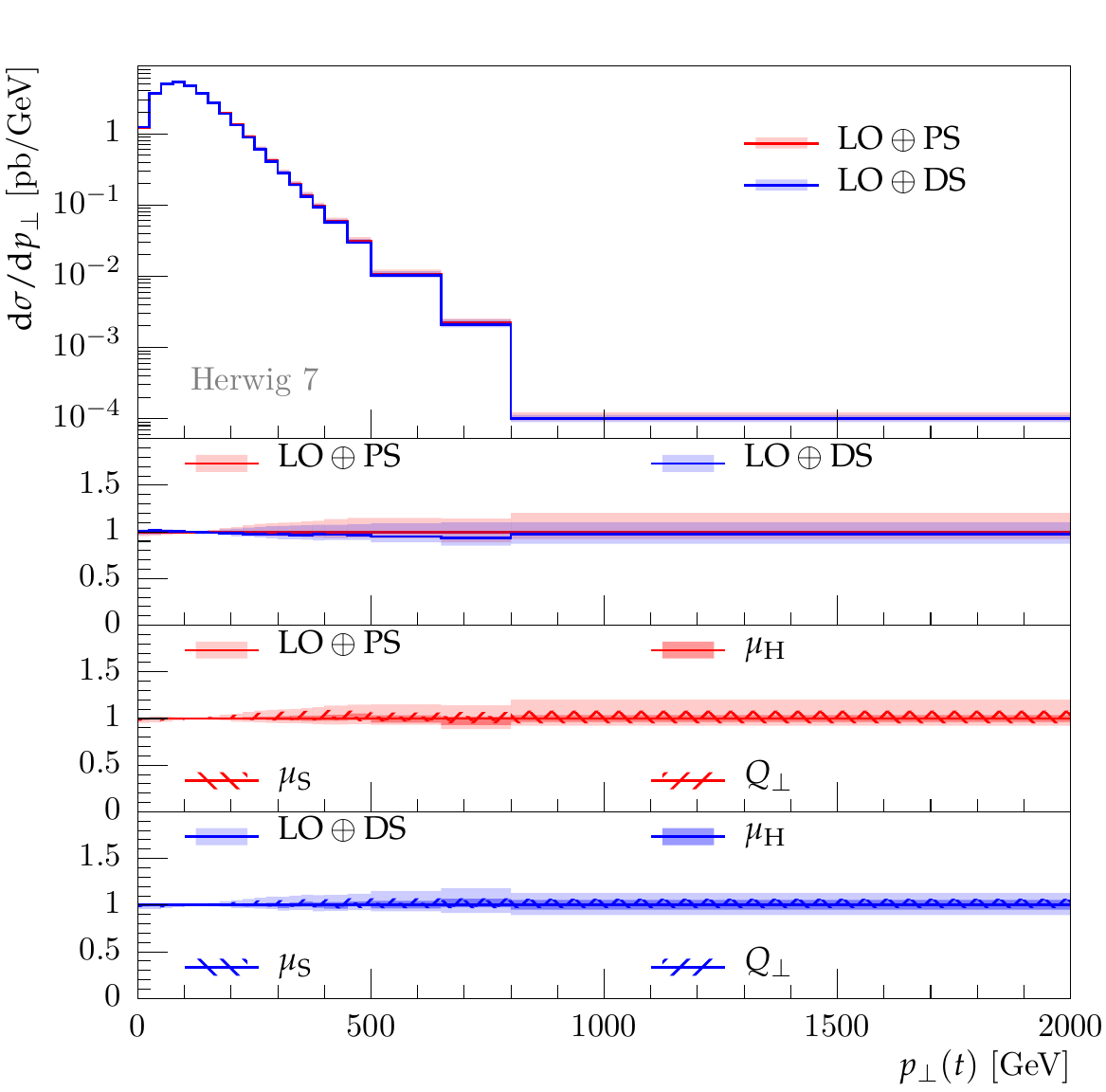}\hspace*{1cm}
  \includegraphics[width=0.4\textwidth]{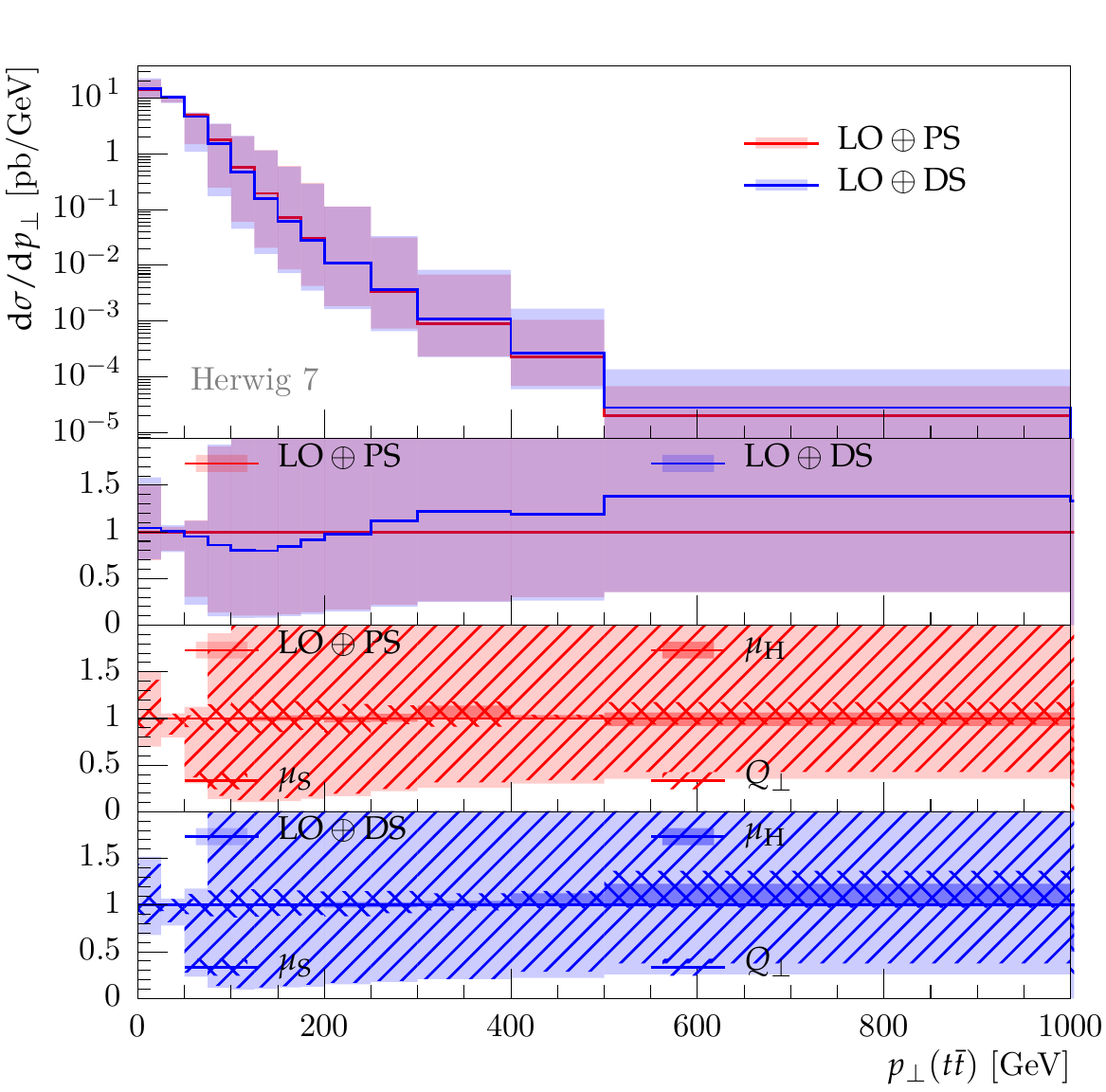}\\
  \includegraphics[width=0.4\textwidth]{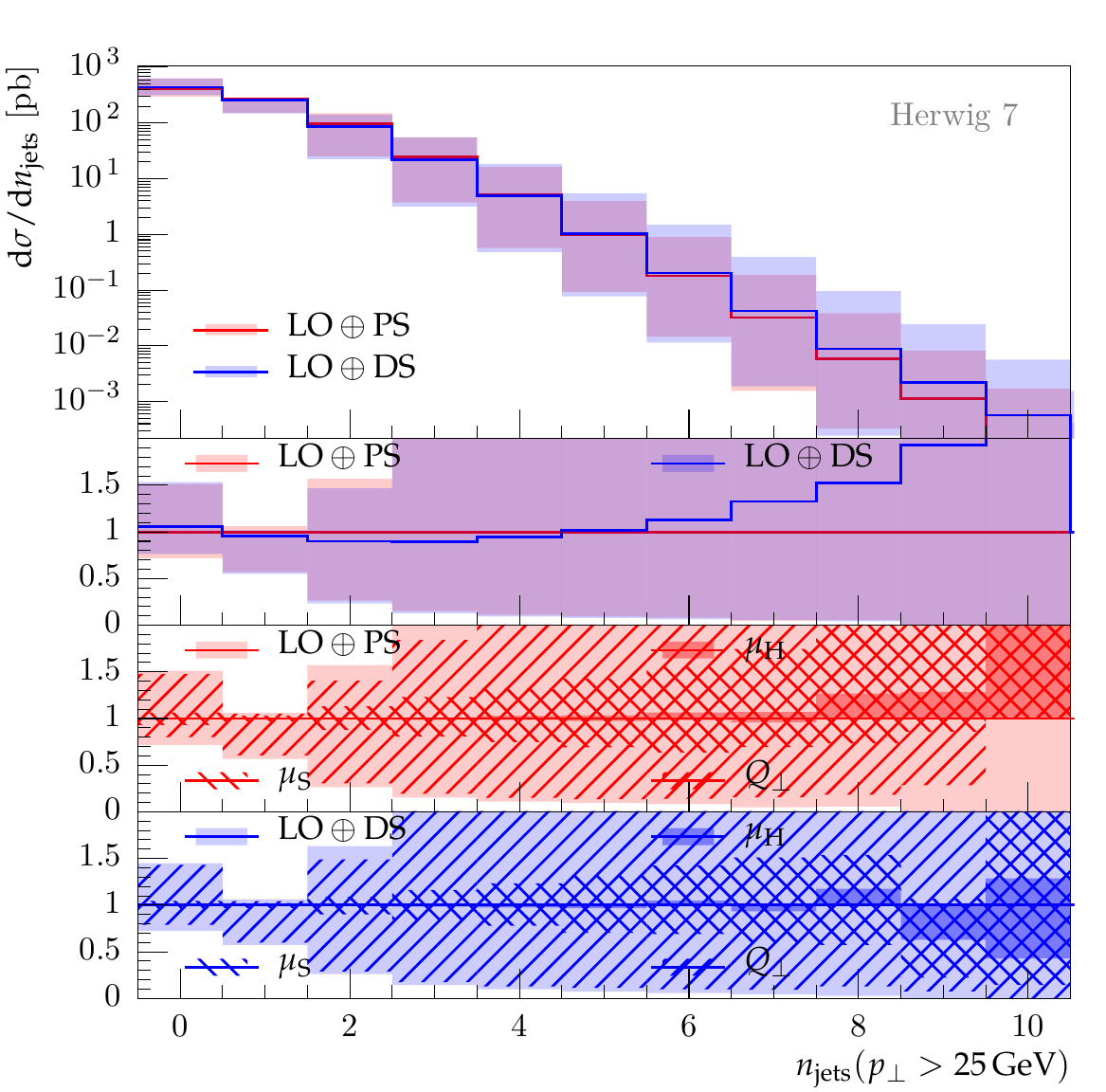}\hspace*{1cm}
  \includegraphics[width=0.4\textwidth]{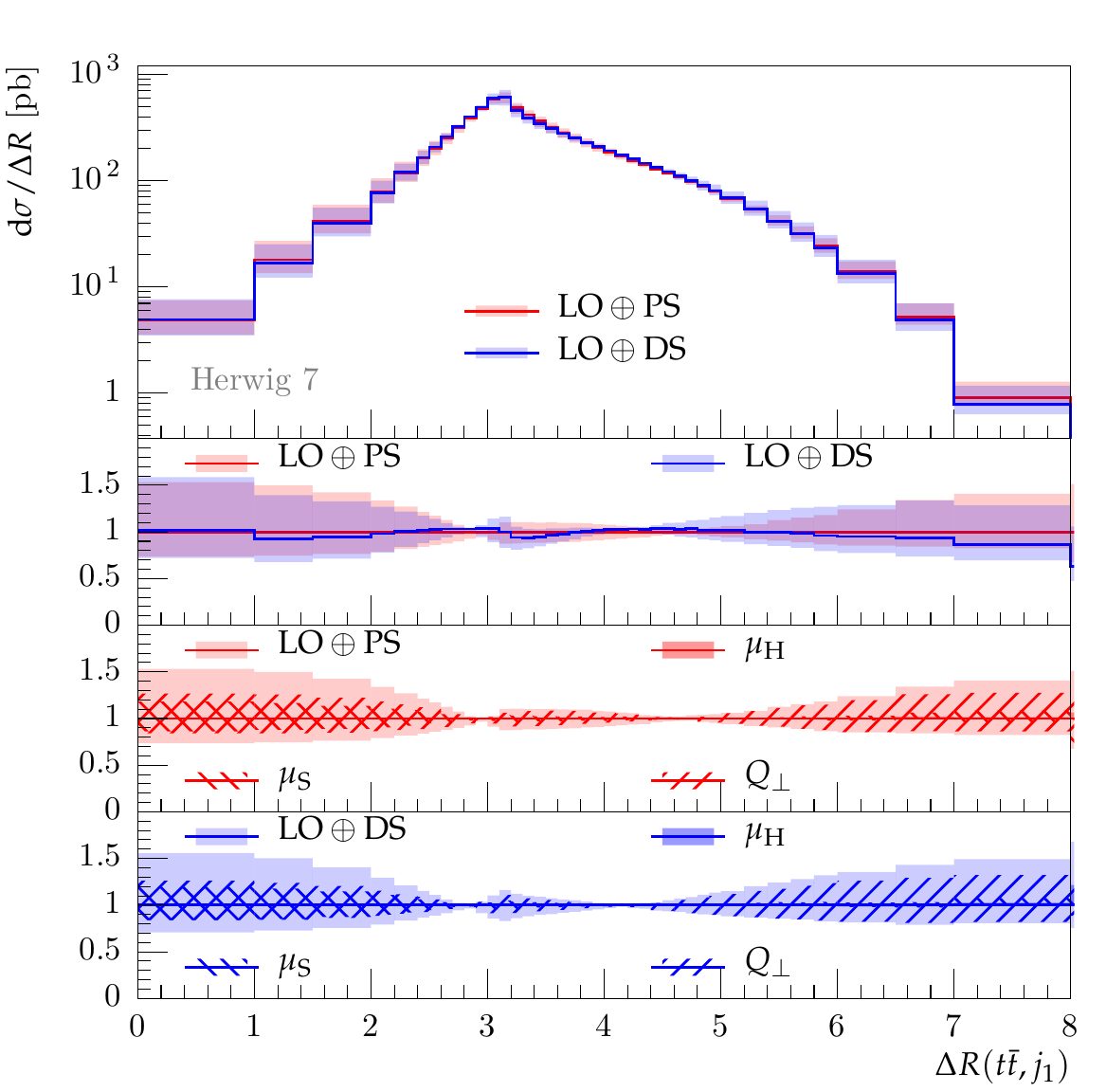}
  \caption{Scale variations for the inclusive top $p_\perp$-spectrum, the top
    pair transverse momentum spectrum, inclusive jet multiplicities and $R$
    distance between the top pair and the hardest jet using LO plus ($\mathrm{LO}\oplus$) parton
    shower simulations at $13\ {\rm TeV}$. In each plot
    the upper
    ratio plot compares the
    envelopes of all variations for the angular-ordered ($\mathrm{PS}$) and dipole ($\mathrm{DS}$)
    showers, with a ratio to the central prediction of the angular-ordered
    shower. The bottom two
    ratio plots in each plot show, for the angular-ordered
    and dipole showers respectively, a breakdown of all 
    variations
    into variations of the (factorization and
    renormalization) scale in the
    hard process ($\hardProcScale$), of the arguments of the running coupling and PDFs in the
    shower ($\showerScale$) and of the hard veto scale ($\vetoScale$).}
  \label{fig:Prod-ScaleVar-LO}
\end{figure*}

Fig.~\ref{fig:Prod-ScaleVar-LO} shows the LO plus
parton-shower predictions
for the transverse momentum distribution of the top quark, $p_\perp(t)$, for both
parton showers. We expect this observable to be well
described by the LO matrix element and that the parton shower should have a limited
impact. As we expect we see that the central lines for the two showers show a good
agreement, to within $10\%$, across the full range of the distribution.
Furthermore the total uncertainty envelopes are similar in size and shape
in all bins.
There is no clear dominant source of uncertainty, with each of the variations
making a small contribution.

Fig.~\ref{fig:Prod-ScaleVar-LO} also shows the LO plus
parton-shower predictions for the
transverse momentum distribution of the $t\bar{t}$-pair, $p_\perp(t\bar{t})$, for
both parton showers.
The $p_\perp(t\bar{t})$ distribution is sensitive to the hardest jet in the event.
We note that using a pure LO ME, {\it i.e.} with no parton shower, this observable is
equal to zero.
At low values of the transverse momentum, $p_\perp(t\bar{t}) < 50\text{GeV}$,
the central lines for the two showers agree within roughly 15$\%$. In this region,
where the hardest jet is relatively soft or collinear to the beam direction
we do indeed expect to see a good agreement between the showers. This is because,
disregarding any differences in the small finite contributions, the divergent
behaviour of the two showers in the infrared limit should be the same.
At higher values of $p_\perp(t\bar{t})$ the showers display a larger
disagreement. We do not expect different parton showers to behave similarly
away from the infrared region, therefore this difference is not concerning.
We also note that the central line of each shower lies within the uncertainty
envelope of the other across the full range of the distribution.
The dominant source of uncertainty in the $p_\perp(t\bar{t})$ distribution is
the variation of the hard veto scale, $\vetoScale$. As discussed above,
the distribution of $p_\perp(t\bar{t})$ is sensitive to the hardest jet in each event.
Given that the hard veto scale sets the maximum allowed scale of the shower
emissions, it is expected that variations of this scale should give rise to significant
differences in this distribution. The reader should also note that there is significant
statistical error on the upper three bins in the results for some of the individual variations.

Furthermore, Fig.~\ref{fig:Prod-ScaleVar-LO} shows the LO plus
parton-shower predictions for
the jet multiplicity, $n_\mathrm{jets}$, distribution for jets with
$p_\perp > 25\text{GeV}$ for both showers.
In general we find that the dipole shower predicts more events with high
jet-multiplicity than the angular-ordered shower.
This can be attributed to differences in the phase-space restrictions in the two
showers, in particular the dipole shower does not have an explicit
angular-ordering restriction on emissions.
Therefore, despite using like-for-like settings in the setups for the two showers,
we do not see good agreement in the upper half of this distribution.
We note that despite the disagreement between the central lines,
in all bins the central line of each shower lies inside the uncertainty envelope
of the other. 
The largest source of uncertainty across the majority of the bins is the
variation of $\vetoScale$. This is because varying $\vetoScale$ directly
changes the available phase space for shower emissions.
This distribution is very sensitive to the parton shower, correspondingly the variation of $\showerScale$ also gives rise to sizeable uncertainties in several bins.

Finally Fig.~\ref{fig:Prod-ScaleVar-LO} shows the LO plus
parton-shower predictions
for the distribution of the separation between the $t\bar{t}$-pair and the
hardest jet in the event for both showers. The separation is defined
as $\Delta R(t\bar{t},j_1) = \sqrt{{\Delta \phi}^2 + {\Delta y}^2}$, where $\Delta \phi$ and
$\Delta y$ denote the difference in the azimuthal angle and rapidity respectively
of the $t\bar{t}$-pair and the hardest jet in the event.
With a pure LO ME and no shower there is no jet and this distribution does not
exist, therefore the predictions are very sensitive to the behaviour of the parton shower.
In the case of an event with only one jet, the distribution is non-zero only in
the region $\Delta R > \pi$.
The distribution in the region $\Delta R > \pi$
is sensitive to the hardest and second hardest jets in the event while 
the distribution in the region $\Delta R < \pi$ is most sensitive to the second hardest
jet in the event.
The central lines exhibit very good agreement across much of the distribution.
The greatest discrepancy is in the uppermost bin in which we still see agreement to
within roughly 20\%. The total uncertainty envelopes are also of a similar shape
and size across the distribution.
The largest uncertainties arise from variations in $\vetoScale$ which reflects the
fact that the distribution is sensitive to the hardest couple of jets in the event.
In the region $\Delta R < \pi$, where the distribution is sensitive to the second
hardest jet, the variation of $\showerScale$ also gives rise to significant uncertainties.
We also see that a full evaluation of the scale variations is required to produce an
accurate estimate of the uncertainties in this region of the distribution.

\begin{figure*}
  \centering
  \includegraphics[width=0.4\textwidth]{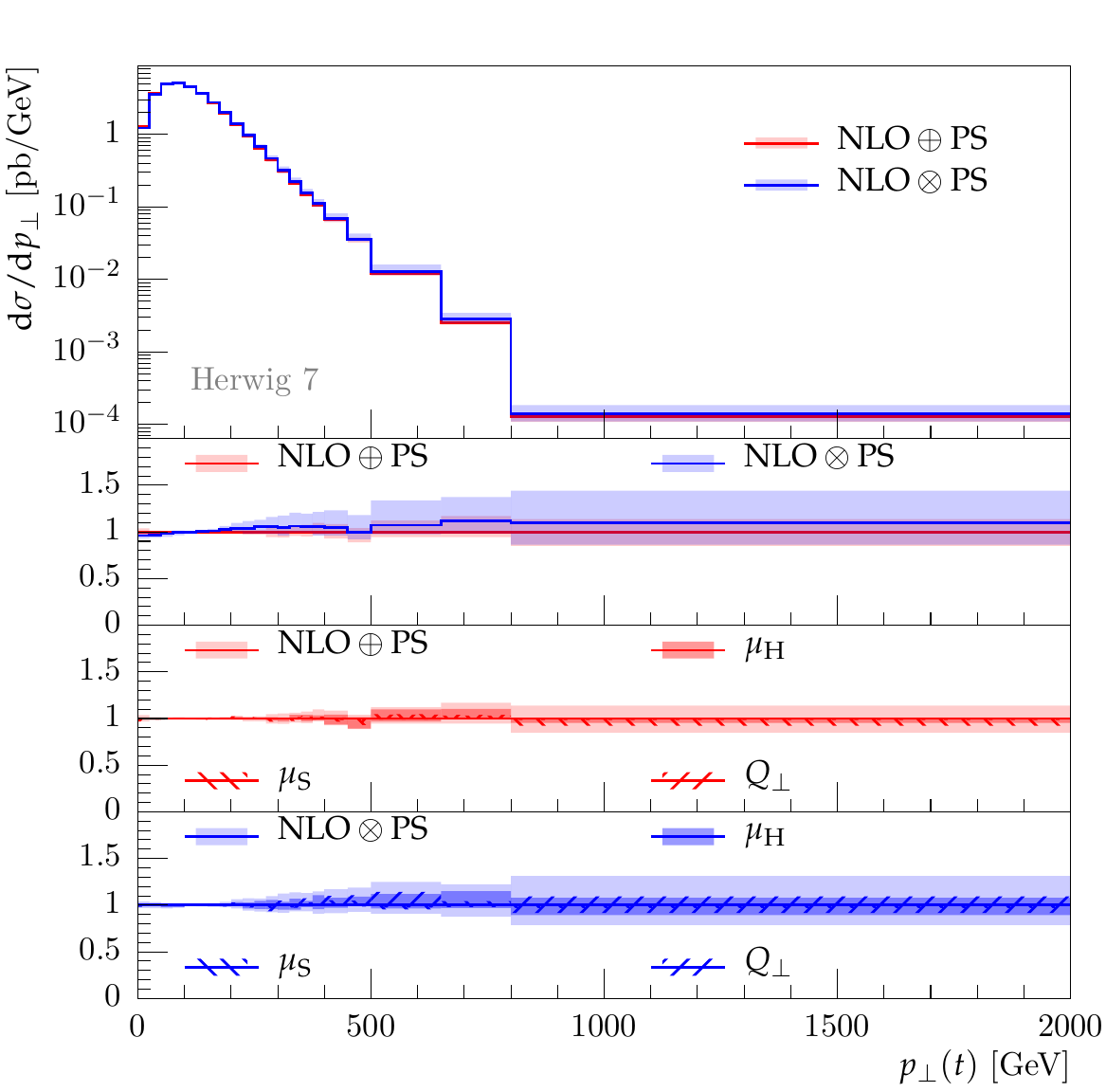}\hspace*{2cm}
  \includegraphics[width=0.4\textwidth]{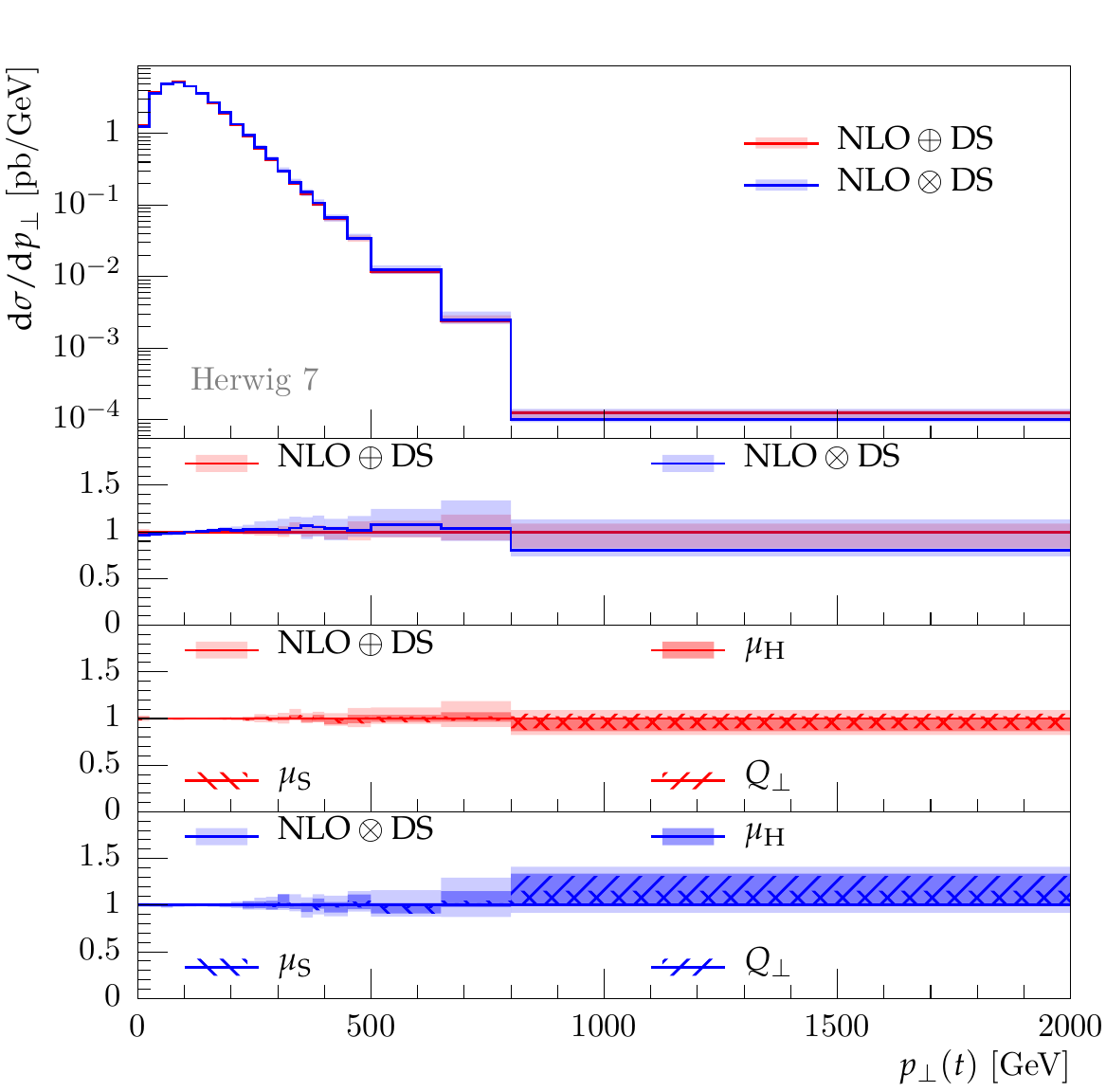}
  \\
  \includegraphics[width=0.4\textwidth]{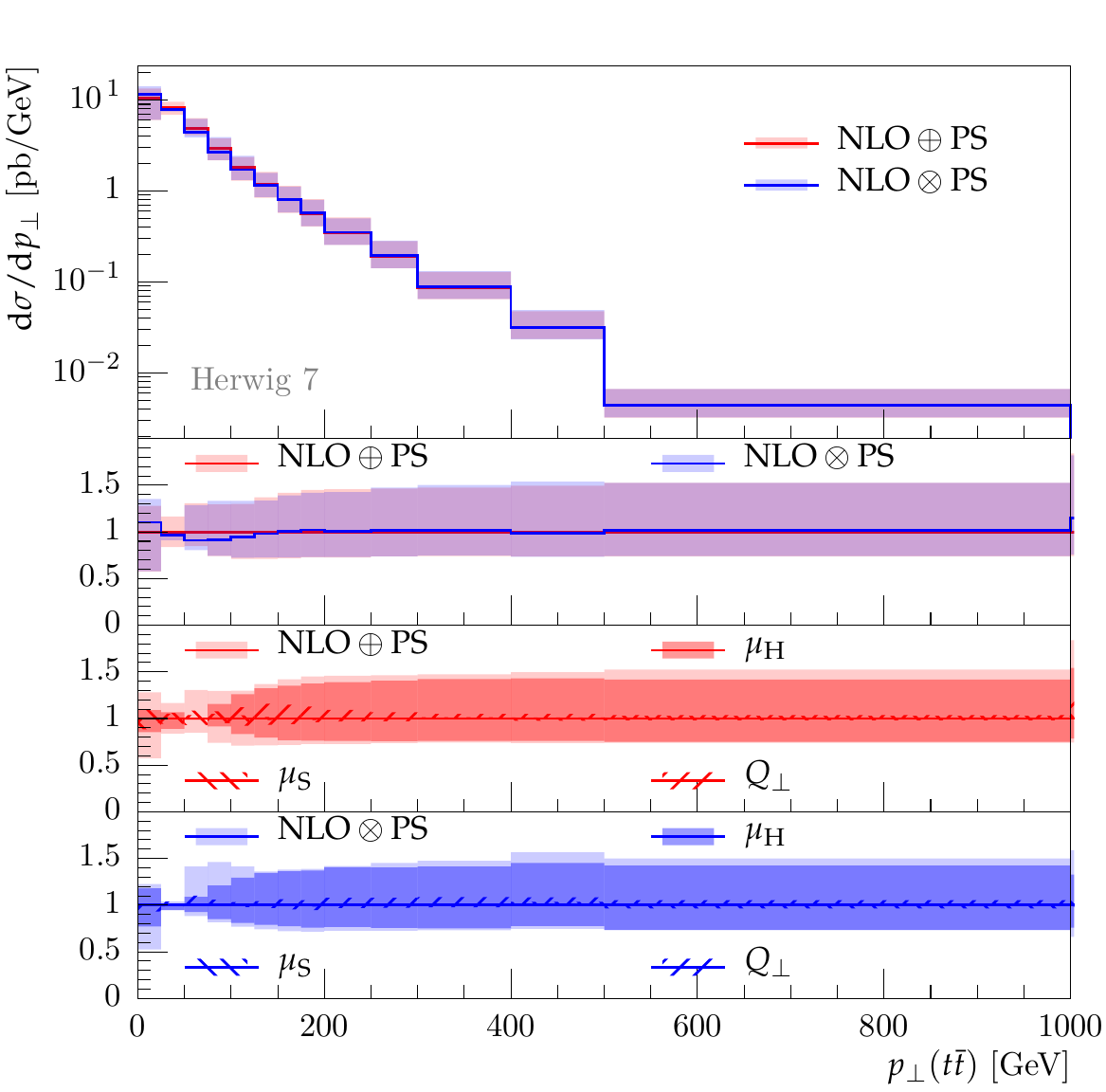}\hspace*{2cm}
  \includegraphics[width=0.4\textwidth]{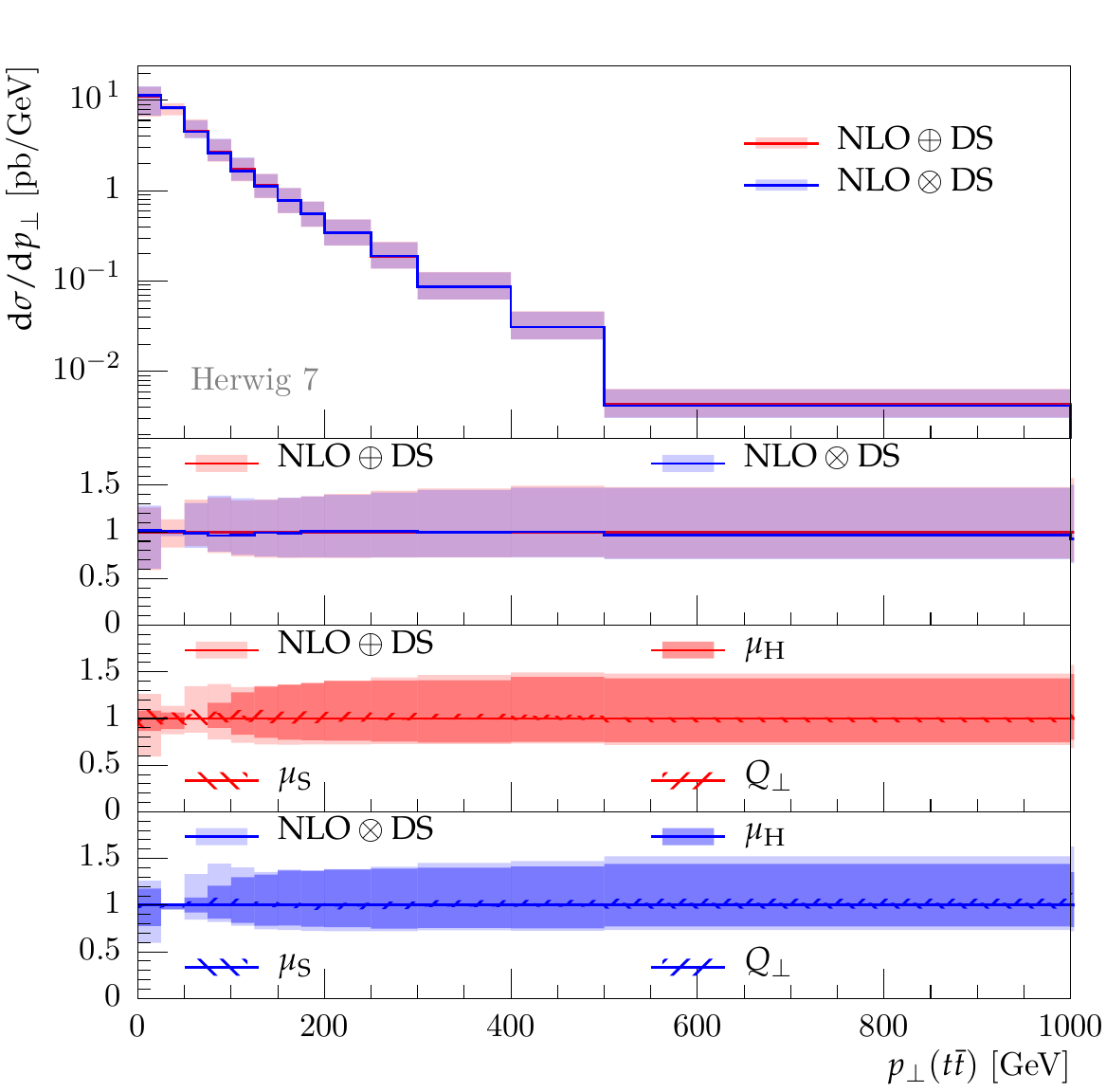}
  \caption{Transverse momenta of the top quark (upper row)
    and the top quark pair (lower row),
    comparing variations for
    NLO matched predictions at
    $13\ {\rm TeV}$ for the angular-ordered ($\mathrm{PS}$,
    left column) and dipole
    showers ($\mathrm{DS}$, right column). The top panels in each plot compare the central
    prediction and overall variation between the MC@NLO-type ($\mathrm{NLO} \oplus$) and
    Powheg-type ($\mathrm{NLO} \otimes$) matching. The
    first ratio plot in each plot allows to directly compare the
    overall variations in both matching
    variants, in a ratio to the central MC@NLO-type prediction, while the lower two
    ratio plots in each plot show a
    breakdown of the variations for both matching
    variants regarding the
    hard process scale($\hardProcScale$),
    the
    shower scale ($\showerScale$) as well as
    the hard veto scale ($\vetoScale$).}
  \label{fig:Prod-ScaleVar-NLO-TopPt-ttbarPt}
\end{figure*}
Fig.~\ref{fig:Prod-ScaleVar-NLO-TopPt-ttbarPt} shows the NLO-matched predictions for
the $p_\perp(t)$ distribution (upper row) obtained using
the angular-ordered (PS, left column) and
dipole showers (DS, right column).  In a NLO-matched sample the top-$p_\perp$ is
formally predicted with
NLO accuracy and any differences between
the MC@NLO-type (NLO$\oplus$, aka subtractive)
and Powheg-type (NLO$\otimes$, aka multiplicative)
matching are due to higher-order effects.  Accordingly we see a good
agreement between the MC@NLO-type and Powheg-type central
lines,
for both showers. In
the angular-ordered shower predictions the central-lines predicted using MC@NLO-type and
Powheg-type matching agree to within roughly 15\% in all bins. The same is true in all but
the highest-$p_\perp$ bin in the results for the dipole shower. In the highest-$p_\perp$ bin the
results agree to within 25\%, however the reader should note that, in this
bin, such a difference is to be attributed to the statistical uncertainty on the
results used to construct the uncertainty envelopes.  As in the LO result,
there is no clear dominant source of uncertainty.

Fig.~\ref{fig:Prod-ScaleVar-NLO-TopPt-ttbarPt} also shows the NLO-matched predictions
for the $p_\perp(t\bar{t})$ distribution (lower row)
using the angular-ordered (PS, left column) and
dipole showers (DS, right column).  In a NLO-matched
sample the distribution is formally predicted with LO accuracy. 
The uncertainty envelope on the
NLO-matched predictions is much smaller than that on the LO plus parton shower
predictions. This is due to the much smaller contribution to the total
uncertainty from the parton showers.  The dominant contribution to the total
uncertainty is the variation of $\hardProcScale$ which reflects that the
predicted distribution is sensitive to the simulation of the hard process.
The results for both showers show agreement between the central 
MC@NLO-type and Powheg-type results to within 10\% across the entire
distribution.

\begin{figure*}
\centering
  \includegraphics[width=0.4\textwidth]{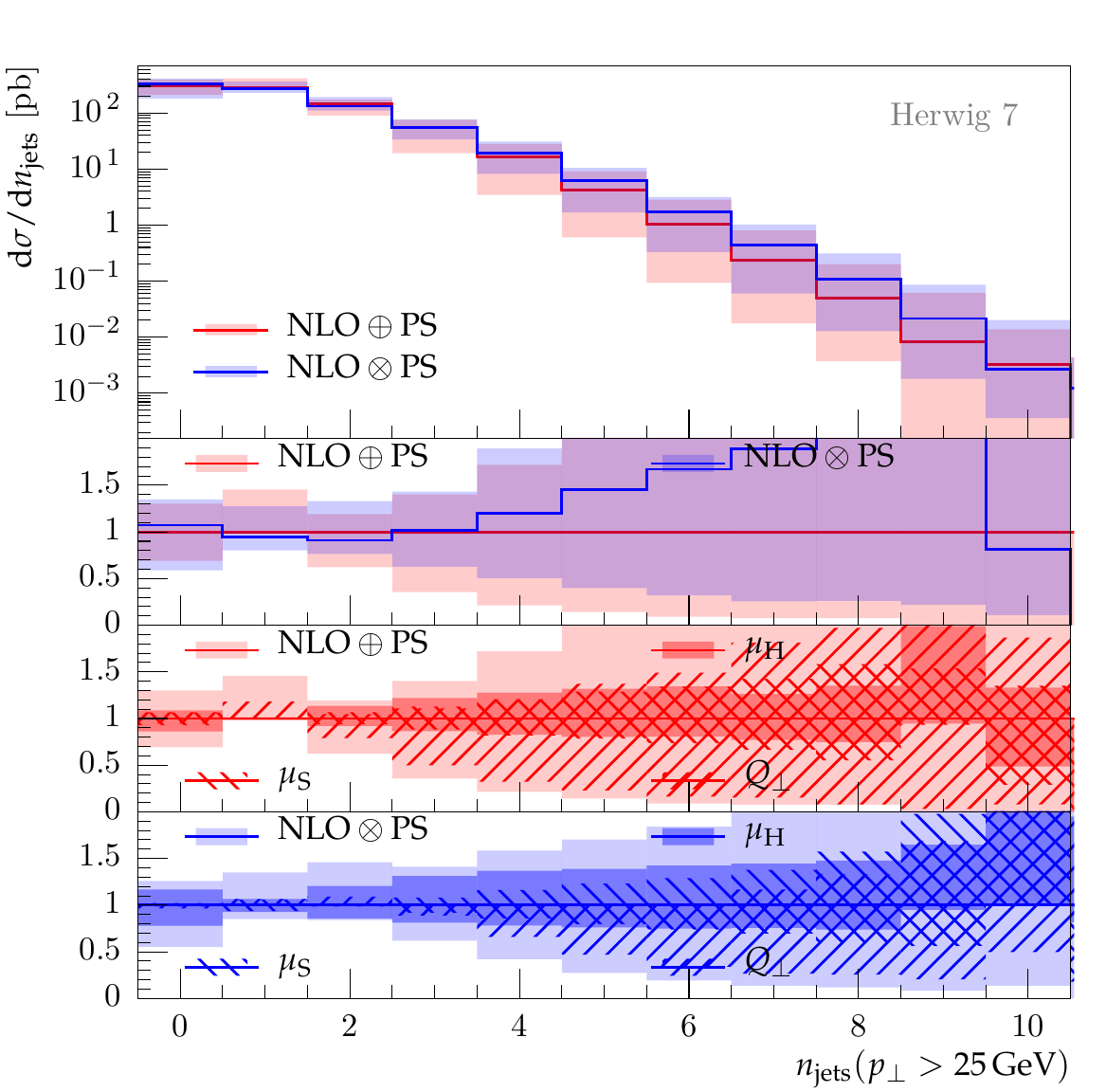}\hspace*{2cm}
  \includegraphics[width=0.4\textwidth]{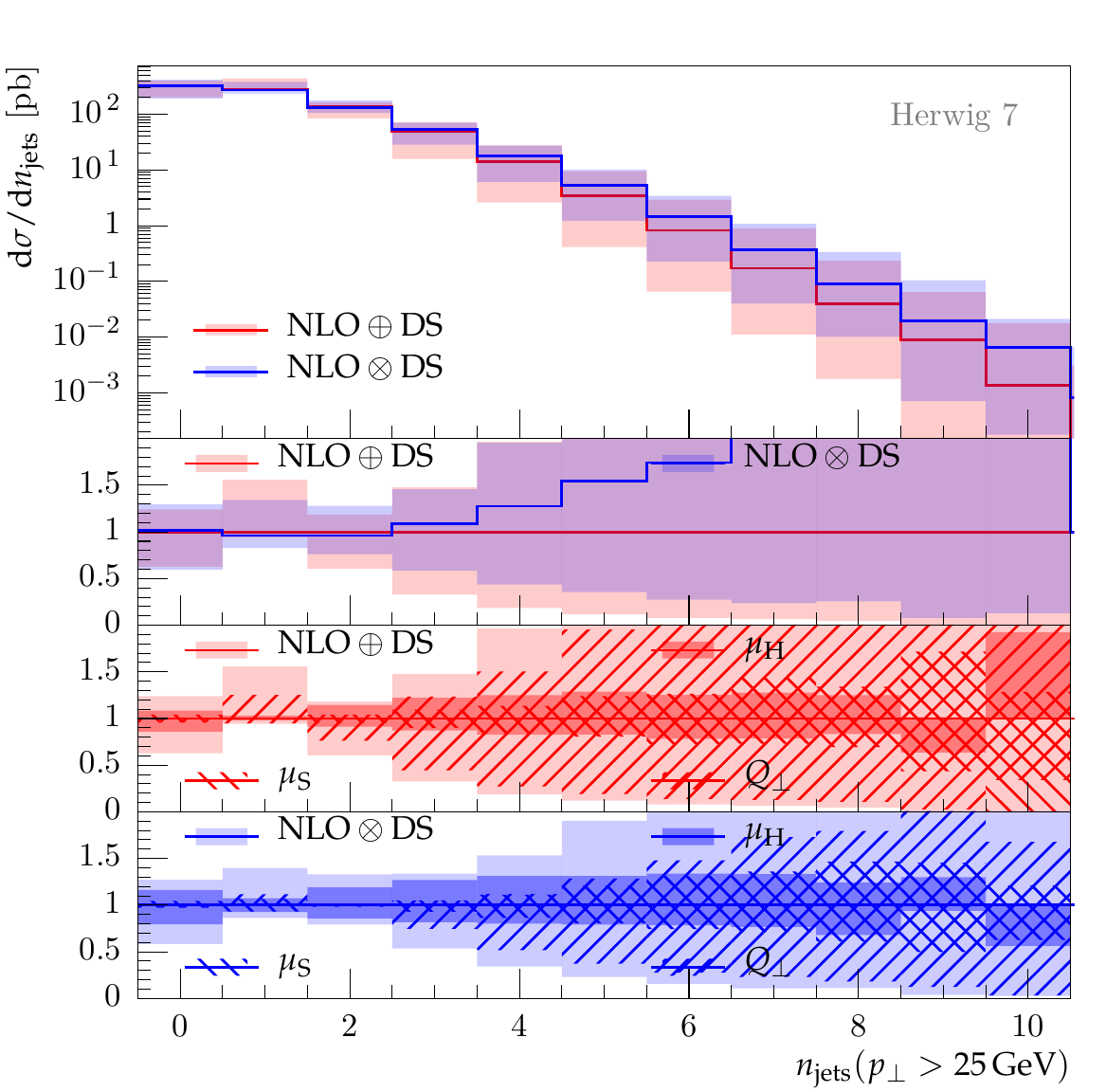}\\
  \includegraphics[width=0.4\textwidth]{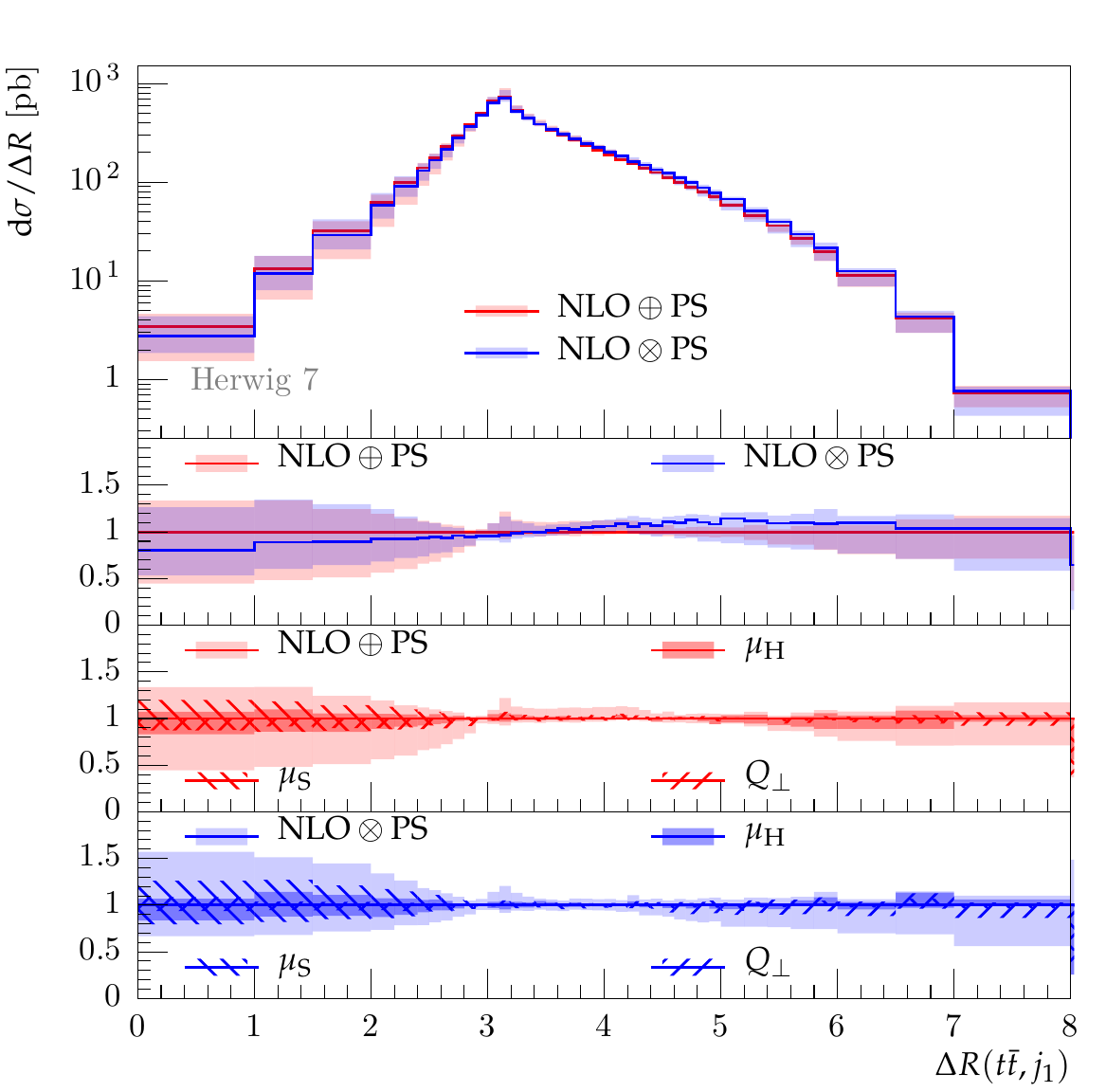}\hspace*{2cm}
  \includegraphics[width=0.4\textwidth]{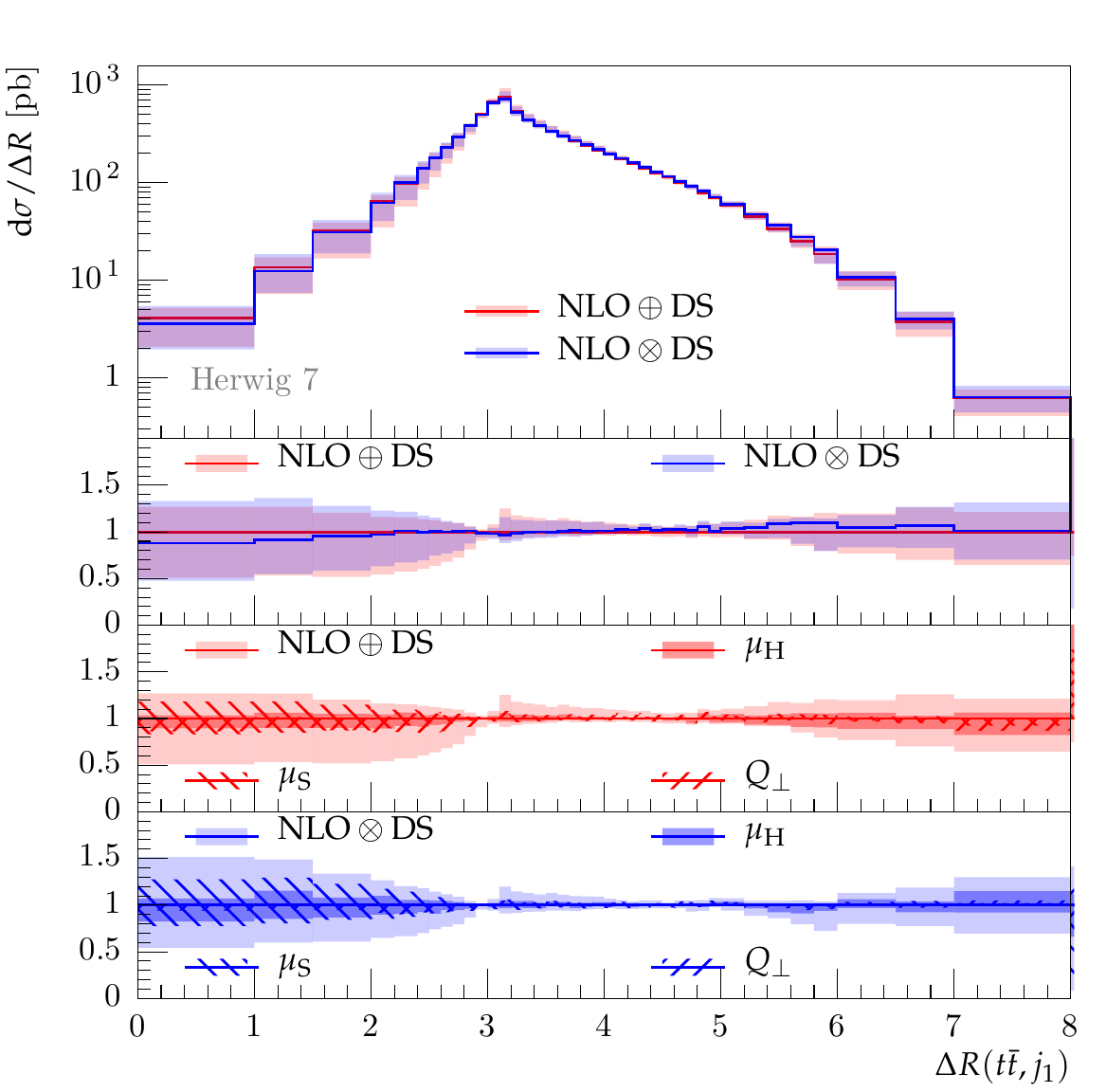}
  \caption{Same as Fig.~\ref{fig:Prod-ScaleVar-NLO-TopPt-ttbarPt}, in this case
    showing the inclusive jet multiplicities and the $R$ distance between the $t\bar{t}$-pair
    pair and the hardest jet. See the text for discussion.}
  \label{fig:Prod-ScaleVar-NLO-nJets25-DeltaR}
\end{figure*}
Fig.~\ref{fig:Prod-ScaleVar-NLO-nJets25-DeltaR} shows the NLO-matched predictions of the
$n_\mathrm{jets}$ distribution using the angular-ordered and dipole
showers, respectively. In a NLO-matched sample the 0-jet and 1-jet rate
predictions are formally accurate to NLO and LO respectively and
higher-multiplicity contributions
are only due to the parton shower.  In the
results from both showers the central predictions obtained using MC@NLO-type and Powheg-type
matching agree to
within roughly 10\% up to and including the 3-jet bin while in
higher-multiplicity bins the Powheg-type prediction rises above the MC@NLO-type
prediction.  Note that there is an exception to this trend in the 10-jet bin
of the angular-ordered shower results, however this fluctuation is
to be attributed to the high statistical uncertainty in this bin.  The MC@NLO-type
matching produces fewer high-multiplicity events than the Powheg-type
matching because of the choice of the hard veto scale, discussed in
detail in Section~\ref{sec:mcatnlo-veto-scale-prod-level}. We see in the ratio
plots that the variations of $\vetoScale$ make a significant contribution to
the total uncertainty envelopes in the MC@NLO-type predictions in high-multiplicity
bins.
In general parton showers are not expected to produce a good description of
hard radiation and therefore one should not expect a parton shower to
accurately predict the jet-multiplicity distribution for high multiplicities.
It follows that in Fig.~\ref{fig:Prod-ScaleVar-NLO-nJets25-DeltaR} we see that the
variations of each of the three scales, \ie also of $\showerScale$ and
$\hardProcScale$, contributes significantly to the total
uncertainty envelope and we see a steady increase in the total uncertainty
with increasing jet-multiplicity.  In general high-multiplicity observables
are better described by multi-jet merging
algorithms.
However, this is beyond
the scope of this paper.

\begin{figure*}
  \centering
    \includegraphics[width=0.4\textwidth]{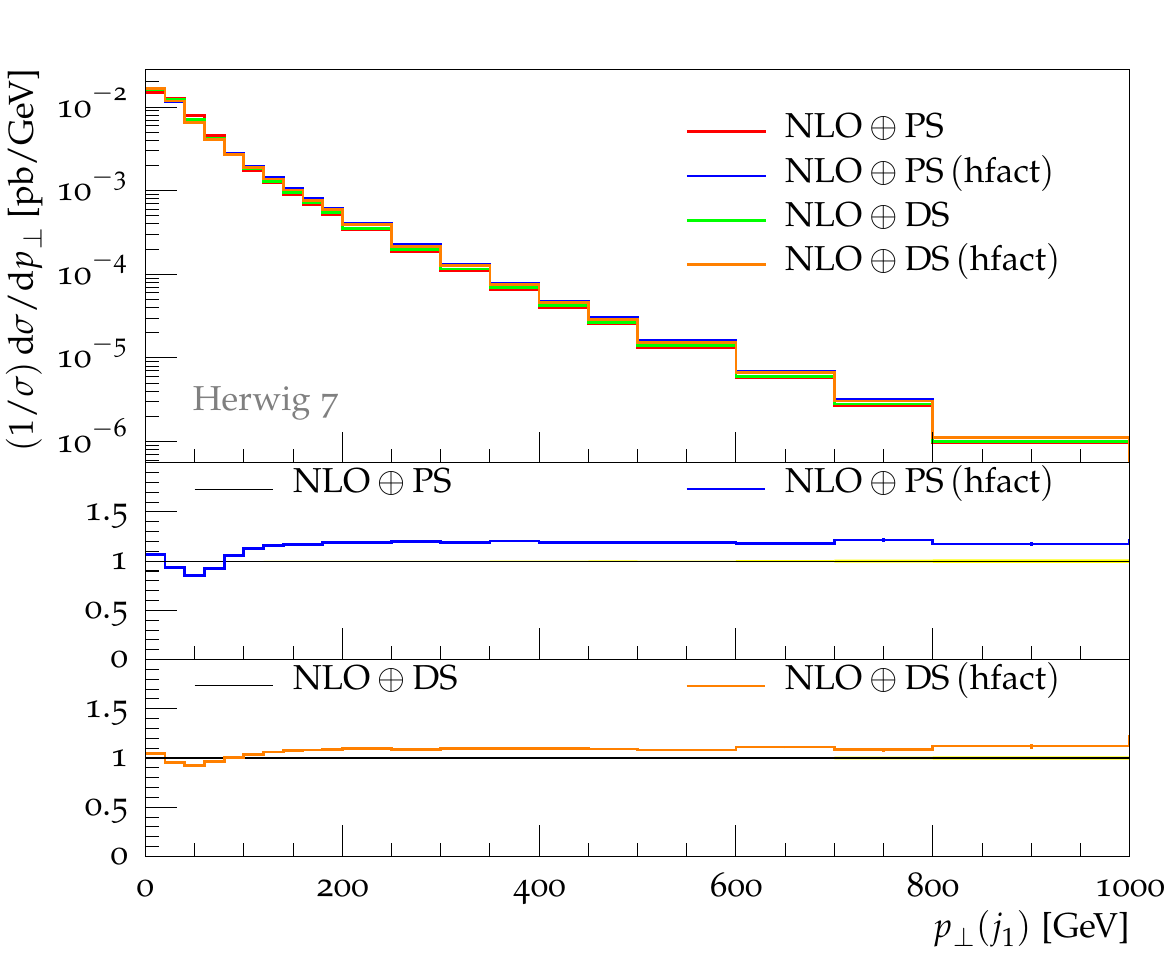}\hspace*{2cm}
    \includegraphics[width=0.4\textwidth]{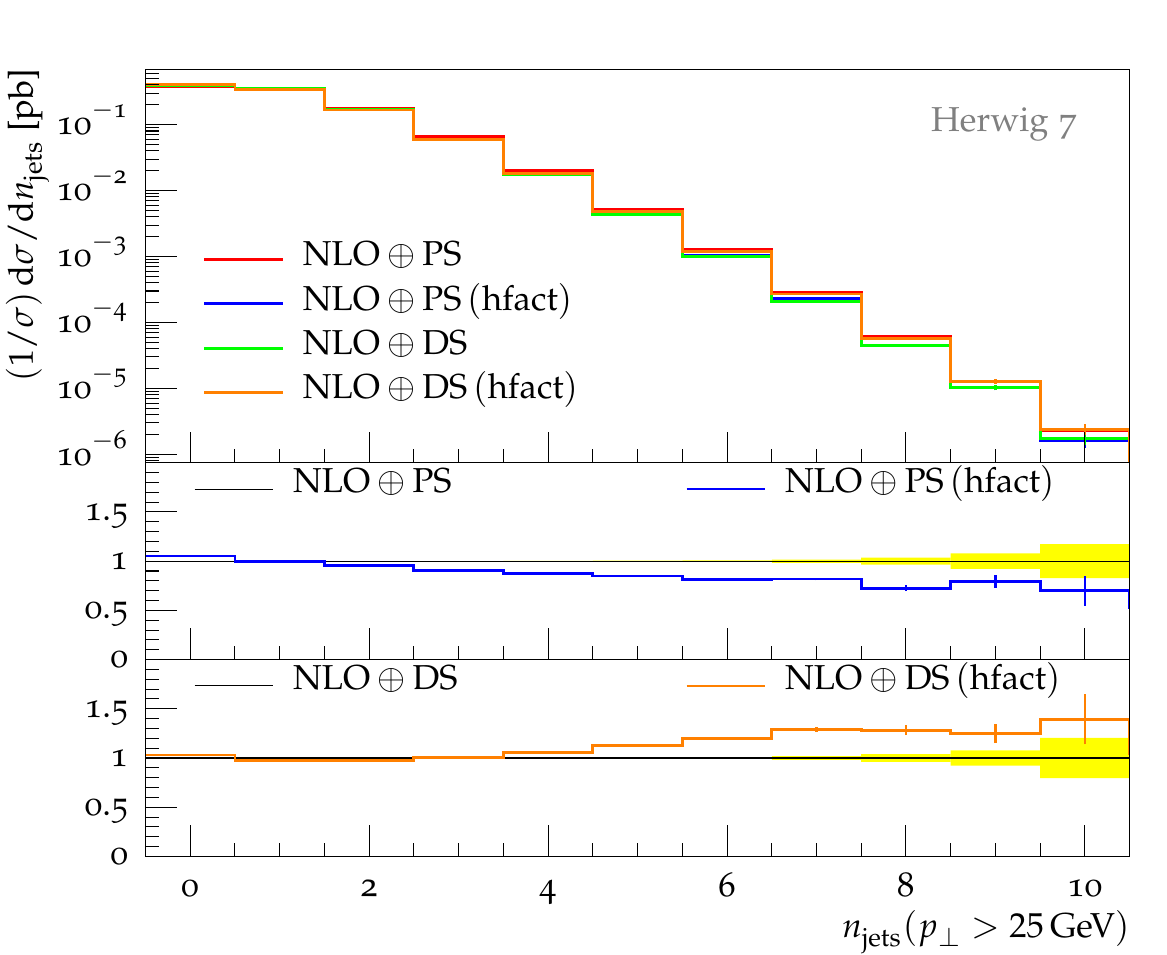}\\
    \includegraphics[width=0.4\textwidth]{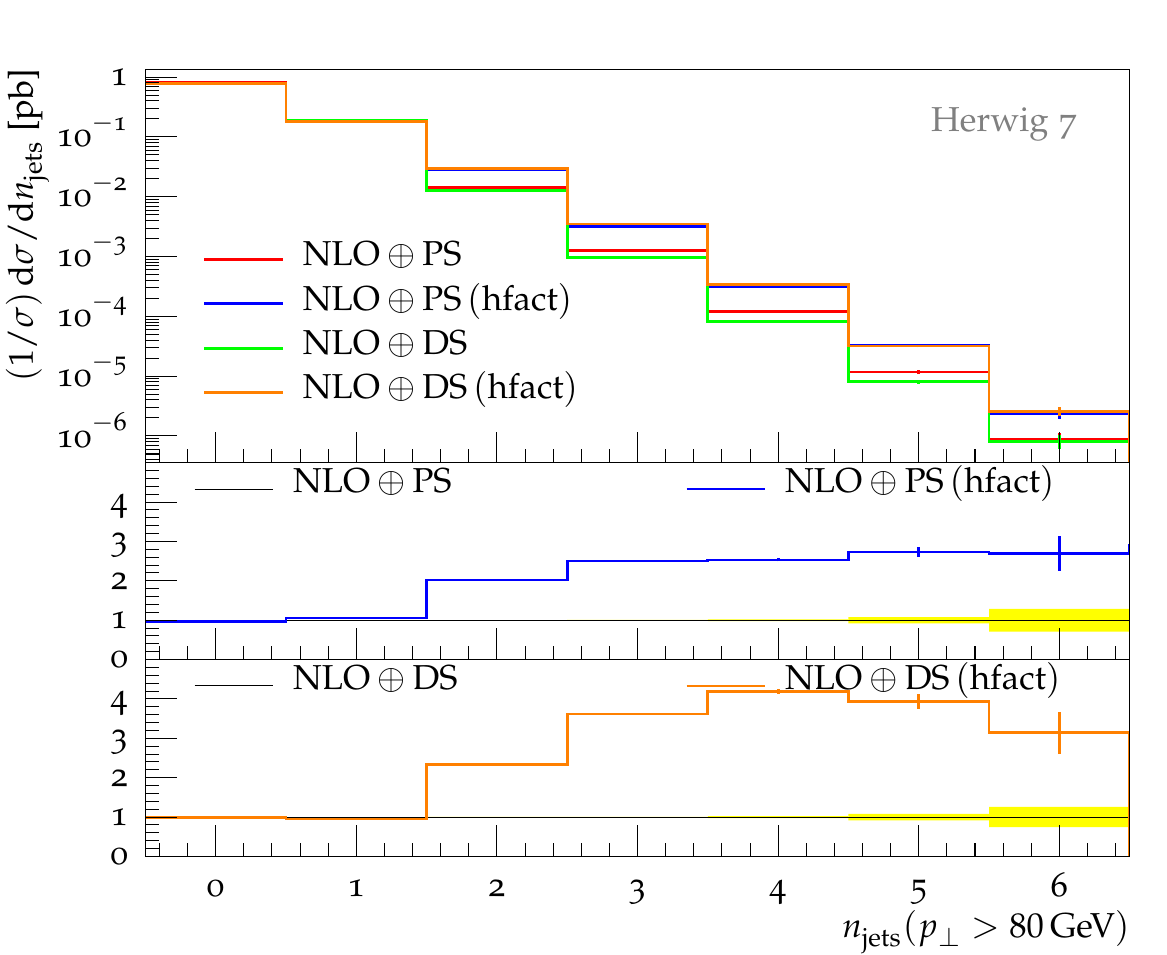}\hspace*{2cm}
    \includegraphics[width=0.4\textwidth]{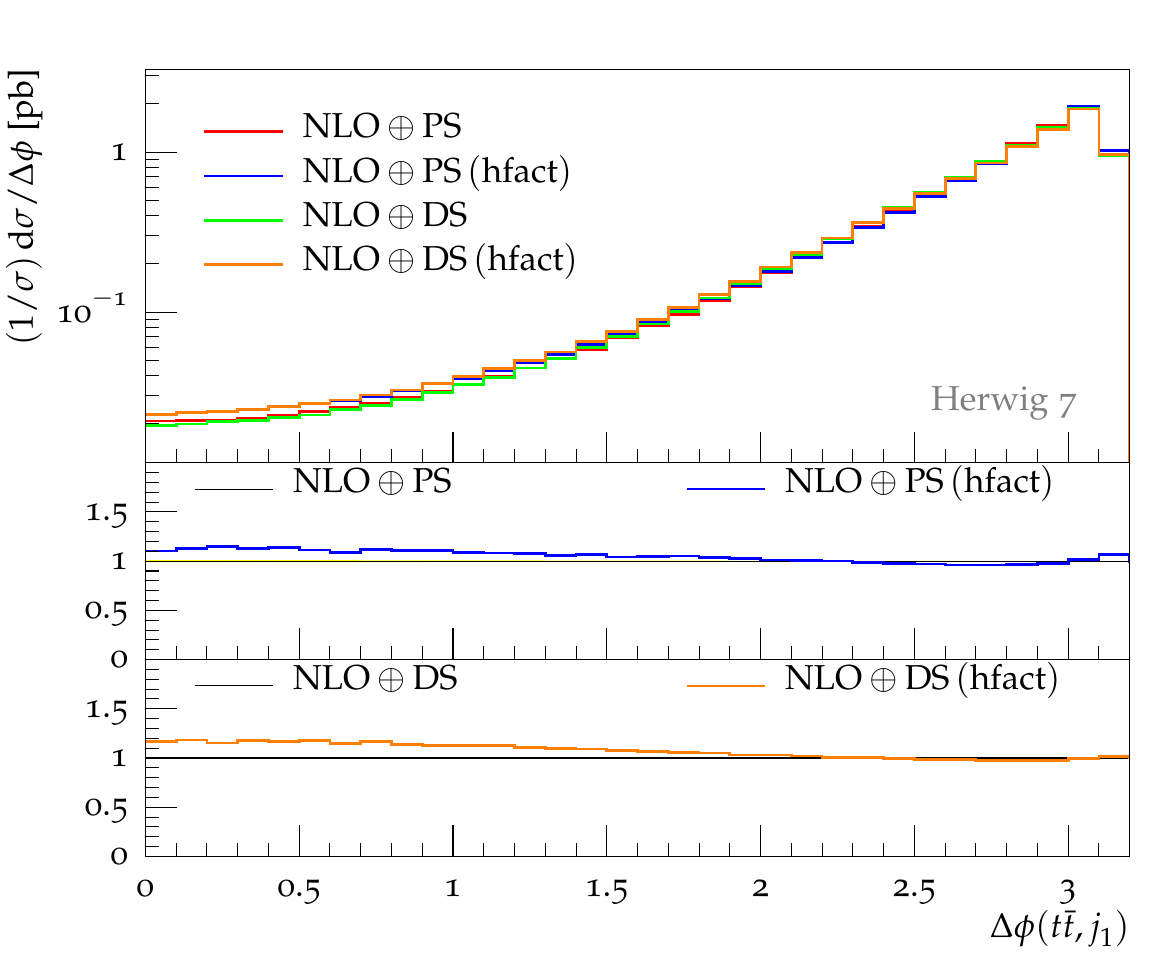}
  \caption{The effect of different profile scale choices for the two shower
    algorithms, angular ordered ($\mathrm{PS}$) and dipole ($\mathrm{DS}$), respectively when using
    MC@NLO-type ($\mathrm{NLO} \oplus$) matching. We compare
    predictions for the default \resummation
    profile versus the broader \hfact profile. From left to right, top to
    bottom, we present the $p_\perp$ spectrum of the hardest jet, the
    inclusive jet multiplicity at a threshold of $25\ {\rm GeV}$ and
    $80\ {\rm GeV}$, respectively, as well as the azimuthal angle distance
    between the top pair and the hardest jet.}
  \label{fig:ProfileScales}
\end{figure*}

In Fig.~\ref{fig:Prod-ScaleVar-NLO-nJets25-DeltaR} we also consider the NLO-matched
predictions for the $\Delta R(t\bar{t},j_1)$ distribution using the
angular-ordered and dipole showers, respectively.  With a pure NLO ME
this distribution would exist only in the region $\Delta R > \pi$ and would be
zero in the region $\Delta R < \pi$.  Therefore in a NLO-matched sample this
observable probes both the hard process and parton shower.
For both showers, the central lines of the MC@NLO-type and
Powheg-type predictions display good agreement across much of the distribution.
The largest discrepancies are around 20\% and are easily accounted for by the
uncertainty envelopes.
This shows that the Powheg-type and 
MC@NLO-type matching schemes produce a similar
description of the hardest few jets using both showers.

Comparing to the LO plus parton shower results we see that in the region $\Delta R > \pi$
the uncertainty due to variations of $\vetoScale$ is much smaller in the
NLO-matched predictions, which reflects that the distribution predicted in this region
is now less sensitive to the parton shower.
The largest contribution to the uncertainty in the region $\Delta R < \pi$,
where the distribution is sensitive to the parton shower, 
is from the variation of $\showerScale$.
The variations of $\hardProcScale$ and $\vetoScale$, which affect the
starting conditions of the parton shower, make smaller but comparable contributions to the
uncertainty in this region of the distribution.

In this section we have compared a selection of distributions predicted
using both parton showers with a LO matrix element and using two NLO-matching schemes.
We used the LO results to highlight differences between the showers
whereas in
the NLO-matched results we focused on the differences between the matching
schemes. We have also highlighted some areas where the limitations of
parton showers must be considered.
In general one must consider which parts of each distribution are well predicted
by the matrix element and which are filled largely or entirely by the parton
shower and one should not expect identical predictions from different
parton showers.
In Monte Carlo studies a thorough evaluation of shower and matching uncertainties
is required to account for these differences.
In Section~\ref{sec:scale-variations-full-process} we investigate the uncertainties
due to scale variations in the prediction of distributions measured
from experiment.

\subsection{Profile Scale Choices in
MC@NLO-type Matching}
\label{sec:MCatNLOProfileScaleChoiceResults}

In Fig.~\ref{fig:ProfileScales} we present results obtained with both showers
using the \resummation and \hfact\ profiles. For clarity we include a separate
ratio plot for each shower which, for each bin, shows the ratio of the result
obtained using the \hfact\ profile to the result obtained using the
\resummation\ profile. This is not intended to be a complete discussion of
profile scales and the 
uncertainties that arise due to
choosing a specific one. We simply
wish to highlight some of the potential effects of the profile scale choice
and present a small selection of observables in which these effects are
important.

We first consider the distribution of the transverse momentum of the hardest
jet, in the top left plot in
Fig.~\ref{fig:ProfileScales}. In both showers we see an
increase in the number of events with a soft $(p_\perp \lesssim \text{20
  GeV})$ hardest jet, a decrease in the number of events with a
moderate-$p_\perp$ $(\text{20 GeV }\lesssim p_\perp \lesssim\text{ 80 GeV})$
hardest jet and an increase in the number of events with a high-$p_\perp$
$(p_\perp \gtrsim \text{80 GeV})$ hardest jet using the \hfact\ profile versus
the \resummation\ profile.
While the \hfact\ profile suppresses hard shower emissions, it does not
apply a hard cut on such emissions as in the \resummation
profile. We therefore expect to see an increase in
the number of events with a high-$p_\perp$ hardest emission.
With the hard process, $pp\to t\bar{t}$, correct to NLO, $p_{\perp,j_1}$
is predicted accurate only to LO and we should expect the shower to have some
moderate impact on this observable. For both showers the differences due to
the profile choice are moderate, $\sim 20\%$.

Next, in the top right plot in 
Fig.~\ref{fig:ProfileScales}, we consider the jet multiplicity,
$n_\mathrm{jet}$, distribution with a minimum jet-$p_\perp$ cut
of 25 GeV and 80 GeV respectively. In general the dipole shower shows an
increase in the number of jets with both of the minimum jet-$p_\perp$ cuts
when using the \hfact\ profile. For the angular-ordered shower we see, in
general, a decrease in the number of low-$p_\perp$ jets when using the
\hfact\ profile. 
On the other hand, the bottom left plot in
Fig.~\ref{fig:ProfileScales} shows an increase in the number of high-$p_\perp$
jets.  The difference in the number of jets with $p_\perp > 80 \text{ GeV}$
due to the profile choice is bigger for the dipole shower than for the
angular-ordered shower.
Successive emissions in the dipole shower decrease in transverse
momentum, therefore an increase in the transverse momentum of the
first shower emission, as we expect with the \hfact\ profile,
increases the phase space available to all emissions that follow.  The
angular-ordering requirement in the angular-ordered shower effectively
puts a cut on the hardness of shower emissions, and through this the
\hfact\ profile can increase the emission phase space only up to a
maximum possible value, such that the effects of the change from the
\resummation\ to \hfact\ profile are expected to be somewhat more
pronounced for the dipole than the angular-ordered shower.  The larger
phase space available to successive dipole shower emissions with the
\hfact\ profile relative to the \resummation\ profile is evident in
the increase in the number of both soft and hard jets. In the case of
the angular-ordered shower we see an increase in the number of hard
jets, however the angular-ordering restriction and the suppression of
soft emissions by the \hfact\ profile lead to a reduction in the
number of low-$p_\perp$ jets.

Finally the bottom right plot in Fig.~\ref{fig:ProfileScales} shows the
distribution of the azimuthal separation of the $t\bar{t}$ pair and the
hardest jet, $\Delta \phi(t\bar{t},j_1)$. At NLO, {\it i.e.} with one QCD
emission from the matrix element, $\Delta
\phi(t\bar{t},j_1)$ is necessarily equal to $\pi$,
therefore the distribution is strongly dependent on the parton shower, in
particular on the hardest few emissions other than the hardest emission.
In the case of the dipole shower, the \hfact\ profile produces a
significant increase in the number of events with small
$\Delta \phi_{t\bar{t},j_1}$ compared to the \resummation\ profile.  In
comparison the angular-ordered shower displays a smaller increase in
the number of events with small $\Delta \phi_{t\bar{t},j_1}$ using the
\hfact\ profile versus using the \resummation\ profile. This is consistent
with what we see in the $n_\mathrm{jet}$ distributions, where using
the \hfact\ profile leads to a larger increase in the number of
high-$p_\perp$ jets in the dipole shower than in the angular-ordered
shower.

\subsection{The Hard Veto Scale in MC@NLO-type Matching}
\label{sec:mcatnlo-veto-scale-prod-level}

\begin{figure*}
    \centering
  \includegraphics[width=0.4\textwidth]{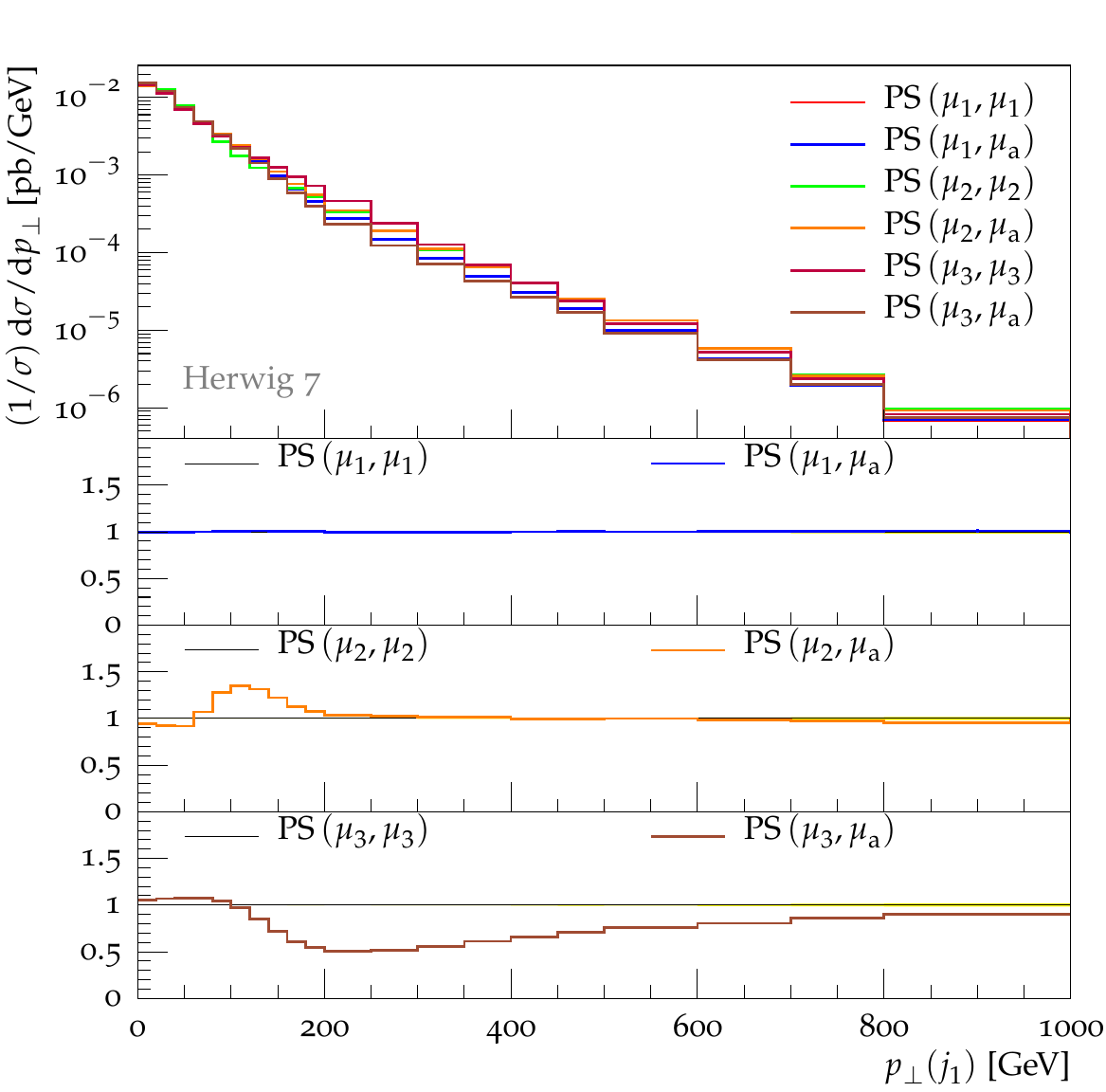}\hspace*{2cm}
  \includegraphics[width=0.4\textwidth]{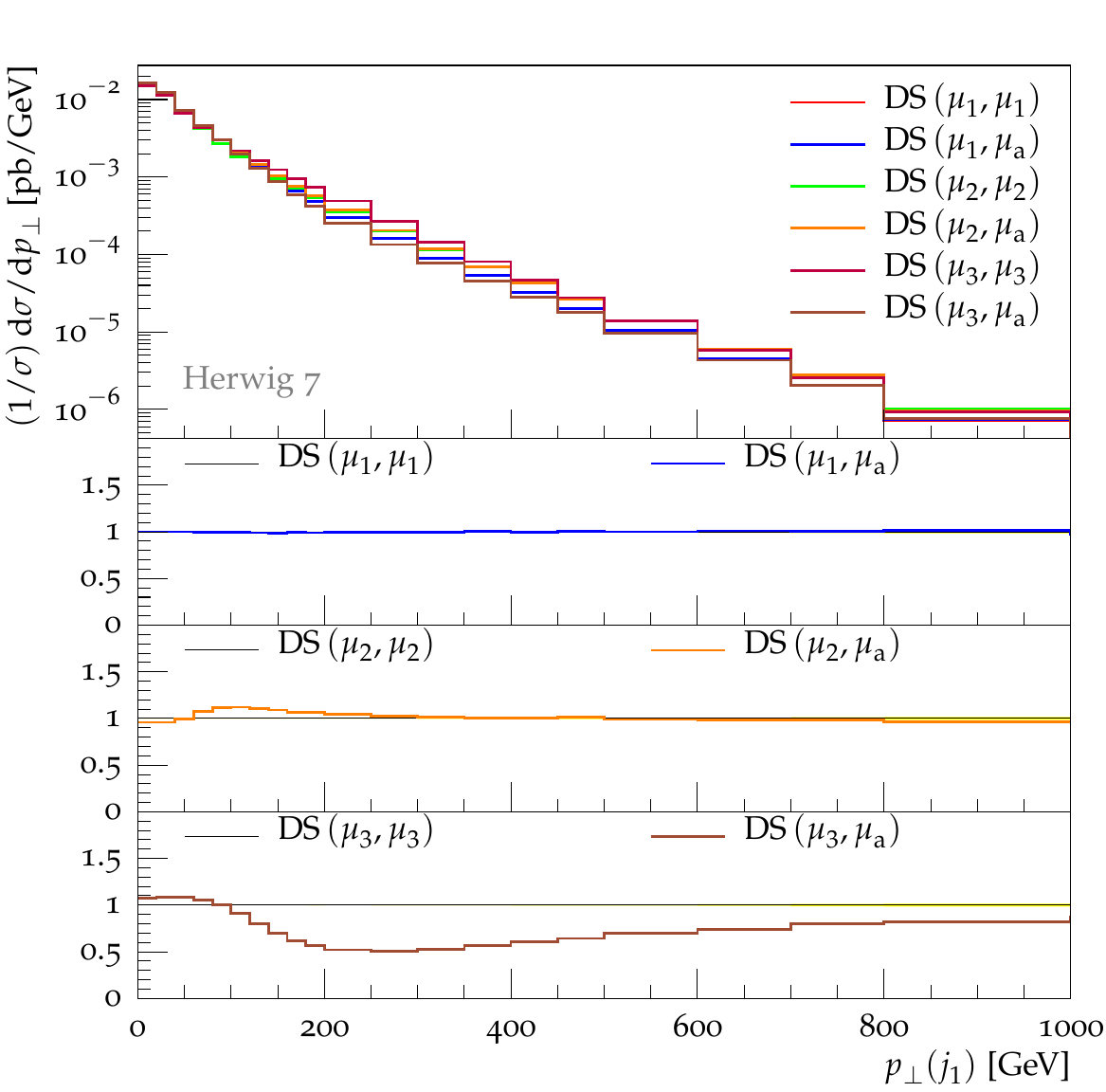}\\
  \includegraphics[width=0.4\textwidth]{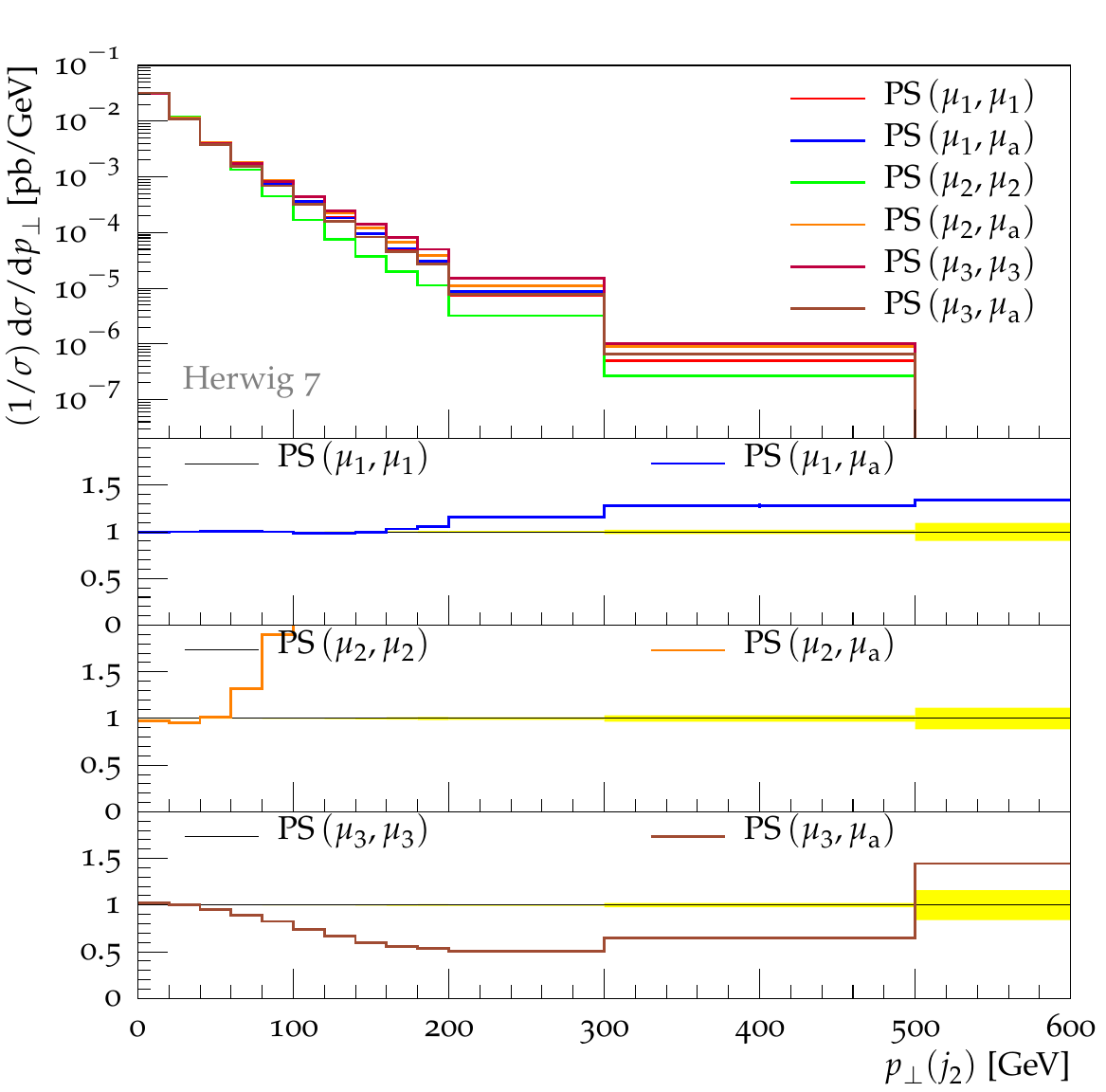}\hspace*{2cm}
  \includegraphics[width=0.4\textwidth]{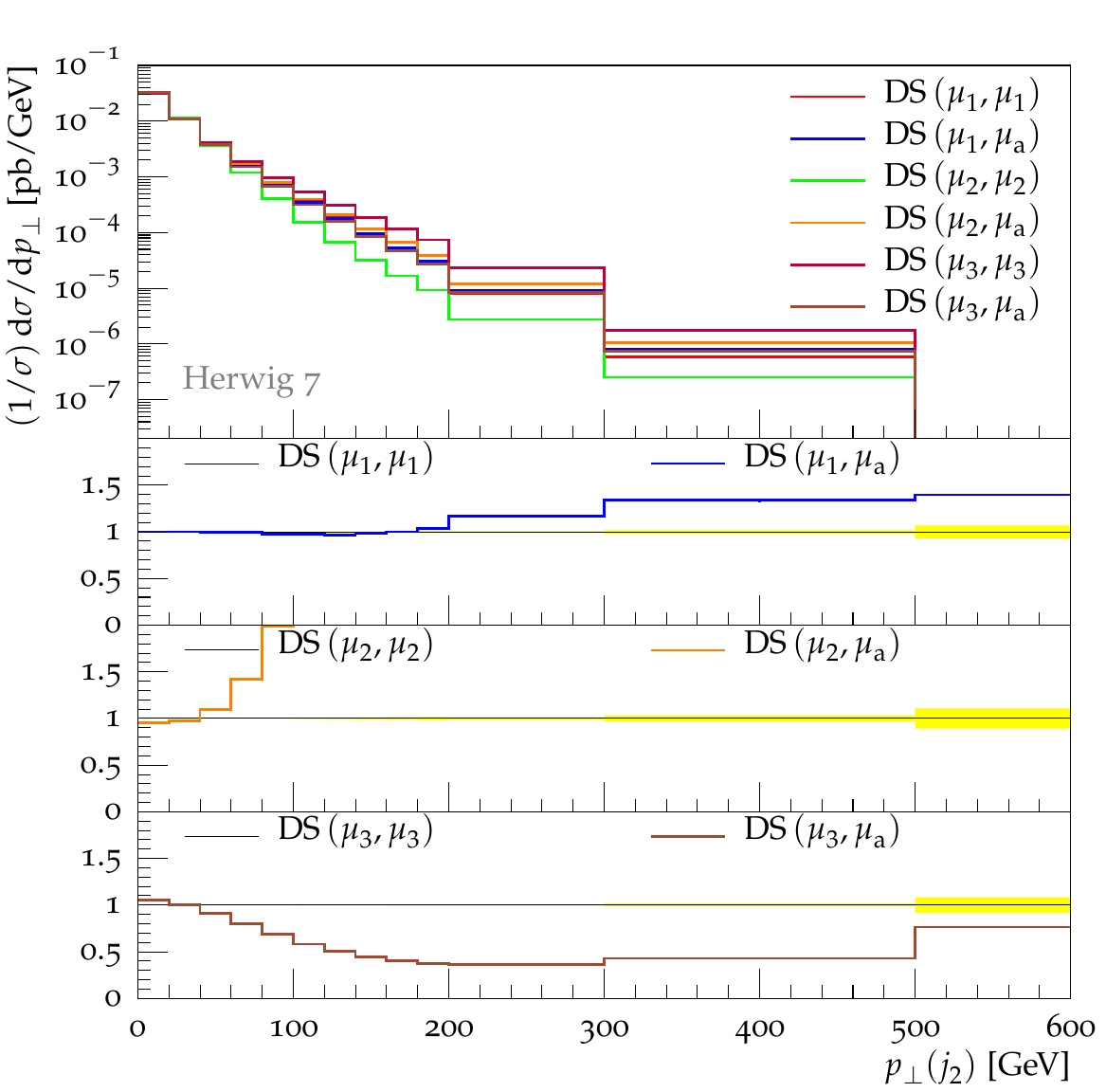}  
  \caption{The effect of different choices of the hard veto scale $\vetoScale$ for the two shower
    algorithms, angular ordered ($\mathrm{PS}$) and dipole ($\mathrm{DS}$), respectively when using
    MC@NLO-type matching. We compare predictions for different choices of the resummation and
    factorization scale choice $\hardProcScale$, using two choices for the hard veto scale in each
    case. The scales are specified in the format $(\hardProcScale,\vetoScale)$ and each of the
    scale choices is defined in the text.
    From top to bottom, left to
    right, we present the $p_\perp$ spectra of the
    hardest and second hardest jets, produced with the angular-ordered and
    dipole shower respectively.}
  \label{fig:Prod-VetoScale-Jet1Pt-Jet2Pt}
\end{figure*}
\begin{figure*}
  \centering
    \includegraphics[width=0.4\textwidth]{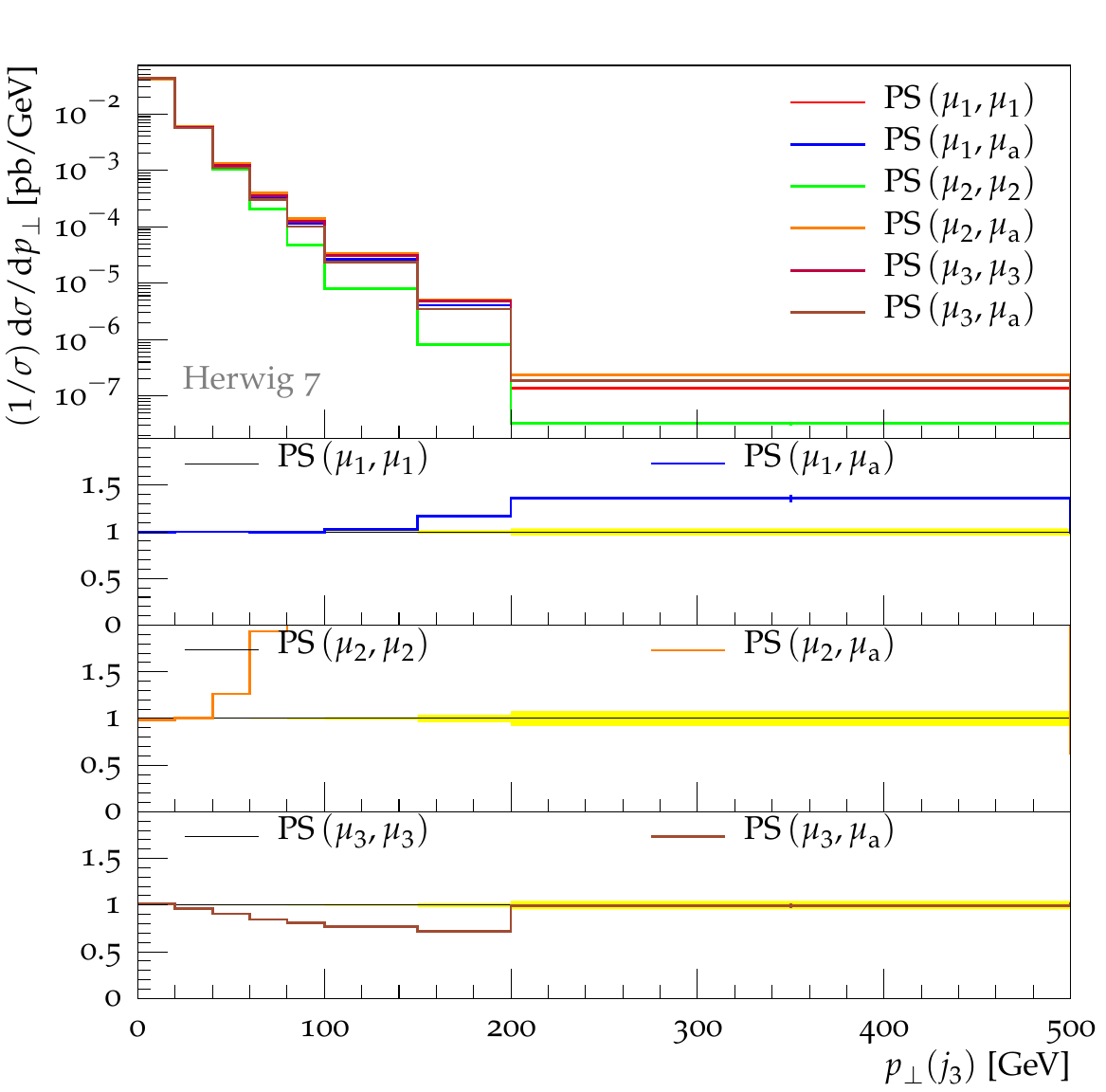}\hspace*{2cm}
    \includegraphics[width=0.4\textwidth]{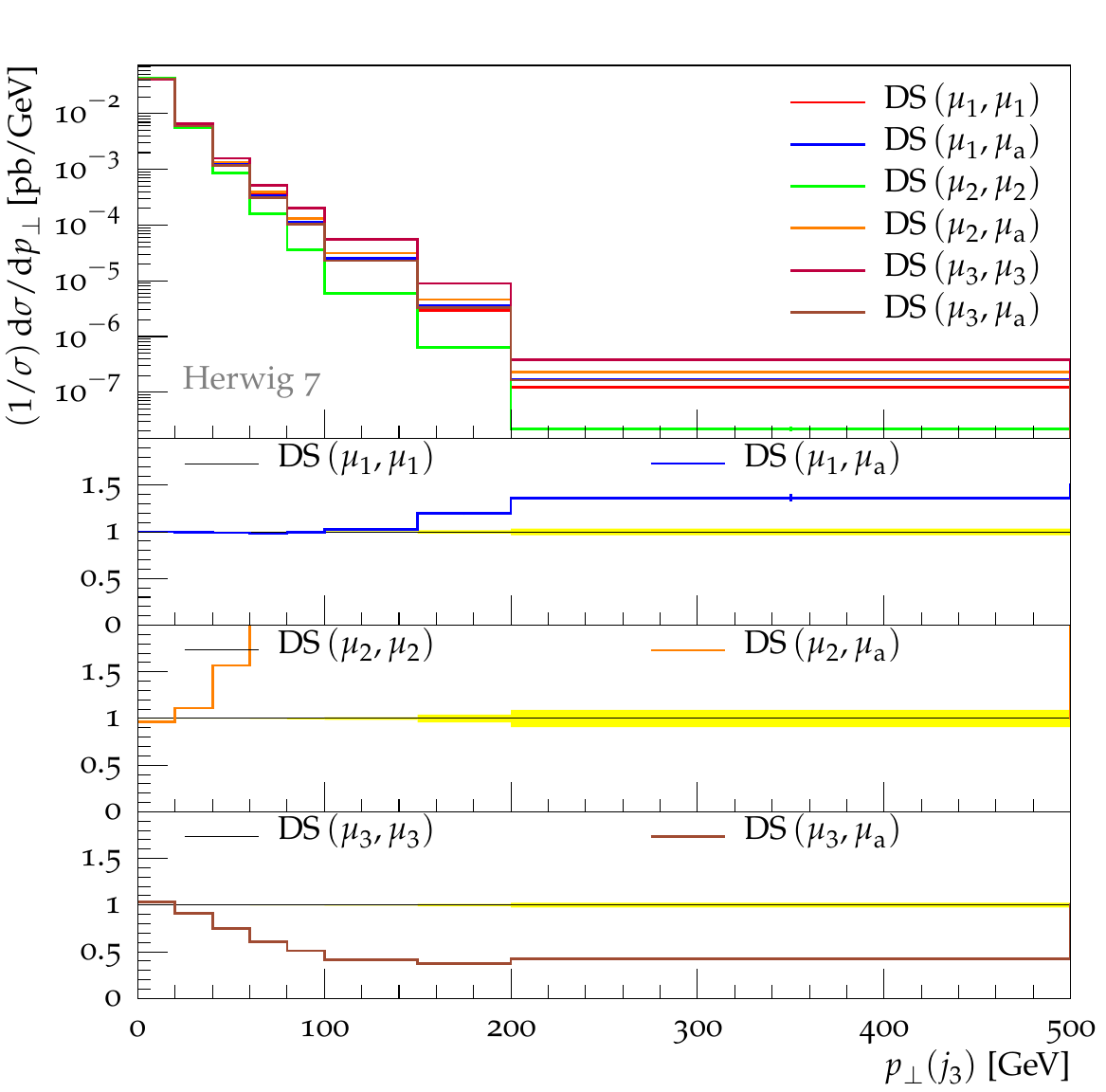}\\
    \includegraphics[width=0.4\textwidth]{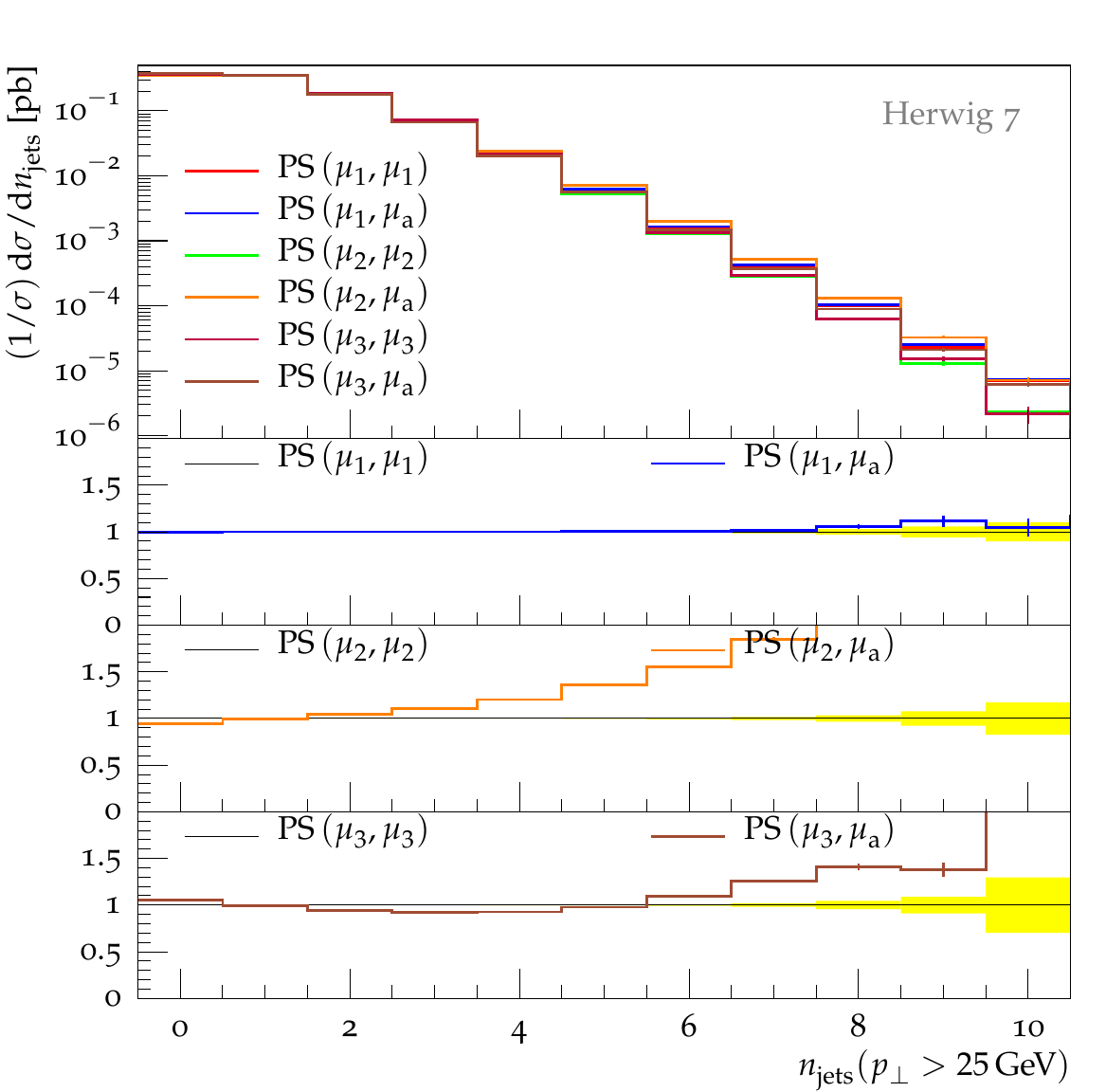}\hspace*{2cm}
    \includegraphics[width=0.4\textwidth]{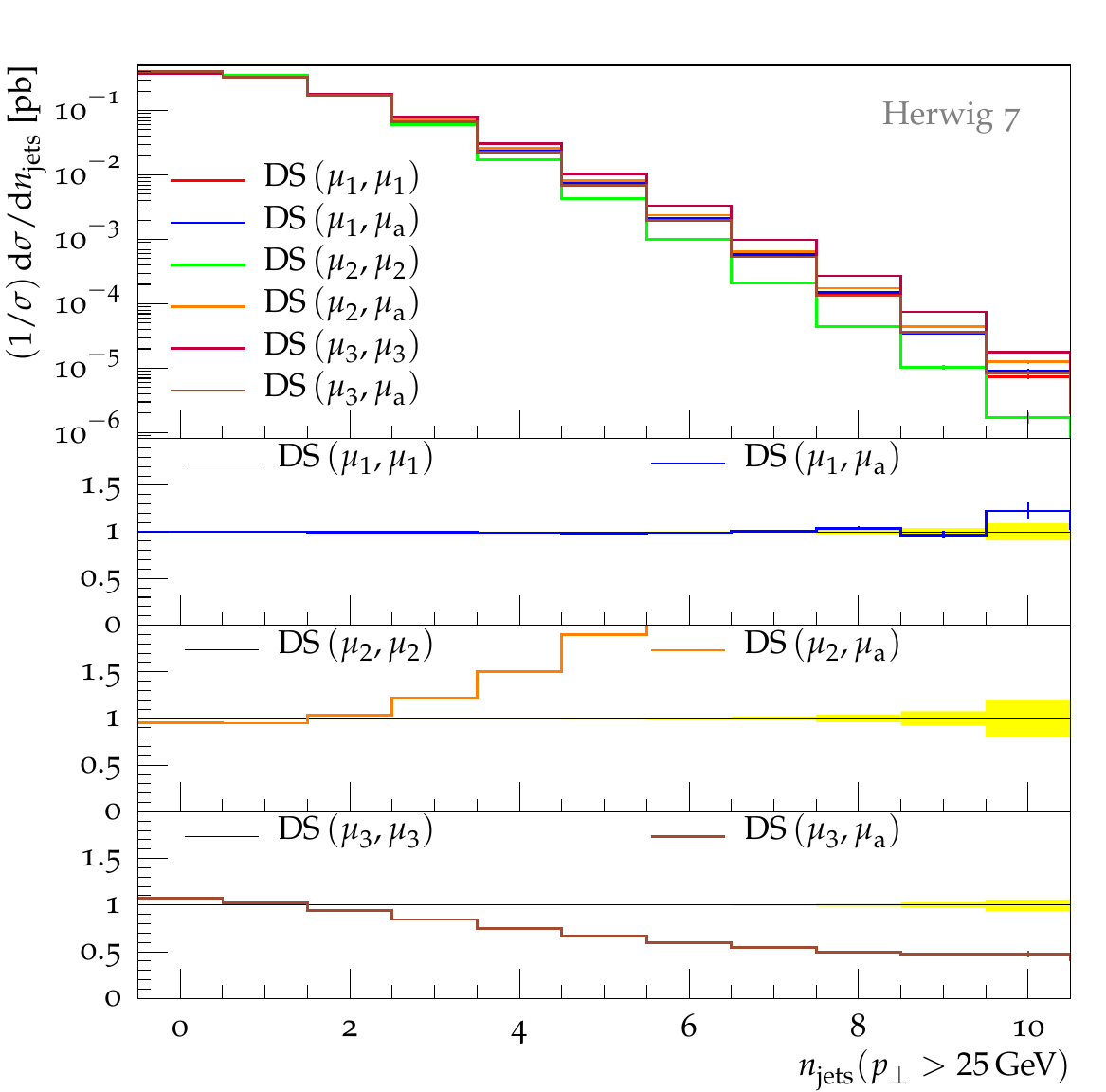}
  \caption{Same as Fig.~\ref{fig:Prod-VetoScale-Jet1Pt-Jet2Pt}, in this case showing the
    $p_\perp$ spectrum of the third hardest jet and the inclusive jet multiplicity distribution
    in the upper and lower row respectively.}
  \label{fig:Prod-VetoScale-Jet3Pt-nJets25}
\end{figure*}

\begin{figure*}
  \centering
    \includegraphics[width=0.4\textwidth]{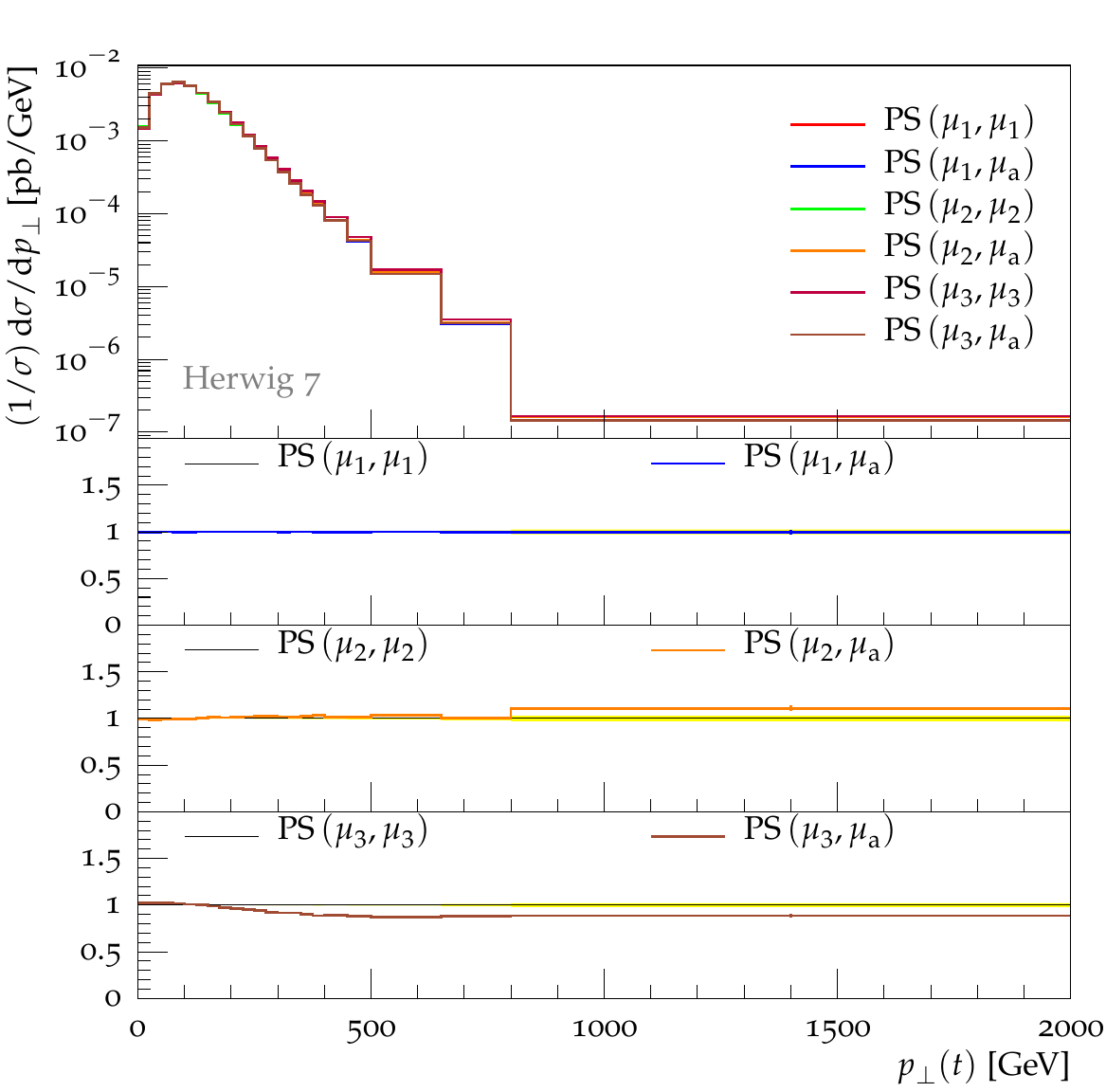}\hspace*{2cm}
    \includegraphics[width=0.4\textwidth]{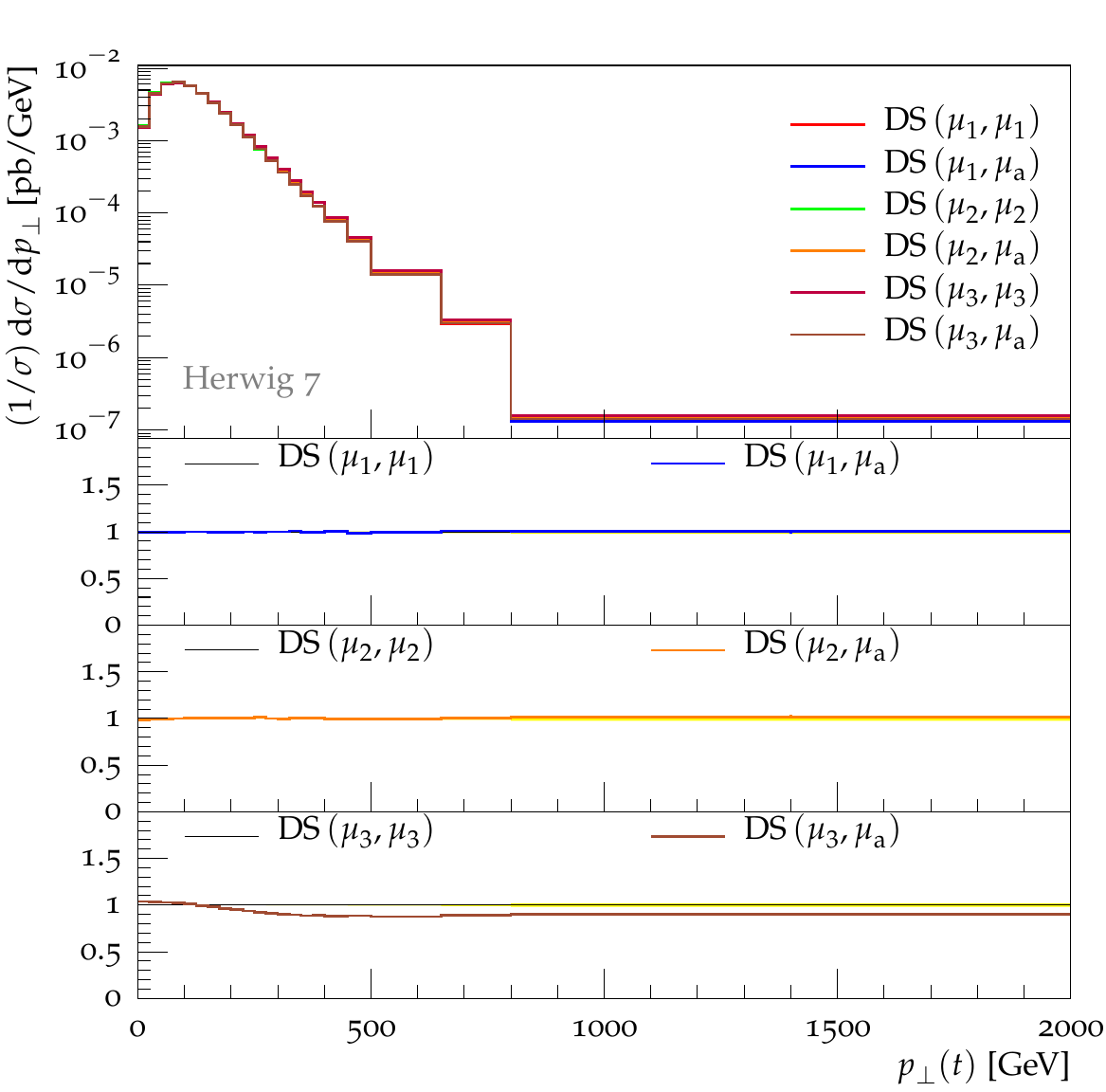}\\  
    \includegraphics[width=0.4\textwidth]{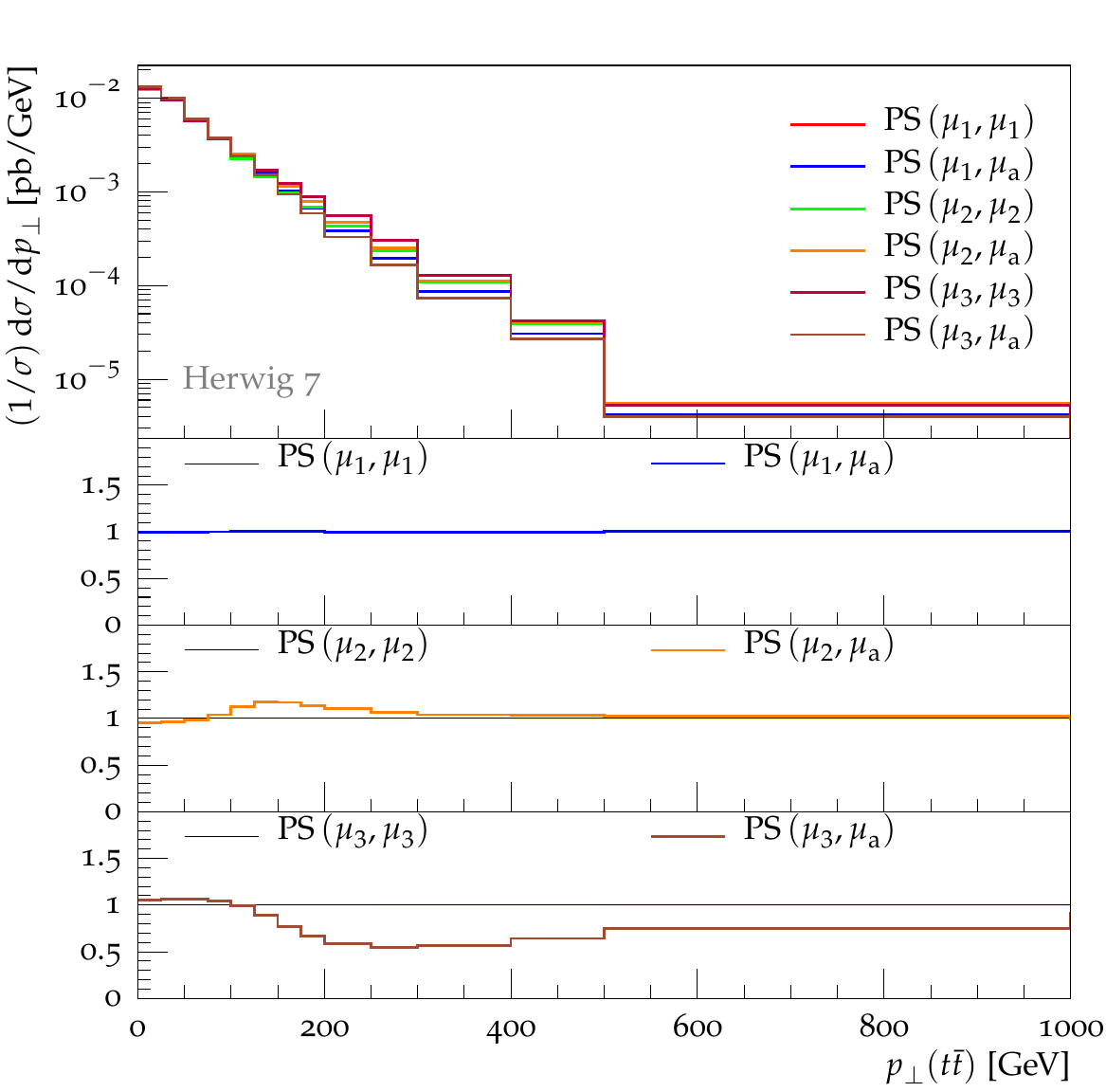}\hspace*{2cm}
    \includegraphics[width=0.4\textwidth]{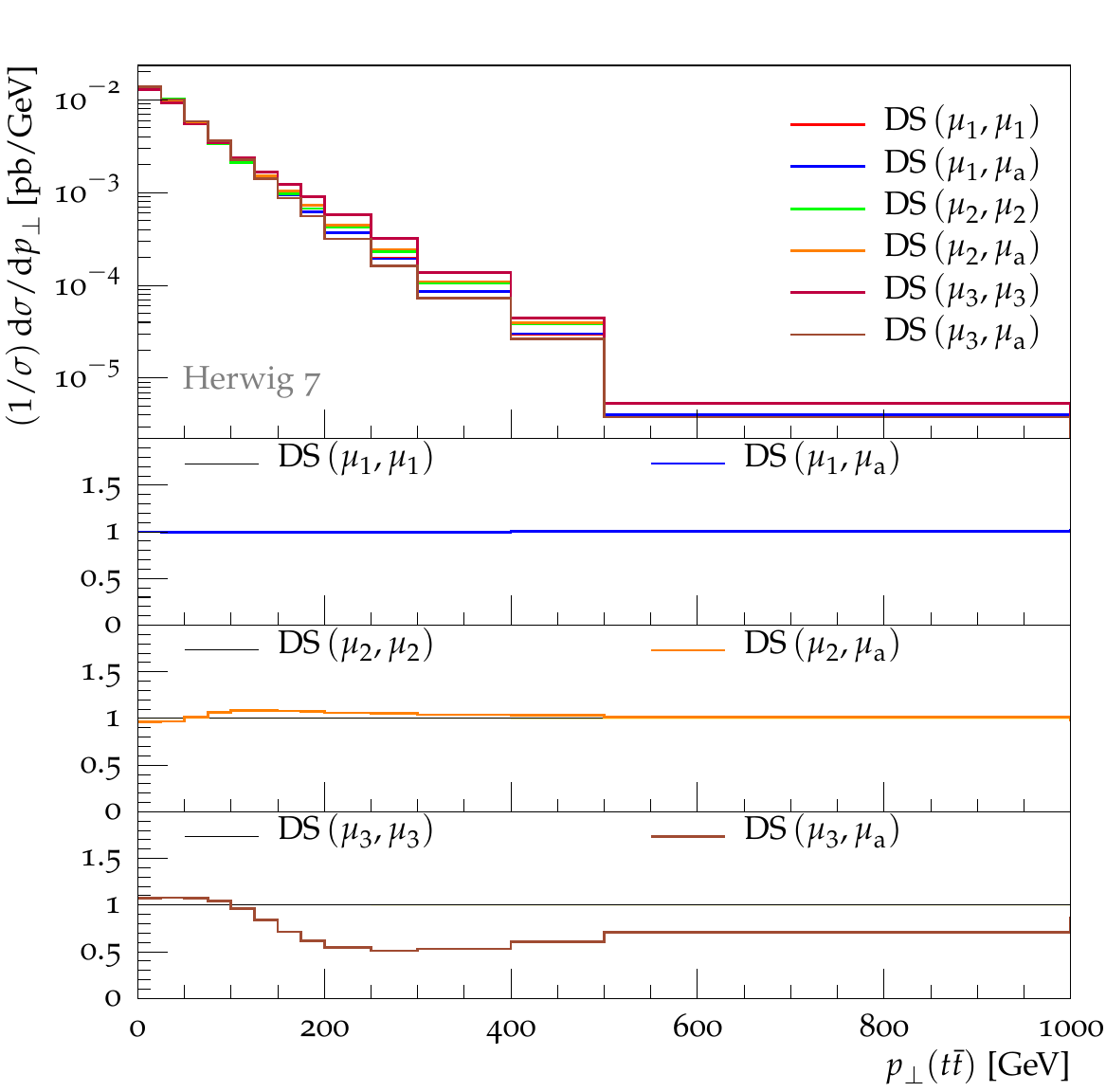}
  \caption{Same as Fig.~\ref{fig:Prod-VetoScale-Jet1Pt-Jet2Pt}, in this case showing the
    $p_\perp$ spectra of the top quark and the $t\bar{t}$-pair in the upper and lower row
    respectively.}
  \label{fig:Prod-VetoScale-TopPt-ttbarPt}
\end{figure*}

In Section~\ref{sec:hardvetoscales} we discussed the role of the hard
veto scale, $\vetoScale$, in MC@NLO-type matching.
In the following we discuss the predictions produced using each
of the three options ($\mu_1$, $\mu_2$, $\mu_3$) for $\hardProcScale$
separately. Given that the $\vetoScale$ directly affects the showering of the
production-level process, we expect to see the largest effects due to the choice
of $\vetoScale$ (which is either
$\vetoScale=\hardProcScale$ or $\vetoScale=\mu_\mathrm{a}$) 
in distributions that reflect the jet activity in each event.
As such these are the distributions that we present for discussion in this section.

Fig.~\ref{fig:Prod-VetoScale-Jet1Pt-Jet2Pt} shows
the transverse momentum distributions of the hardest jet, $p_\perp(j_1)$,
and second hardest jet, $p_\perp(j_2)$, in events showered using the
angular-ordered (PS) and
dipole showers (DS). The scale choices are specified
in the format $(\hardProcScale,\vetoScale)$. 
Similarly, the transverse momentum distributions of the third hardest
jet and the jet multiplicity distributions are shown in
Fig.~\ref{fig:Prod-VetoScale-Jet3Pt-nJets25},
where only jets with transverse momentum greater than
25 GeV are counted in the multiplicity distributions.
Finally, Fig.~\ref{fig:Prod-VetoScale-TopPt-ttbarPt} shows the transverse
momentum distributions of the top quark, $p_\perp(t)$, and the
$t\bar{t}$-pair, $p_\perp(t\bar{t})$, in events showered using the
angular-ordered and dipole showers. In MC@NLO-type events the hard
process, $pp\to t\bar{t}$, is formally accurate to NLO in QCD,
therefore the $p_\perp(t)$ distribution is formally accurate to NLO
whereas the $p_\perp(t\bar{t})$ distribution is accurate only to LO.
Accordingly the dependence of the $p_\perp(t)$ distribution on
$p_\perp(j_1)$ is expected to be modest while the $p_\perp(t\bar{t})$
distribution should be closely related to $p_\perp(j_1)$. Indeed for a
pure NLO cross section we would have the simple one-to-one
relationship, $p_\perp(t\bar{t}) = p_\perp(j_1)$.

We first consider the choice $\hardProcScale = \mu_1$, which in $\clS$-events is identical
to $\mu_\mathrm{a}$, and compare the results for $\vetoScale = \mu_1$ to those for
$\vetoScale = \mu_\mathrm{a}$.

In $\clH$-events with a low or moderate-$p_\perp$ NLO real emission, $\mu_1$ is
larger than $\mu_\mathrm{a}$, however the difference is small enough that we do
not see any corresponding effects at low or moderate-$p_\perp$ in the 
jet-$p_\perp$ distributions in
Fig.~\ref{fig:Prod-VetoScale-Jet1Pt-Jet2Pt} and Fig.~\ref{fig:Prod-VetoScale-Jet3Pt-nJets25}.
It is only in $\clH$-events with the very hardest NLO emissions that
$\mu_\mathrm{a}$ is significantly larger than $\mu_1$. This
is evident from the increase in the $p_\perp(j_2)$ and 
$p_\perp(j_3)$ distributions at high-$p_\perp$ using 
$\vetoScale = \mu_\mathrm{a}$ compared to $\vetoScale = \mu_1$.
The fact that we do not see any difference at high-$p_\perp$ in the
$p_\perp(j_1)$ distribution indicates that this region of the distribution
is filled by high-$p_\perp$ NLO emissions in $\clH$-events.

As we would expect given the discussion above, looking at
Fig.~\ref{fig:Prod-VetoScale-TopPt-ttbarPt}, for $\hardProcScale = \mu_1$ we see no
significant differences due to the choice of $\vetoScale$ in the
$n_\mathrm{jets}$, $p_\perp(t)$ or $p_\perp(t\bar{t})$ distributions.

In summary, $\mu_1$ and $\mu_\mathrm{a}$ are identical in $\clS$-events 
and are similar in most $\clH$-events, which is why we see 
varying differences in jet activity due to the choice of
$\vetoScale$.

Next we consider $\hardProcScale = \mu_2$ for which $\mu_\mathrm{a} > \hardProcScale$ in all events.
In $\clS$-events we have $\mu_\mathrm{a} = 2 \mu_2$ and in $\clH$-events with a 
low-$p_\perp$ NLO first emission we have 
$\mu_\mathrm{a} \sim \sqrt{8/3}\ \mu_2$. 

The larger hard veto scale in such events
explains the increase that we see in the $p_\perp(j_1)$ distributions
in Fig.~\ref{fig:Prod-VetoScale-Jet1Pt-Jet2Pt}
at around $75 \text{ GeV } < p_\perp(j_1) < 250$ GeV. 
The fact that this increase in the rate drops off at around 250 GeV, above which
the distributions using the two different options for $\vetoScale$ become very
similar, suggests that jets harder than this are primarily produced as a
high-$p_\perp$ real emission in $\clH$-events.
In Fig.~\ref{fig:Prod-VetoScale-Jet1Pt-Jet2Pt} and Fig.~\ref{fig:Prod-VetoScale-Jet3Pt-nJets25} 
we observe a large increase in the number of moderate and high-$p_\perp$ second and
third jets for $\vetoScale = \mu_\mathrm{a}$ compared to $\vetoScale = \mu_2$.
The simple fact that $\mu_\mathrm{a} > \mu_2$ in all events means we expect to
see such an increase at moderate values of the jet-$p_\perp$.
In $\clH$-events the difference between $\mu_\mathrm{a}$ and
$\mu_2$ grows with the transverse momentum of the NLO emission. This explains
why using $\vetoScale = \mu_\mathrm{a}$, as opposed to $\vetoScale = \mu_2$,
gives rise to an increase in the $p_\perp(j_2)$ and $p_\perp(j_3)$ 
distributions at high-$p_\perp$ that grows with the jet-$p_\perp$.
In Fig.~\ref{fig:Prod-VetoScale-Jet3Pt-nJets25} we see 
a large increase in the number of events with high jet-multiplicities 
for $\vetoScale = \mu_\mathrm{a}$ compared to $\vetoScale = \mu_2$. This corresponds 
to the increase that we see in the $p_\perp(j_2)$ and $p_\perp(j_3)$ 
distributions. 

The moderate difference in the $p_\perp(j_1)$ distribution is
not evident in the $p_\perp(t)$ distributions, in
Fig.~\ref{fig:Prod-VetoScale-TopPt-ttbarPt}.
However, it is evident in the 
$p_\perp(t\bar{t})$ distribution, which is very sensitive to the hardest 
emission.

In summary, $\mu_\mathrm{a}$ is larger than $\mu_2$ in all events therefore we
see an increase in jet activity using $\vetoScale = \mu_\mathrm{a}$ compared to
$\vetoScale = \hardProcScale = \mu_2$.

Finally, we consider the results for $\hardProcScale = \mu_3$, the invariant mass of
the $t\bar{t}$ pair, which is a large scale compared to $\mu_1$ and $\mu_2$.

The $p_\perp(j_1)$ distributions in Fig.~\ref{fig:Prod-VetoScale-Jet1Pt-Jet2Pt} display a significant decrease for
$p_\perp(j_1) > 100 \text{ GeV }$ using
$\vetoScale = \mu_\mathrm{a}$ compared to $\vetoScale = \mu_3$. This indicates that for
the choice $\vetoScale = \hardProcScale = \mu_3$, the hardest jet is predominantly produced as
the first shower emission, as opposed to NLO emission in $\clH$-events,
up to a much higher scale $p_\perp(j_1)$ than for either
$\vetoScale = \hardProcScale = \mu_1$ or $\vetoScale = \hardProcScale = \mu_2$.
The $p_\perp(j_2)$ and $p_\perp(j_3)$ distributions in
Fig.~\ref{fig:Prod-VetoScale-Jet1Pt-Jet2Pt} and Fig.~\ref{fig:Prod-VetoScale-Jet3Pt-nJets25}
also display a decrease in the rate for the choice $\vetoScale = \mu_\mathrm{a}$
compared to using $\vetoScale = \mu_3$. This is expected 
given that $\mu_\mathrm{a} < \mu_3$.
In Fig.~\ref{fig:Prod-VetoScale-Jet3Pt-nJets25} the dipole shower with $\vetoScale = \mu_\mathrm{a}$
displays a decrease in the number of high-multiplicity events compared to using
$\vetoScale = \mu_3$. This is in straightforward agreement with the decreases seen in
the jet-$p_\perp$ distributions.
As we expect, for the two choice of $\vetoScale$,
we also see a large difference in the $p_\perp(t\bar{t})$
distribution, in
Fig.~\ref{fig:Prod-VetoScale-TopPt-ttbarPt}, which
matches the difference in the $p_\perp(j_1)$ distribution. 

The jet-multiplicity distribution predicted using the angular-ordered
shower, displays a less consistent change between the scale choices
$\vetoScale = \mu_\mathrm{a}$ and $\vetoScale = \mu_3$.  In fact for
$n_\mathrm{jets} > 5$ we actually see an increase in the distribution
using $\vetoScale = \mu_\mathrm{a}$ compared to $\vetoScale = \mu_3$.
This is consistent with the behaviour seen in the jet-$p_\perp$
distributions, in which we see the difference due to the choice of
$\vetoScale$ reduce considerably between the $p_\perp(j_1)$ and
$p_\perp(j_3)$ distributions.  As in the dipole shower predictions,
for the two choices of $\vetoScale$, we see a large and corresponding
difference in the $p_\perp(t\bar{t})$ and $p_\perp(j_1)$
distributions.

We also see a small change, due to the choice of $\vetoScale$,
in the $p_\perp(t)$ distribution, for both showers. As discussed above, the impact of the hardest
emission on this distribution is a NLO effect, however the difference in the
$p_\perp(j_1)$ distribution due to the hard veto scale choice is large enough to
induce a sizeable difference in the $p_\perp(t)$ distribution.

In this discussion we have compared the effect of using $\vetoScale =
\hardProcScale$ and $\vetoScale = \mu_\mathrm{a}$ for three different choices of
$\hardProcScale$.  As there is no first principles choice for the scale $\vetoScale$, the aim of
this discussion is to highlight that when we use MC@NLO-type matching we have to
make a choice for this scale. We have shown that in general
using a smaller hard veto scale reduces the predicted jet activity in
an event, whereas using a larger hard veto scale generally increases
the predicted jet activity. We use $\mu_\mathrm{a}$ to reflect the
transverse momenta of the objects outgoing from the hard process.  We
leave further investigation of potential scale choices to future work.
We return to this topic in
Section~\ref{sec:mcatnlo-veto-scale-full-process} in which we
investigate the effect of the choice for $\vetoScale$ on the prediction of distributions
measured in experiment.
As far as the corrections to the decay, and similar variations therein
are considered we cannot find any significant impact on the
observables considered here, which are mostly insensitive to changes
in the decay system.

\section{Boosted Topologies}
\label{sec:Boosted}

\begin{figure}
  \centering
  \includegraphics[width=0.4\textwidth]{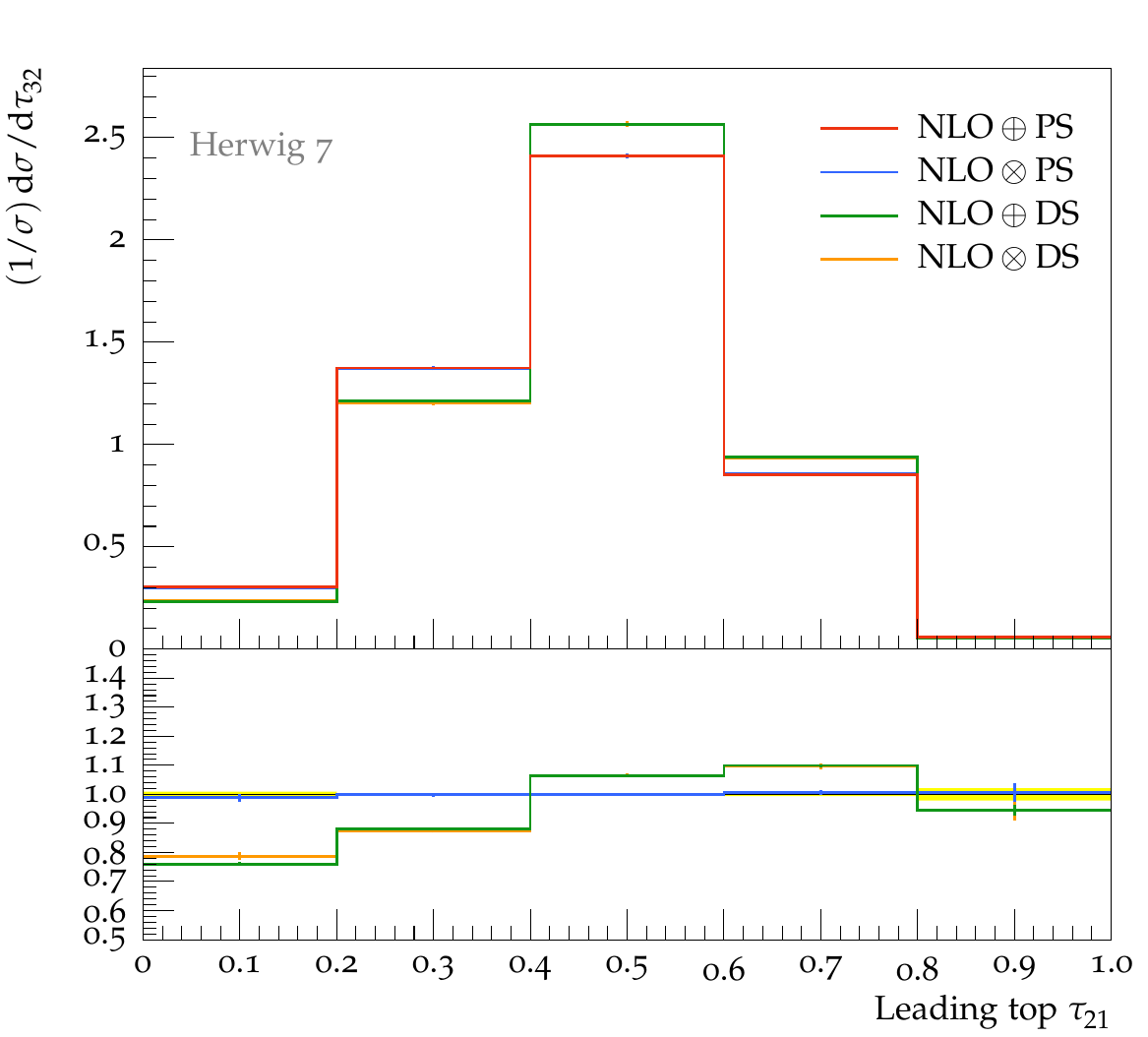}\\
  \includegraphics[width=0.4\textwidth]{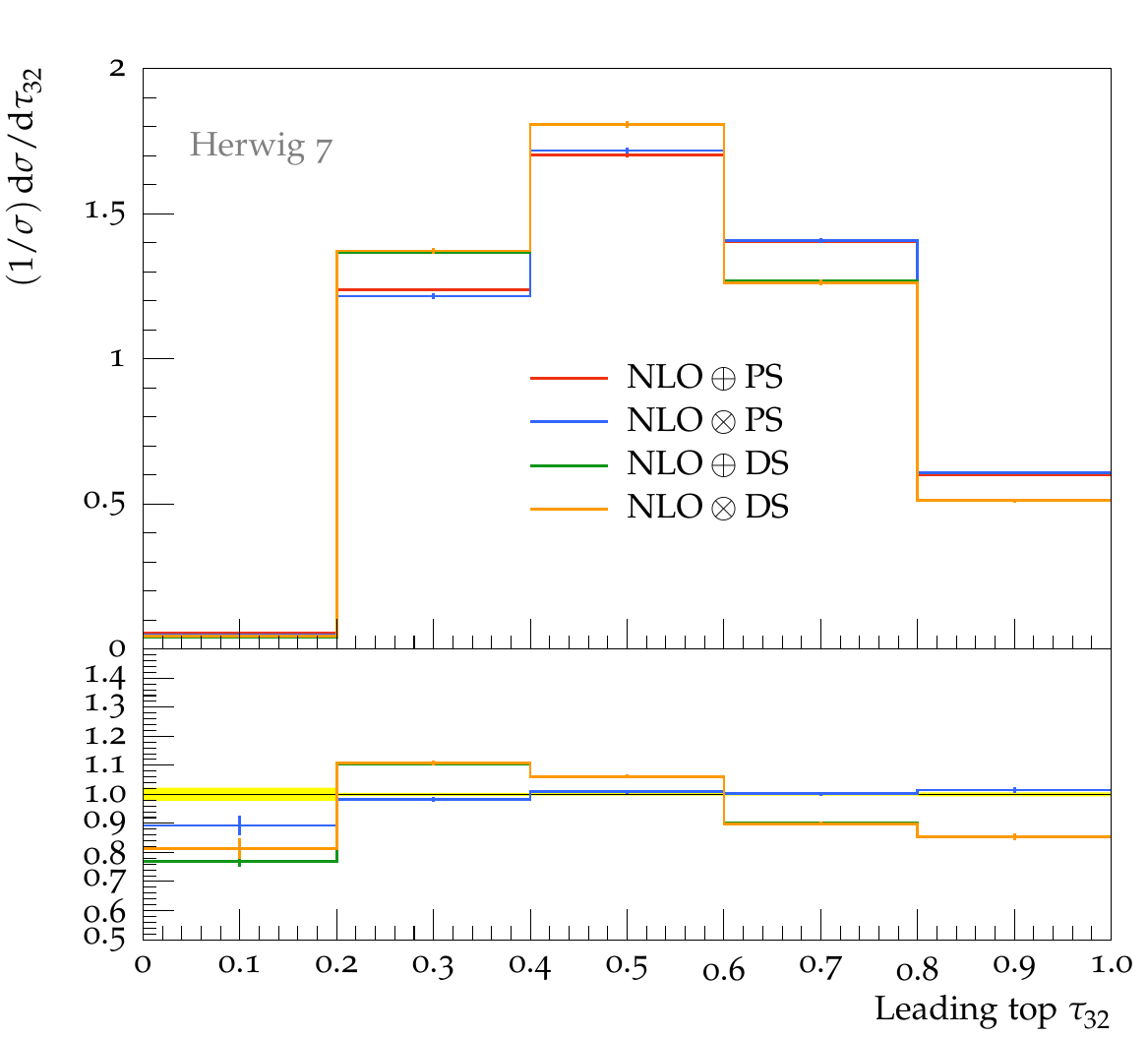} \\
  \includegraphics[width=0.4\textwidth]{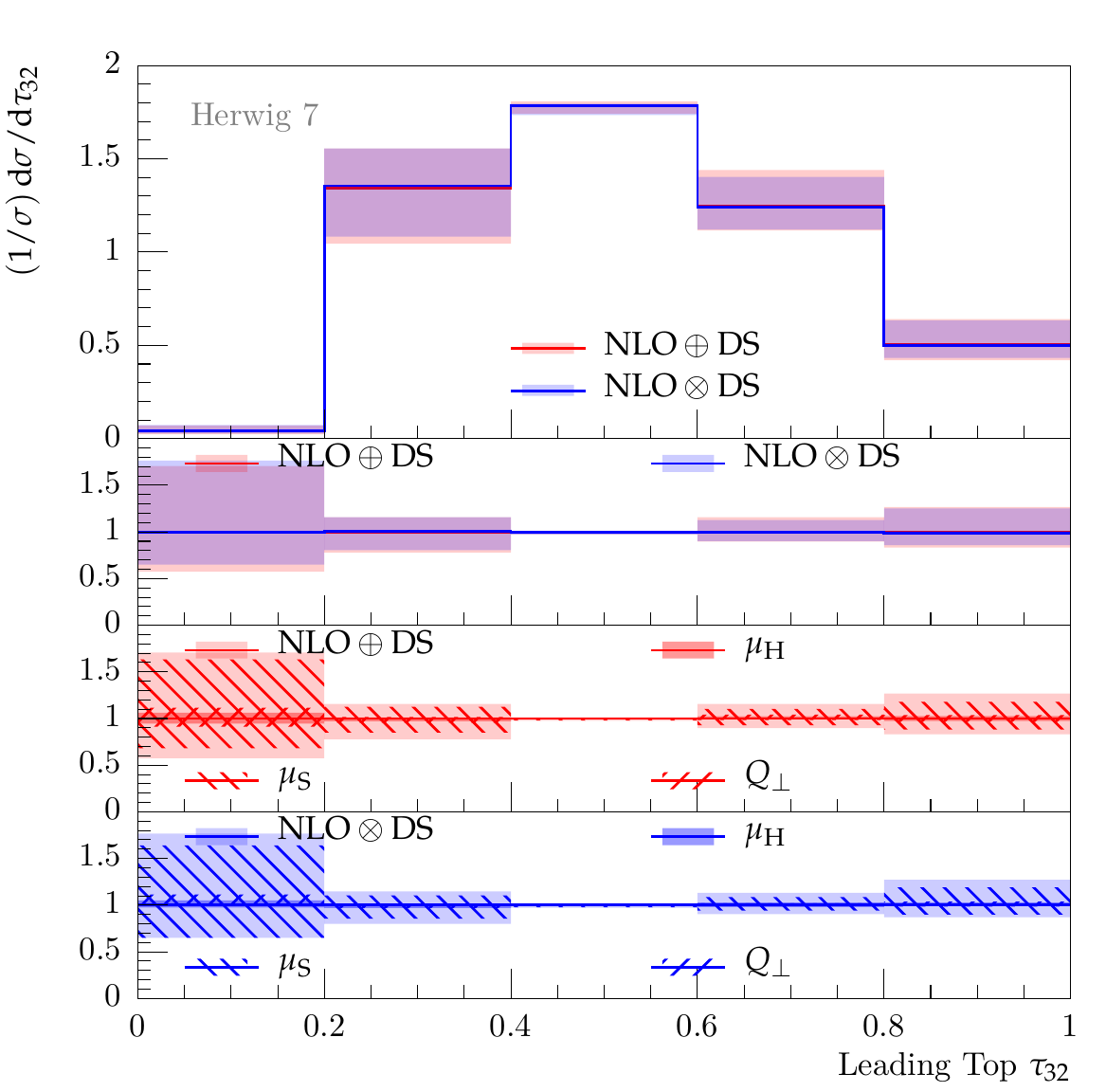}
  \caption{The N-subjettiness ratios  $\tau_{21}$ (top) and $\tau_{32}$ (middle and bottom)
    for the large-radius jet 
    associated with the highest momentum top-quark in boosted $t\bar{t}$  events. The top and middle plots 
    show comparisons of the angular-ordered ($\mathrm{PS}$) and 
    dipole ($\mathrm{DS}$) showers with both the MC@NLO-type (NLO$\oplus$)
    and Powheg-type (NLO$\otimes$) matching schemes. The bottom plot shows 
    the effects of scale variations on $\tau_{32}$ using the dipole shower
    with each of the NLO matching schemes.}
  \label{fig:boosted-subjet}
\end{figure}

The energy and luminosity provided at the LHC allow studies of
top quarks with transverse momenta much higher than the top mass. In such
cases the decay products of the top quark are generally not well separated.
The $b$~quark, and decay products from the $W$~boson are often collimated,
forming a single large jet referred to as a `boosted' top jet. This 
topology has several distinct difficulties compared to the lower momentum
 cases. 

Firstly, large-radius jets originating from top quarks need to be discriminated from
large-radius jets originating from other coloured particles or from the decays of
W and Z bosons. This discrimination, referred to as tagging, typically makes use
of the substructure of the large jet. The three pronged nature of the top-quark
decay leaves a characteristically three-pronged structure within the large jet
which is not usually found in boson decays or pure QCD jets. In practice 
many different techniques are used to tag large jets as originating from 
a top quark. Whether it is through machine learning applied directly
to jet-algorithm inputs or techniques based directly on high-level observables
designed to provide substructure information these taggers all ultimately make
use of the distribution of energy within a jet to perform tagging. The 
performance of taggers is often estimated from simulation and it is therefore
important to understand the impact of the various choices made in the Monte Carlo
simulation on the
description of the substructure of large jets originating from boosted top quarks.

As a probe of the sensitivity of jet substructure to the Monte Carlo approach
we examined the N-subjettiness\,\cite{Thaler:2011Nsub} of boosted top quarks
produced with \Herwig\,7. N-subjettiness measures the degree to which the constituents of
a subjet are collimated along its N primary axes. Ratios of N-subjettiness
values for different values of N are often used to tag large-radius jets. The
ratios of 2-subjettiness to 1-subjettiness ($\tau_{21}$) and 3-subjettiness
to 2-subjettiness ($\tau_{32}$) were compared using different
\Herwig\,7 
settings as shown in Fig.~\ref{fig:boosted-subjet}.
We expect radiation at intermediate scales to have the largest impact
as hard emissions are typically outside the radius of the jet, while very soft
emissions have little impact on the overall substructure. Variation of the shower 
scale, $\mu_s$, is found to have the largest contribution to the uncertainty envelope, whereas the 
contributions from the other scale variations are negligible. The choice of matching
scheme is also found to have very little impact, except for
the lowest $\tau_{32}$ bin. On
the other hand, comparing the dipole shower and
angular-ordered shower algorithms shows more significant differences,
comparable to the uncertainty envelopes produced by scale variations.

A second difficulty for high momentum top quark simulation is that of CPU time
required to populate the high-$\text{p}_{\perp}$ region of phase space
targeted by these analyses. Given the steeply falling $t\bar{t}$ 
cross-section as a function of the transverse momentum of the top quark, 
analyses targeting boosted topologies are typically targeting $\sim 1$\% of the total
phase space or less. Simulating the inclusive phase-space can therefore be
very inefficient, requiring orders of magnitude more events to be fully simulated
than are actually of interest. In order to reduce the resources required for
simulation a simple cut mechanism is available in the \Matchbox
framework of \Herwig~7.
This mechanism makes several options available to improve the efficiency
of producing high-$\text{p}_{\perp}$ top quarks.

Example runs with cuts on the transverse momentum of the leading top quark 
of 200, 300 and 600~GeV were performed to test the efficiency. They showed no significant
change in the distributions of weights. For a centre-of-mass energy of
13~TeV kinematic bins well beyond the cut values increased their overall 
statistics by factors of $\sim5$, $\sim20$ and $\sim500$ respectively for the same total 
number of events, and relative errors were reduced accordingly. No appreciable impact on the
computing time per event was found, allowing a significant reduction
in computing power to achieve the same or better statistical power. 
A code snippet of an input card to produce similar cuts is provided 
as an example in Appendix~\ref{app:boostedCut}.

\section{Data Benchmarks}
\label{sec:data-benchmarks}
  
After investigating several aspects related to the predictions provided by our
improved simulation we now turn to an in-depth comparison to experimental data
in order to quantify how the different algorithms and their intrinsic
uncertainties compare to existing collider data. We use existing and publicly
available \texttt{Rivet} analyses, for which the collision energy, $\sqrt{s}$,
at which each experimental result was measured and the final-states included
are summarised in the following text. Specific details of the experimental
analyses are available in the references provided. All of the measurements
presented in this section are taken in the `combined channel', {\it i.e.} including
both electron and muon final states. Unless otherwise stated, the hard process
scale used to generate events is
\begin{equation}
  \hardProcScale = \frac{m_{\perp,t} + m_{\perp,\bar{t}}}{2} .
\end{equation}
This scale was chosen because it was found to give rise to reasonable predictions of
several observables sensitive to jet activity using MC@NLO-type matching.
In particular we compared predictions of several observables included in the
publicly available \texttt{Rivet} analyses for Refs.~\cite{Aad:2014iaa,Aad:2015eia}
obtained using
$\hardProcScale=\mu_{1,2,3}$, {\it i.e.} the three scales defined in Section~\ref{sec:hardvetoscales}.
We use the default choice, $\vetoScale = \hardProcScale$, for the hard veto scale in all runs
apart from those in which this is the scale of interest. Similarly, the \resummation profile scale is used in all runs unless otherwise stated.

The default angular-ordered and dipole
shower tunes of \Herwig~7.1.1 are used in all runs with the respective
showers.  The PDF set used is again\linebreak \texttt{MMHT2014nlo68cl} while
$\alpha_S$ is defined separately by using the tuned value for each
shower. We use a five-flavour scheme in the runs using the angular-ordered
shower, with massless incoming bottom quarks, and the four-flavour scheme in
runs using the dipole shower, which treats partons of a given flavour as having the
same mass in both the initial and final states. The masses of the bottom quark and top quark are
set to 4.2 GeV and 174.2 GeV, respectively, while all other quarks are
considered to be massless.

All distributions that are not normalised to their integral are scaled to the
appropriate next-to-next-to-leading order cross section, as described for the
investigation of the production-level process in
Section~\ref{sec:UncertaintyBenchmarks}. The NNLO cross sections are
$173.60\ {\rm pb}$ and $247.74\ {\rm pb}$ for $7\ {\rm TeV}$ and $8\ {\rm
  TeV}$ collisions, respectively.

\subsection{Scale Variations}
\label{sec:scale-variations-full-process}

In Section~\ref{sec:scale-variations-prod-level} we discussed the uncertainty
on predictions of distributions in the production-level process due to scale
variations in the simulation. In this section we complete this discussion by
looking at the uncertainty on predictions of distributions in the
full-process, including top quark decays and hadronization. We perform the
same scale variations as in Section~\ref{sec:scale-variations-prod-level} and
the reader is referred to that discussion for details. We highlight that the
veto scale in the showering of decay processes is fixed at the mass of the
decayed particle and is not varied.  We compare predictions obtained using the
angular-ordered and dipole showers with several experimental
measurements.

\begin{figure}
  \centering
  \includegraphics[width=0.4\textwidth]{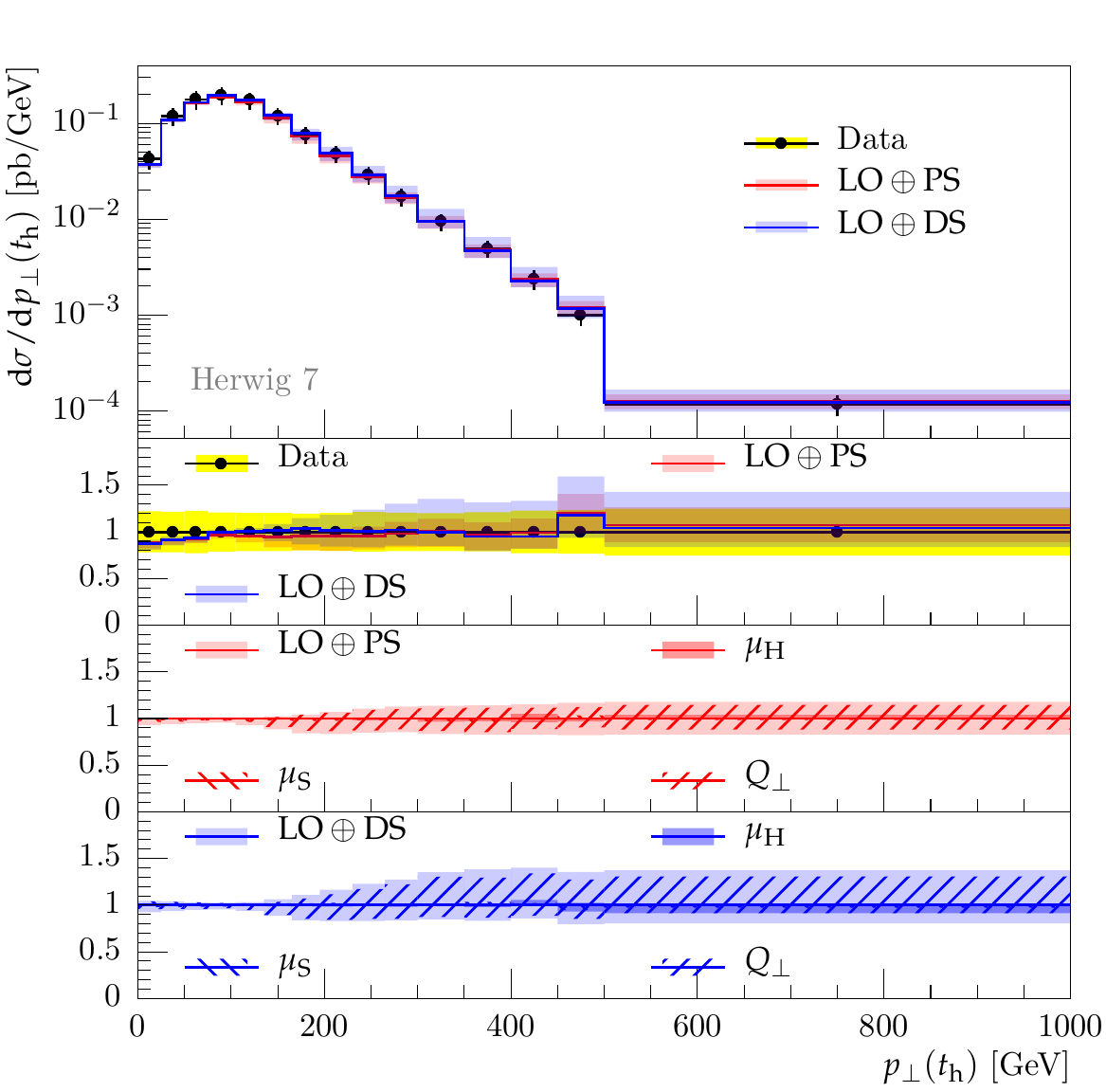}\\
  \includegraphics[width=0.4\textwidth]{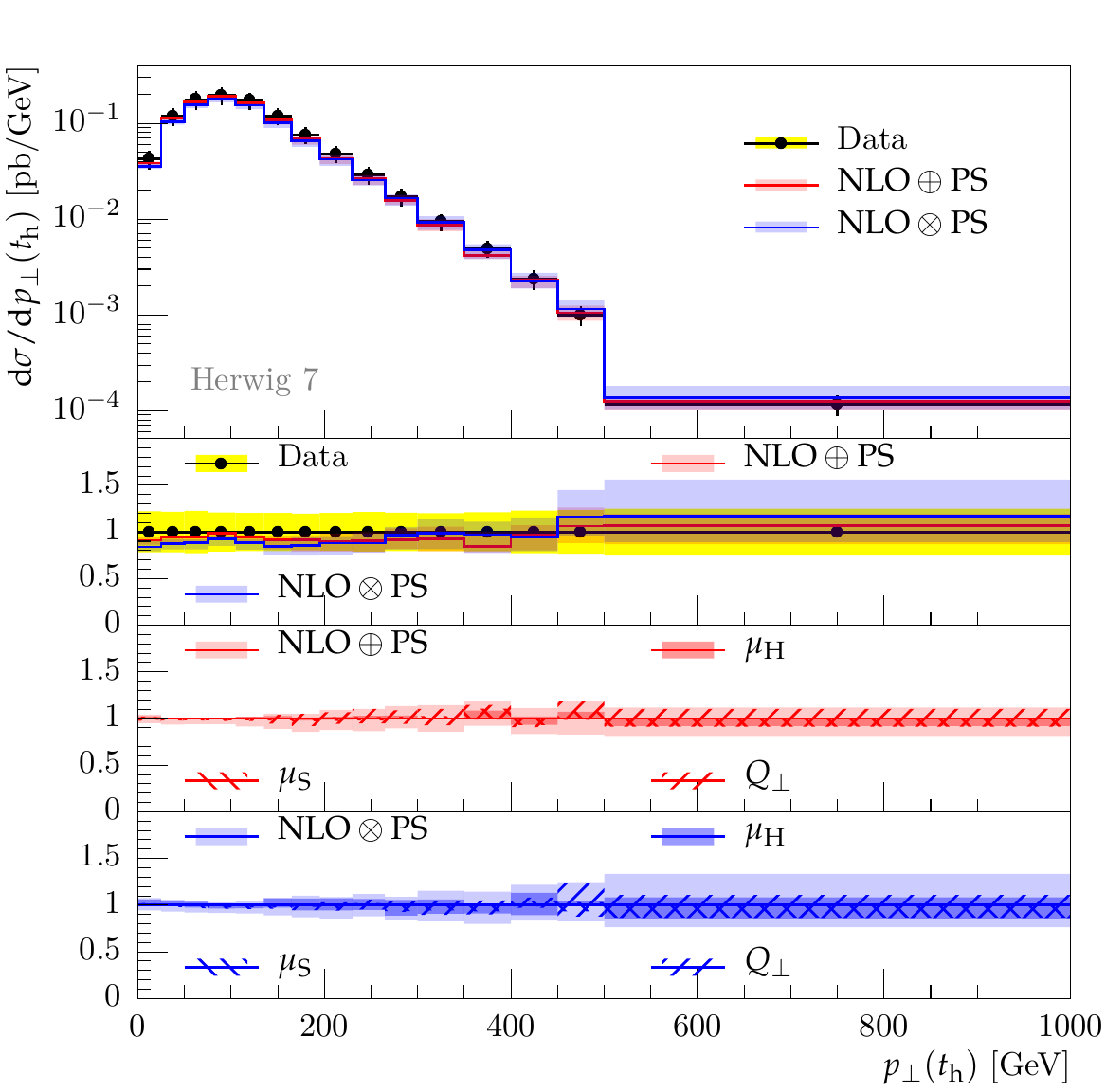}\\
  \includegraphics[width=0.4\textwidth]{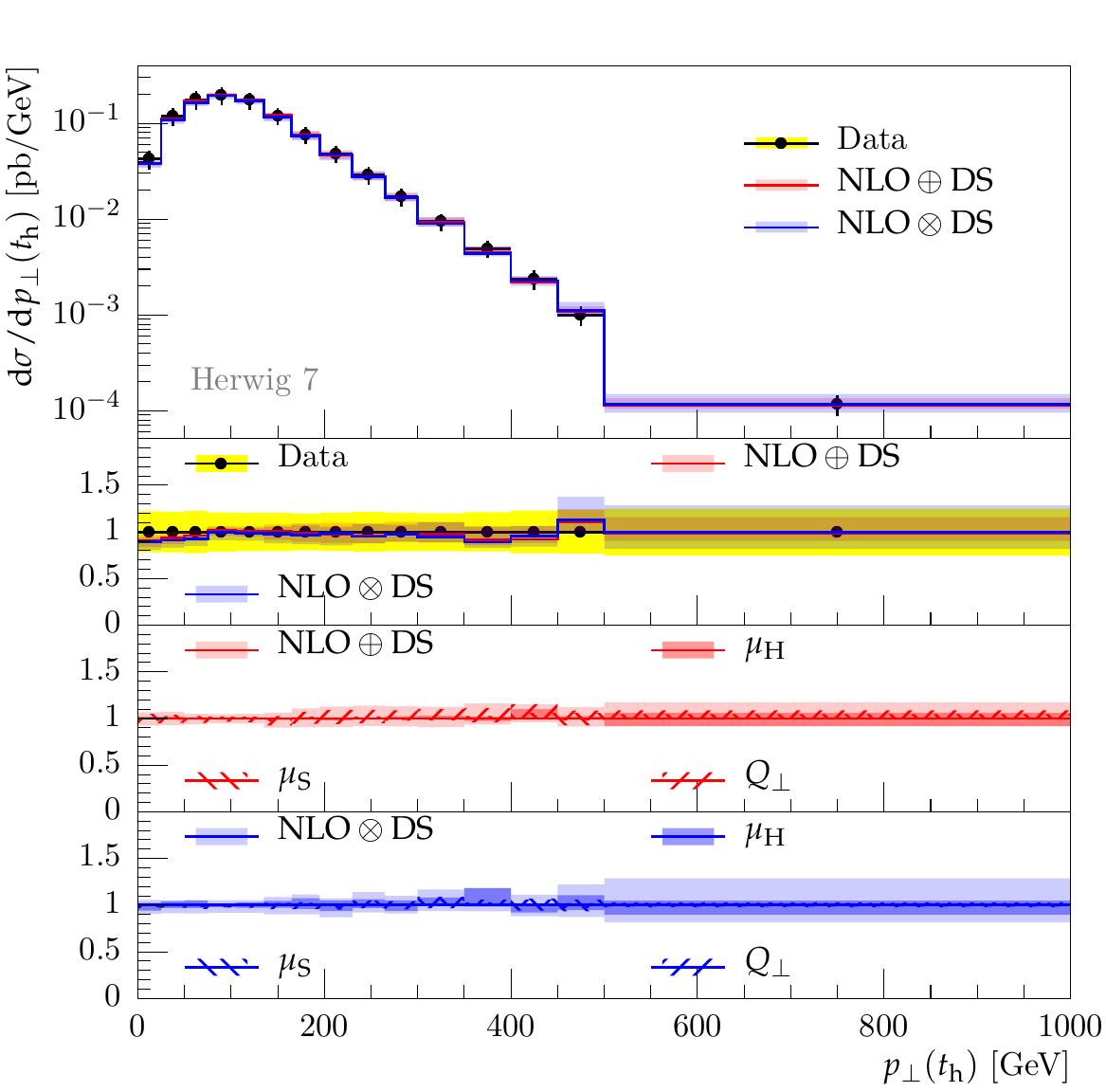}
  \caption{The transverse momentum of the reconstructed hadronically decaying
    top quark measured by ATLAS in semileptonic $8\ {\rm TeV}$ $pp\to
    t\bar{t}$ events~\cite{Aad:2015mbv}. The top plots shows leading-order production
    with angular-ordered~(PS) and dipole~(DS) parton showers, the middle plot
    NLO production matched to the
    angular-ordered parton shower while the bottom plot
    shows NLO production matched to the dipole shower. Two
    NLO matching schemes,  MC@NLO-type ($\mathrm{NLO} \oplus$) and
    Powheg-type ($\mathrm{NLO} \otimes$), are used.
  }
  \label{fig:FullProc-ScaleVar-TopPt}
\end{figure}
We first look at two observables for which we have considered analogous
results in the production-level discussion, namely the $p_\perp(t)$ and jet
multiplicity distributions. In Fig.~\ref{fig:FullProc-ScaleVar-TopPt} we show
predictions for the distribution of the transverse momentum of the hadronically
decaying top quark as measured by the ATLAS collaboration~\cite{Aad:2015mbv}
in semileptonic $pp \to t\bar{t}$ events at a centre-of-collision energy of
$8\ {\rm TeV}$.  The LO matrix element plus parton shower and NLO-matched
predictions for the angular-ordered and the dipole showers are included.
The dominant source of uncertainty on the LO predictions is the variation of
$\vetoScale$. This is in contrast to the production-level result in which
there was no dominant source of uncertainty. We have confirmed that this
difference is due to the different choice of the central hard process scale,
which in turn is used as the central hard veto scale. This distribution is
only moderately sensitive to the parton
shower.
However, with a larger central
hard veto scale the upper variation allows the production of jets that are
hard enough to affect the distribution and give rise to a larger
uncertainty envelope. The larger uncertainty on the dipole shower prediction
reflects the less restricted emission phase space.  The
central MC@NLO-type and Powheg-type predictions display good agreement for both showers
and there is no single dominant source of uncertainty in the NLO-matched
results.

\begin{figure}
  \centering
  \includegraphics[width=0.4\textwidth]{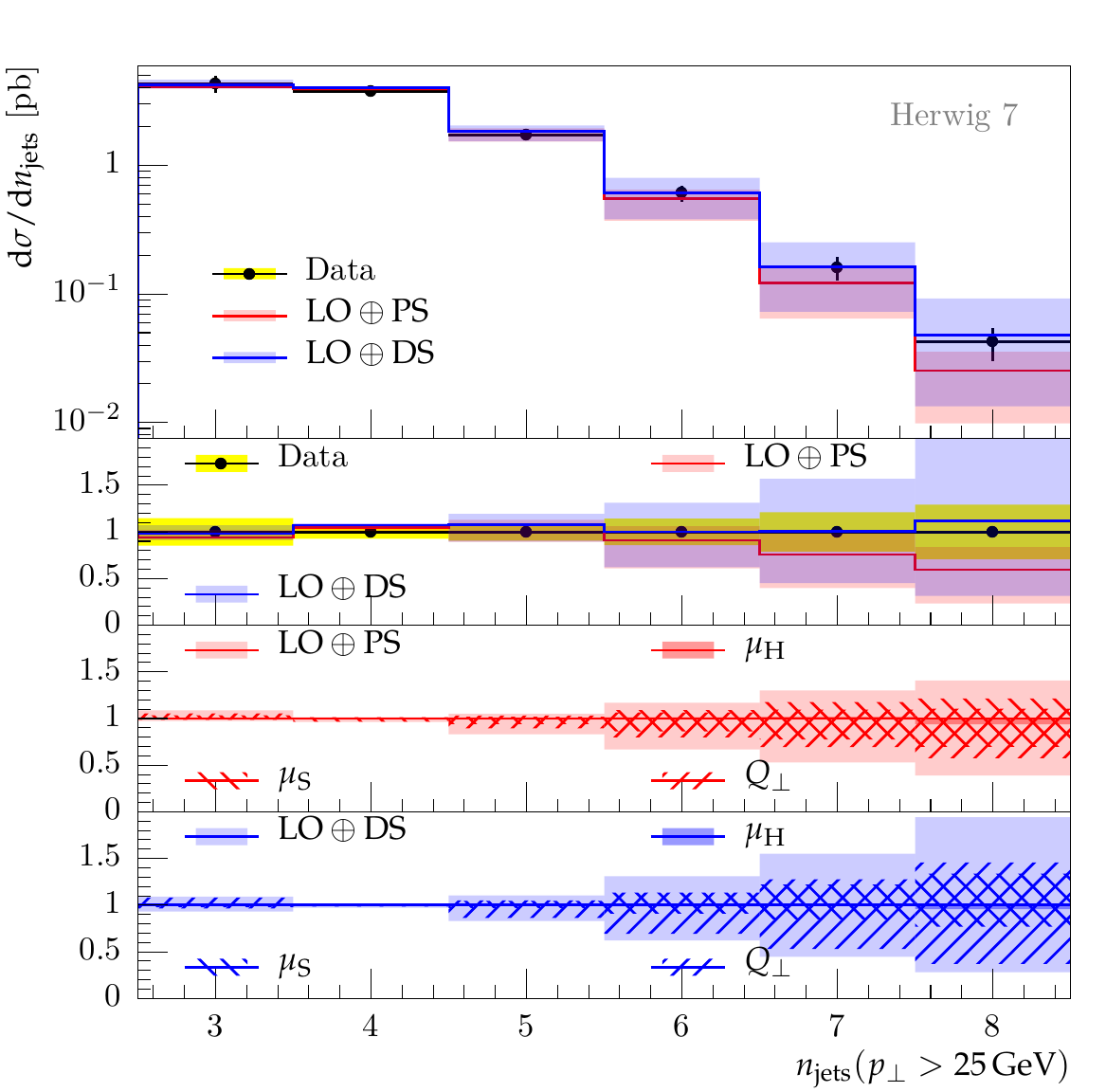}\\
  \includegraphics[width=0.4\textwidth]{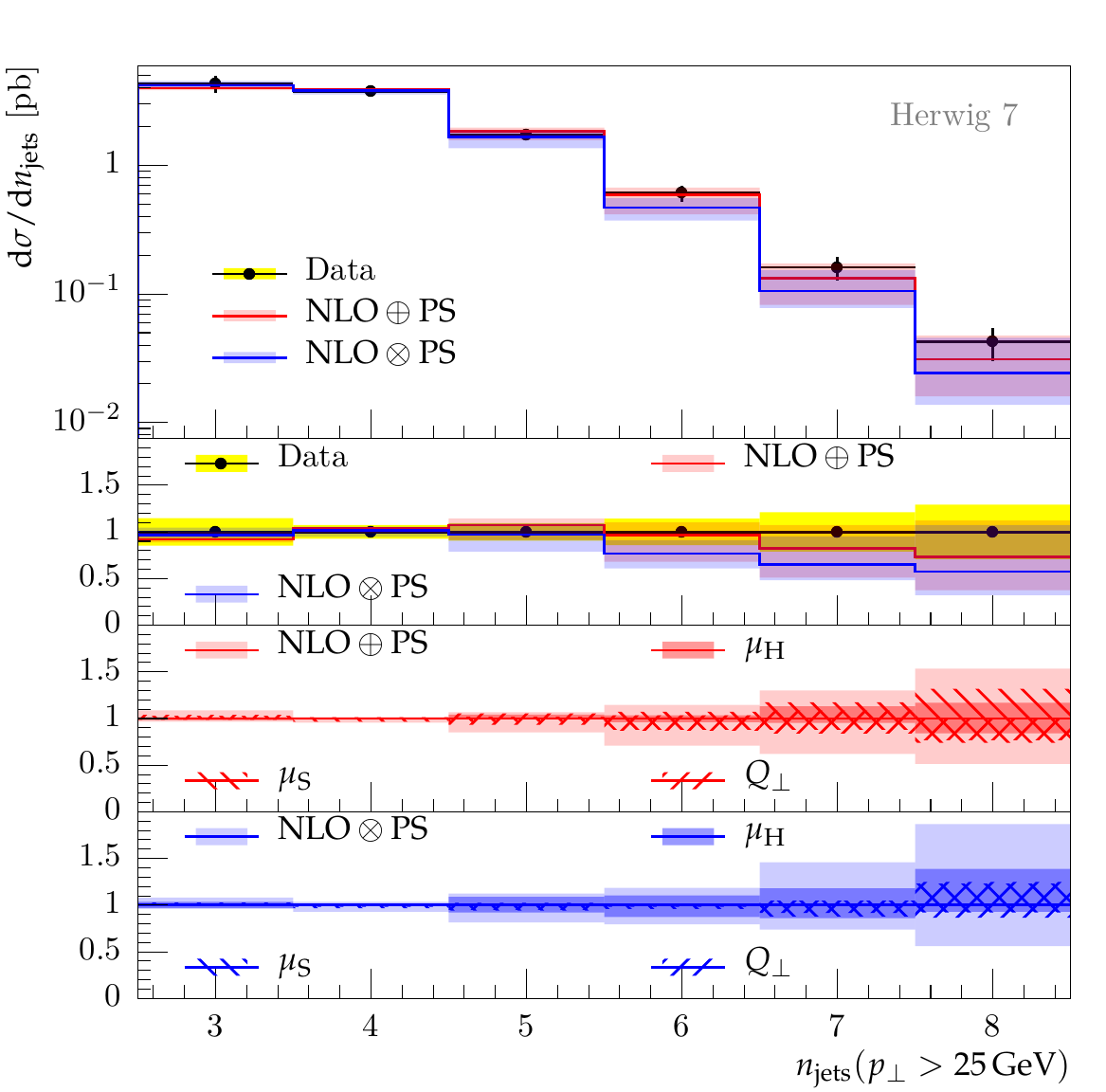}\\
  \includegraphics[width=0.4\textwidth]{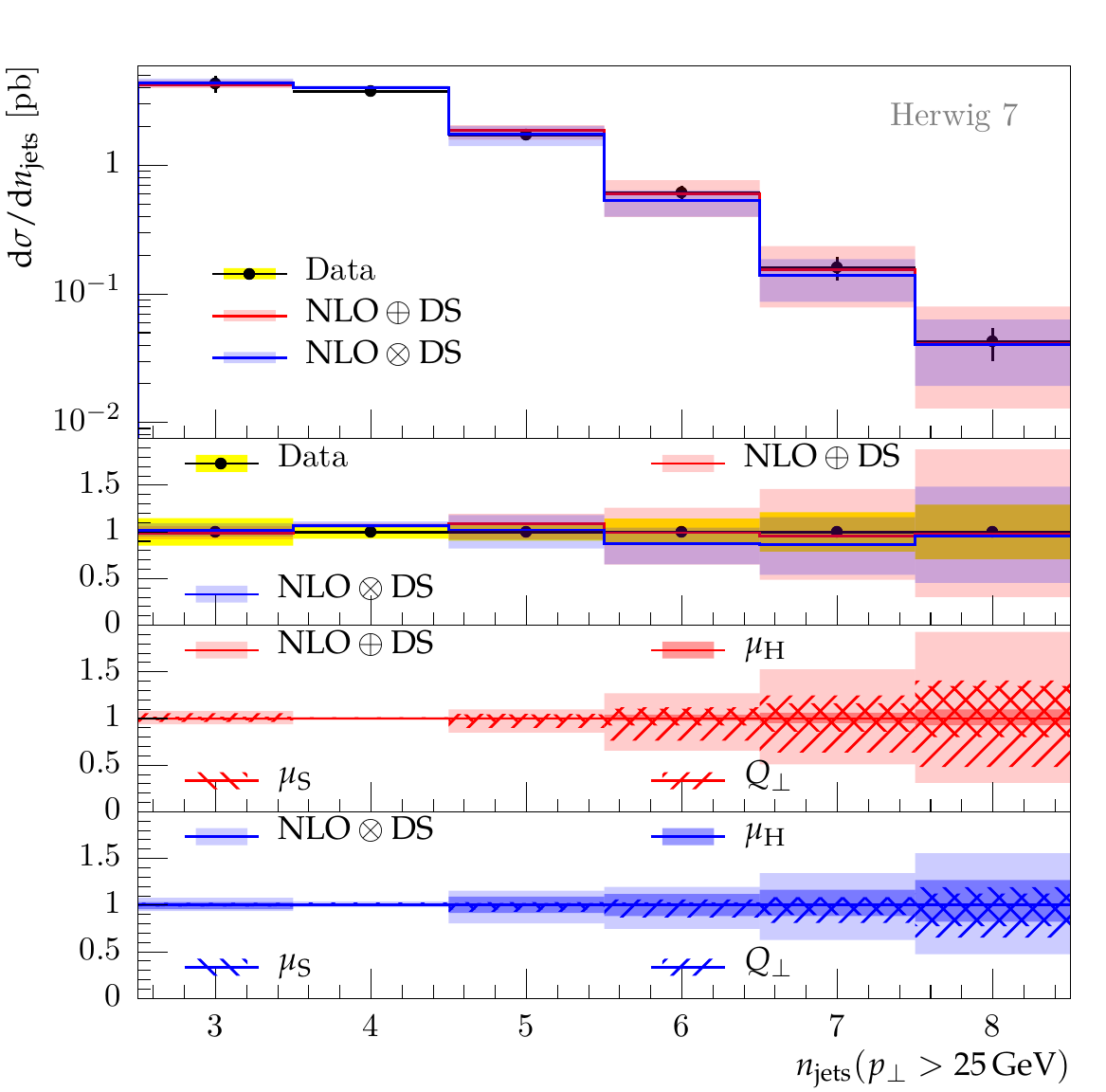}
  \caption{The multiplicity distribution of jets with transverse momentum greater than 25\,GeV, measured
    in semileptonic 7\,TeV $pp\to t\bar{t}$ events by
ATLAS~\cite{Aad:2014iaa}.
    The theoretical predictions are the same as those described in the caption of
    Fig.\ref{fig:FullProc-ScaleVar-TopPt}.
  }
  \label{fig:FullProc-ScaleVar-nJets25}
\end{figure}

In Fig.~\ref{fig:FullProc-ScaleVar-nJets25} we show predictions for the jet
multiplicity distribution for jets with $p_\perp$ greater than $25\ {\rm GeV}$
measured by the ATLAS collaboration~\cite{Aad:2014iaa} in semileptonic $pp
\to t\bar{t}$ events at a centre-of-collision energy of $7\ {\rm TeV}$.
Comparing the LO predictions we find that, in general, the dipole shower
predicts larger multiplicities than the angular-ordered shower and,
particularly at higher multiplicities, the uncertainty on the dipole shower
prediction is greater than that on the angular-ordered shower prediction. Both
of these observations again reflect the differences in the phase-space
restrictions of the two showers, in particular that the dipole shower has no
explicit angular-ordering restriction on emissions. The variations of the
scales directly related to the shower, $\vetoScale$ and $\showerScale$, both
contribute to the total uncertainty which demonstrates the dependence of this
distribution on the parton shower. The larger contribution is from the
variations of $\vetoScale$ and reflects that this scale directly affects the
phase space available to shower emissions.

The MC@NLO-type and Powheg-type predictions display reasonable agreement with each other for
both showers. This was not the case in the production-level results and
reflects the different choice for the central hard process scale for the production-level and full
process predictions. 
In the angular-ordered shower predictions there is no
single dominant source of uncertainty using either matching scheme. In the
dipole shower predictions using MC@NLO-type matching the uncertainty due to the
variation of $\vetoScale$ is significant and reflects the discussion in
Section~\ref{sec:mcatnlo-veto-scale-prod-level} and
Section~\ref{sec:mcatnlo-veto-scale-full-process} on the choice of the hard veto
scale in MC@NLO-type matching.  We note that while we do not expect parton showers
to produce good predictions of high jet multiplicities, evident in the
increasing uncertainty with increasing jet multiplicity, we do find that in
all bins the uncertainty envelopes on the LO and NLO-matched predictions all overlap with
the experimental error bars.

\begin{figure}
  \centering
  \includegraphics[width=0.4\textwidth]{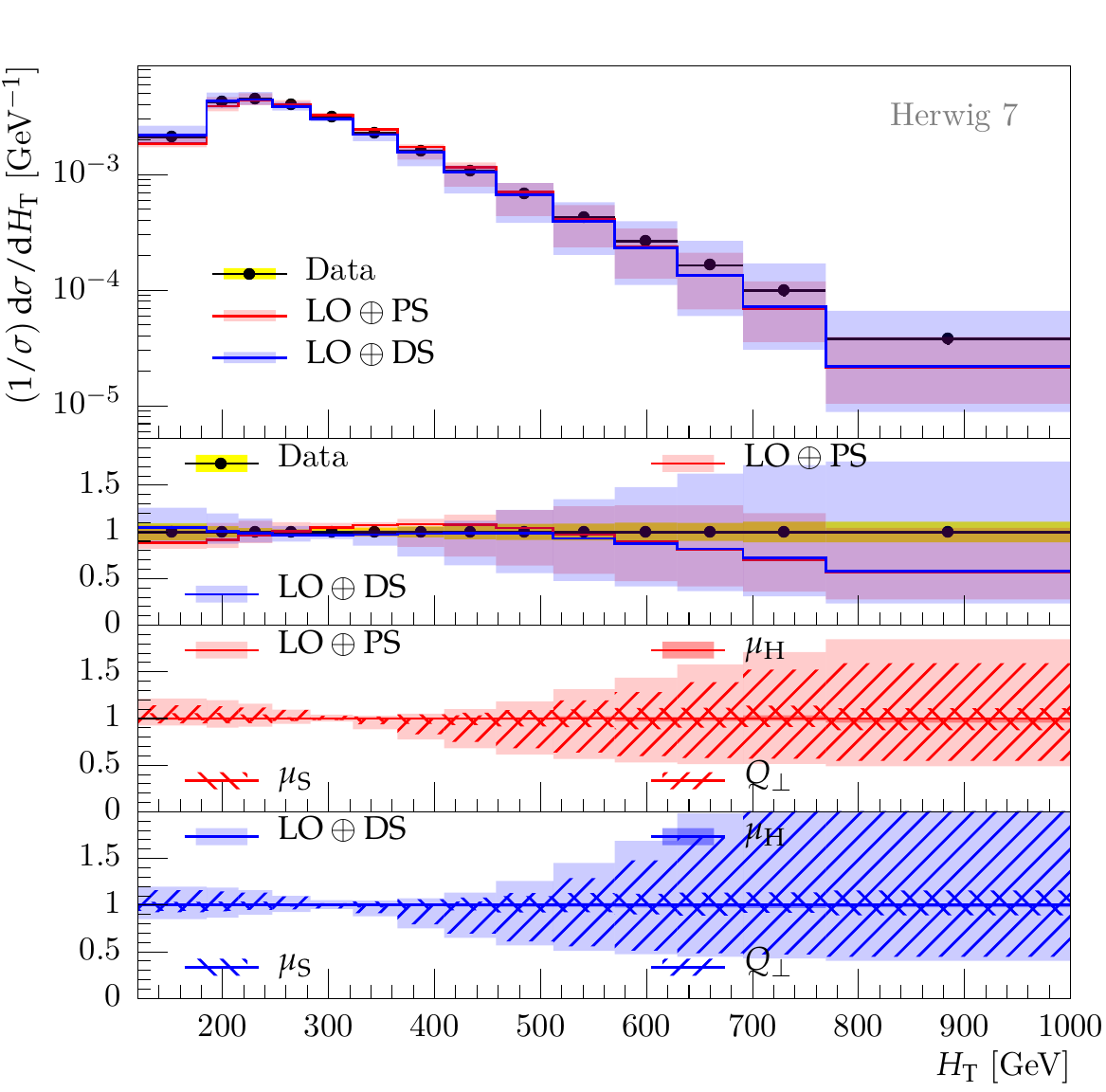}\\
  \includegraphics[width=0.4\textwidth]{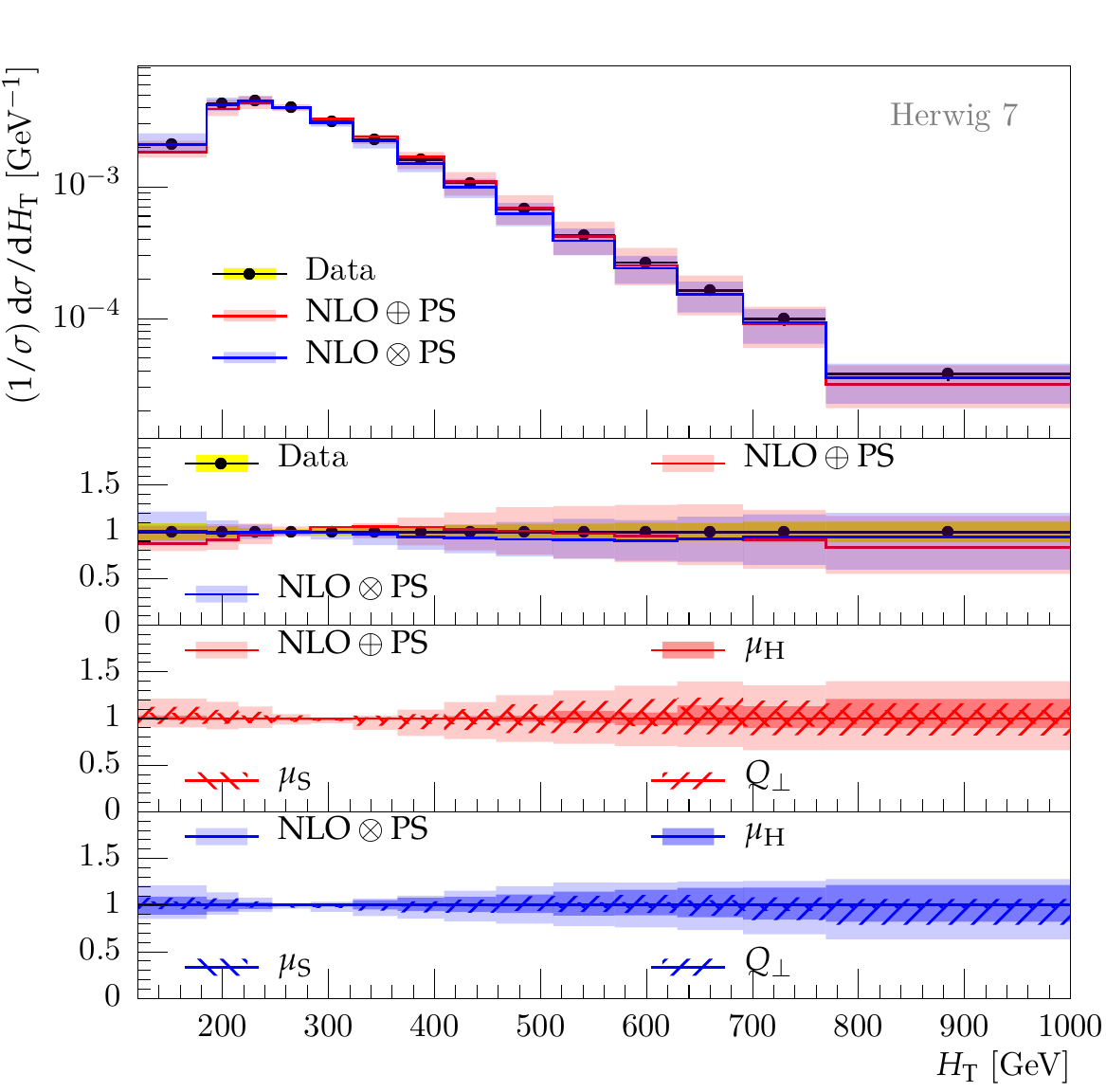}\\
  \includegraphics[width=0.4\textwidth]{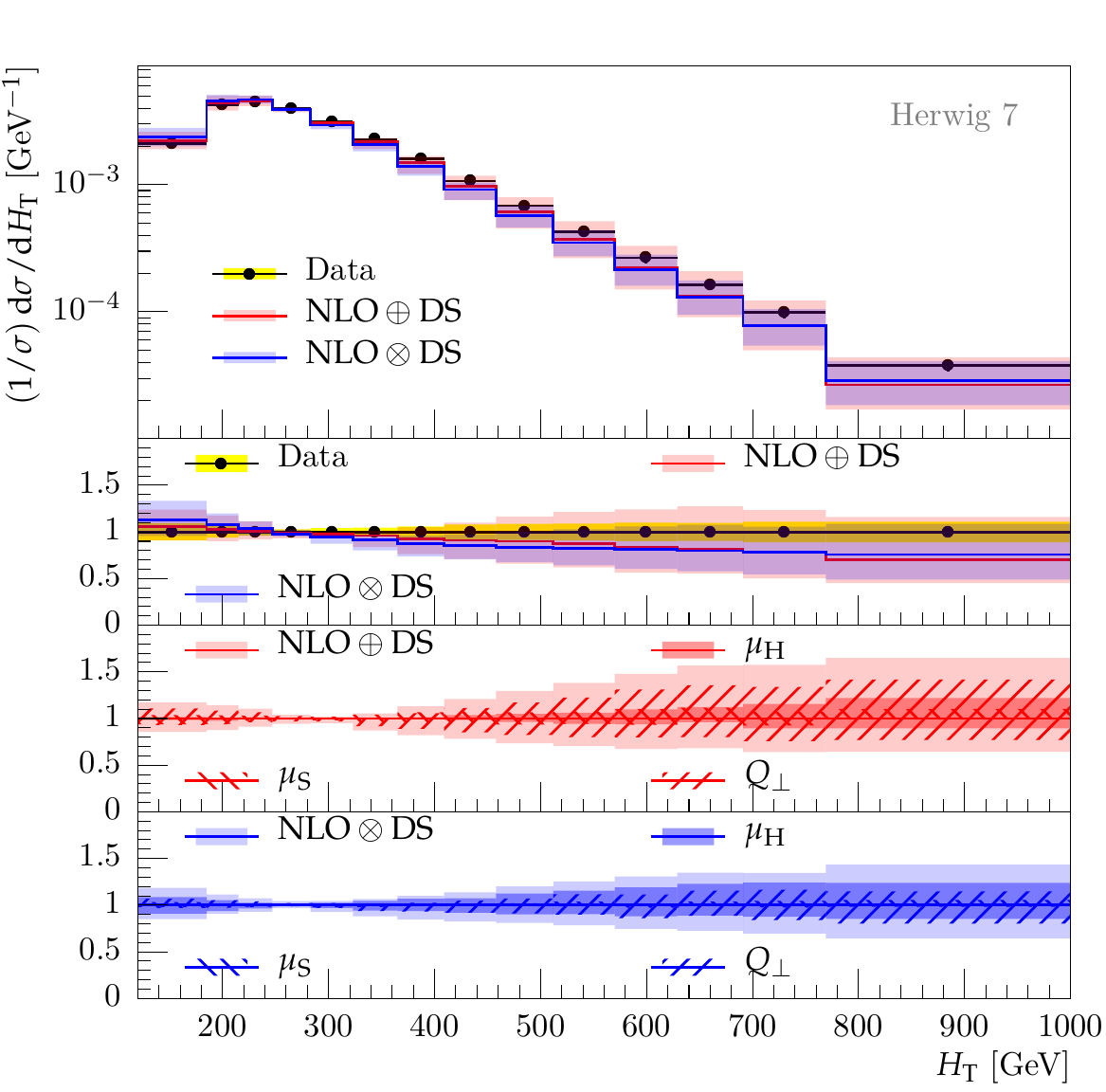}
  \caption{The $H_\mathrm{T}$ distribution measured in semileptonic 8\,TeV
    $pp\to t\bar{t}$ events by
CMS~\cite{Khachatryan:2016oou}.
    The theoretical predictions are the same as those described in the caption of
    Fig.\ref{fig:FullProc-ScaleVar-TopPt}.
  }
  \label{fig:FullProc-ScaleVar-HT}
\end{figure}

Fig.~\ref{fig:FullProc-ScaleVar-HT} shows predictions for the $H_\mathrm{T}$
distribution measured by CMS~\cite{Khachatryan:2016oou} in semileptonic
$t\bar{t}$ events at a centre-of-collision energy of $8\ {\rm TeV}$.  The
observable $H_\mathrm{T}$ is defined as the scalar sum of the transverse
momentum of all jets in each event
\begin{equation}
  H_\mathrm{T} = \sum_{\mathrm{jets}} p_{\perp, \mathrm{jet}} \ .
\label{eq:htcms}
\end{equation}
Considering the LO plus parton shower results, the central predictions of the
angular-ordered shower and the dipole shower display different shapes in the
lower bins however they come into good agreement in the upper bins. It is
clear that the uncertainty band on the dipole shower prediction is larger than
on the angular-ordered shower prediction, and in both cases it is dominated by
the variation of $\vetoScale$ which directly affects the predicted jet
activity.  The larger uncertainty in the dipole shower prediction due to
variations in $\vetoScale$ again reflects the differences in the phase space
of the two showers.
For each shower, the uncertainty on the prediction obtained using MC@NLO-type matching is
slightly larger than that on the prediction obtained using Powheg-type matching. There is no
clear single dominant source of uncertainty on the angular-ordered shower
predictions. As in the $n_\mathrm{jets}$ distributions, the largest
contribution to the uncertainty envelope on the dipole shower prediction using
MC@NLO-type matching is due to the variation of $\vetoScale$.

Of the variations considered in 
this study, only the variation of $\showerScale$ directly affects the 
simulation of a given individual decay process. However, some decay-related observables, such 
as measures of the separation of the decay products from different particle decays,
are sensitive to the hard 
process. It is therefore important to investigate the size of the 
uncertainty on such observables due to the variations of
all three scales considered.

\begin{figure}
  \centering
    \includegraphics[width=0.4\textwidth]{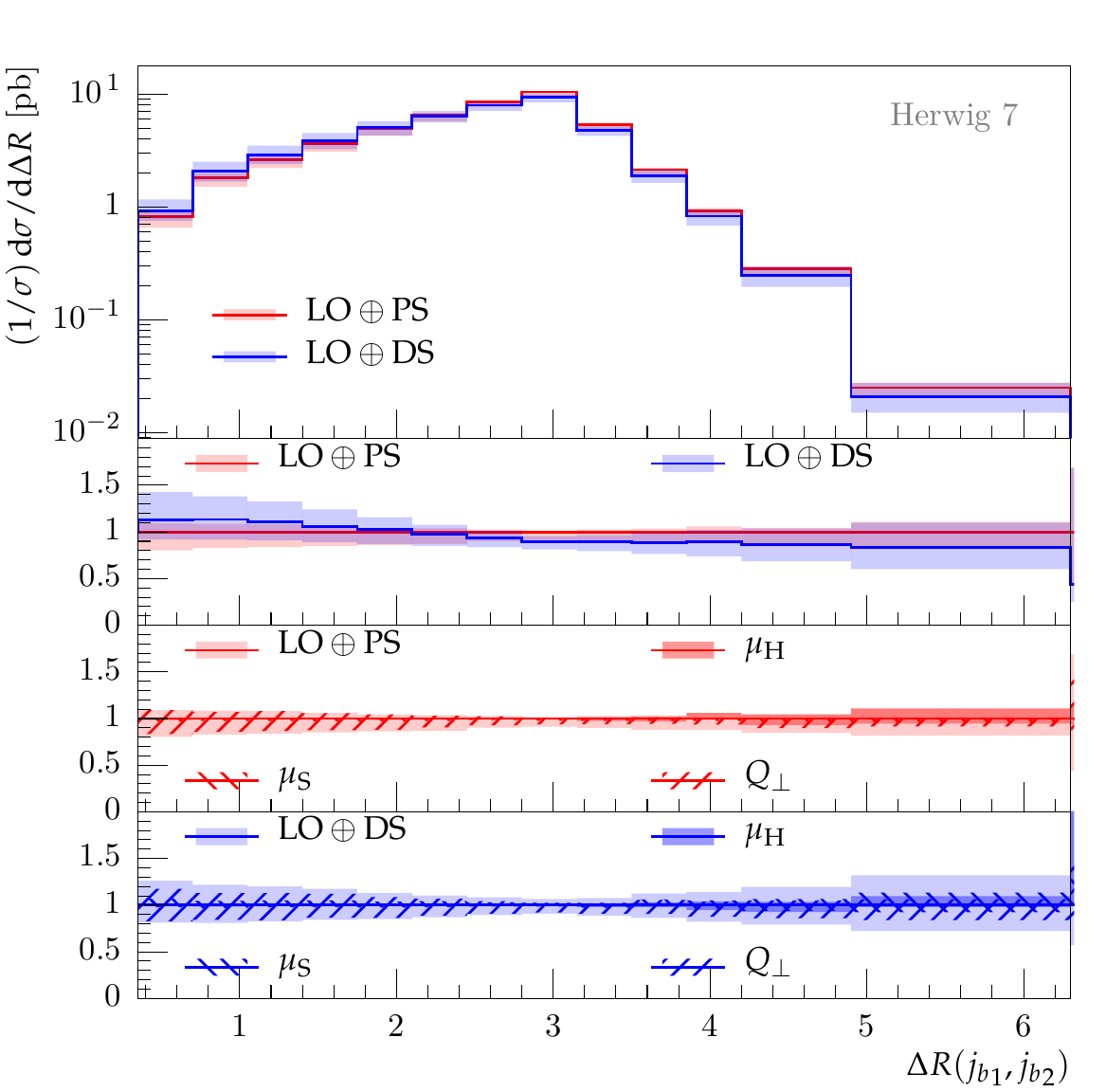}\\
    \includegraphics[width=0.4\textwidth]{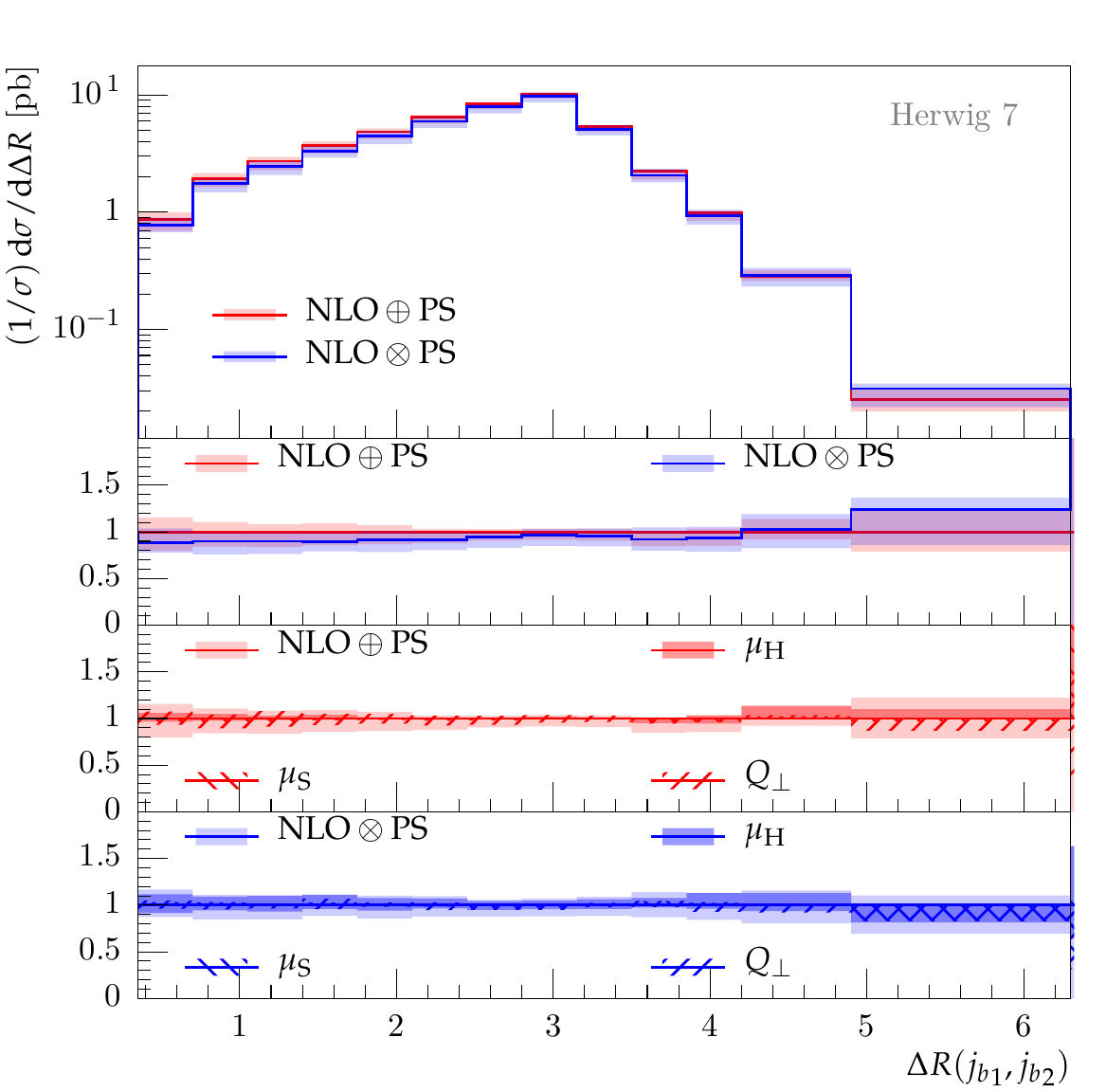}\\
    \includegraphics[width=0.4\textwidth]{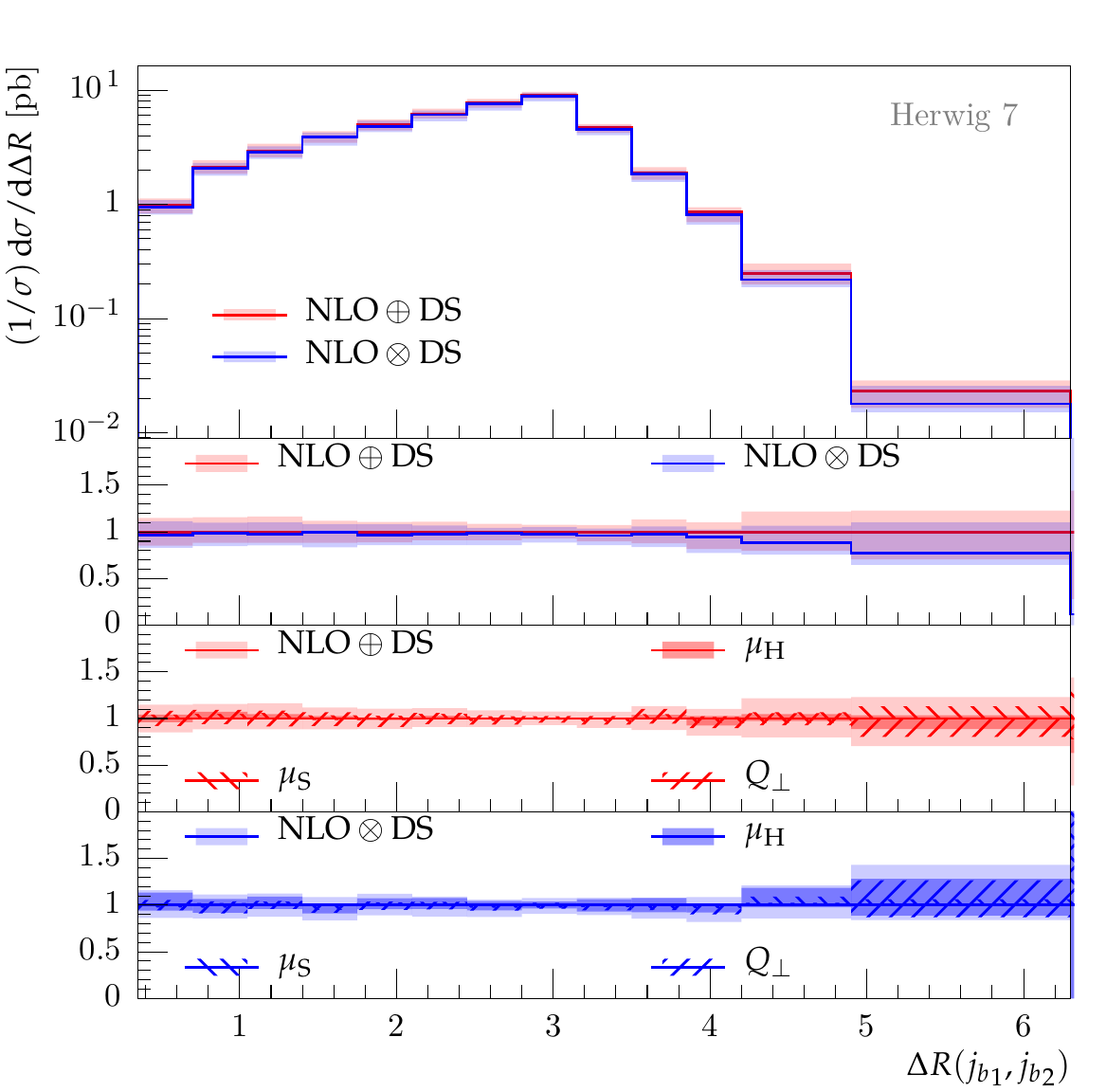}  
  \caption{The $\Delta R (b_1, b_2) = \sqrt{{\Delta \phi}^2 + {\Delta
        \eta}^2}$ distribution, described in the text, simulated for
    semileptonic $8\ {\rm TeV}$ $pp\to t\bar{t}$ events.}
  \label{fig:FullProc-ScaleVar-DeltaR-jetb1-jetb2}
\end{figure}

In Fig.~\ref{fig:FullProc-ScaleVar-DeltaR-jetb1-jetb2} we show predictions of the
separation of the two hardest b-tagged jets in semi-leptonic $pp\to t\bar{t}$
events at a centre-of-collision energy of $8\ {\rm TeV}$. The separation is
defined as $\Delta R ({j_b}_1, {j_b}_2) = \sqrt{{\Delta \phi}^2 + {\Delta
    \eta}^2}$, where $\Delta \phi$ and $\Delta \eta$ denote the difference in
the azimuthal angle and pseudorapidity respectively of the hardest and
second-hardest bottom-tagged jets. 
This observable is sensitive to both the
simulation of the decay and to the direction of the top quarks that decay to
produce the bottom quarks. We measure this distribution using a purpose-built
analysis in which we require at least one final-state dressed lepton, electron
or muon, with $p_\perp > 30\text{GeV}$ and $\lvert \eta\rvert < 4.2$. Dressed
leptons are created by clustering each bare lepton with any photons within a
cone of $\Delta R = 0.1$ around the lepton.  We also require at least two
light-flavour jets and two bottom-tagged jets with $p_\perp > 30\text{GeV}$
and $\lvert \eta \rvert < 4.2$. Additionally we implement a minimum missing
transverse energy cut of 30\,GeV, where the transverse energy of each visible
outgoing particle is defined as $E_\perp = E \sin(\theta)$ where $E$ and
$\theta$ denote the energy and polar angle of the particle respectively,
measured in the lab frame.

As we do not use the benchmark settings to produce predictions intended for
comparison to experimental data, it is not informative to compare the predictions of the
parton showers and matching schemes.  The dominant source of uncertainty on
the LO predictions in the region $\Delta R < \pi$ is the variation of
$\vetoScale$. This is because the relative orientation of the top quarks, and
hence the separation of the bottom-tagged jets, is sensitive to hard radiation
from the production process.  The uncertainty envelopes on the NLO-matched
predictions are in general smaller than those on the LO predictions, and there
is no single dominant source of uncertainty.  This is because the the hardest
jet from the production process is simulated to LO, rather than
parton-shower, accuracy.

To summarise, following our detailed discussion of scale variations in the
production-level process in Section~\ref{sec:scale-variations-prod-level}, we
have identified three observables measured in experiments and compared
predictions obtained using \Herwig~7 including a full evaluation of the
uncertainties due to scale variations. In general, the total uncertainty
envelope, from the complete set of all 27 scale variation
combinations, is not accurately reproduced by the variation of any single
scale. Therefore a full evaluation of the scale variations is required to
produce a good estimate of the uncertainty on predictions of experimental
observables.
In addition we have also considered one observable, the separation of the
bottom-tagged jets, that is sensitive to the simulation of both the production
process and the decay processes.  The uncertainty on this prediction due to
scale variations is small and our findings suggest that most of the
uncertainty is due to the sensitivity to the production process.  With
relatively few experimental analyses that measure decay-process sensitive
observables currently available, the evaluation of the uncertainty on
predictions of such observables is an area for future investigation.

\subsection{The Hard Veto Scale in Data Predictions
  Using MC@NLO-type Matching}
\label{sec:mcatnlo-veto-scale-full-process}

\begin{figure*}
  \centering
  \includegraphics[width=0.4\textwidth]{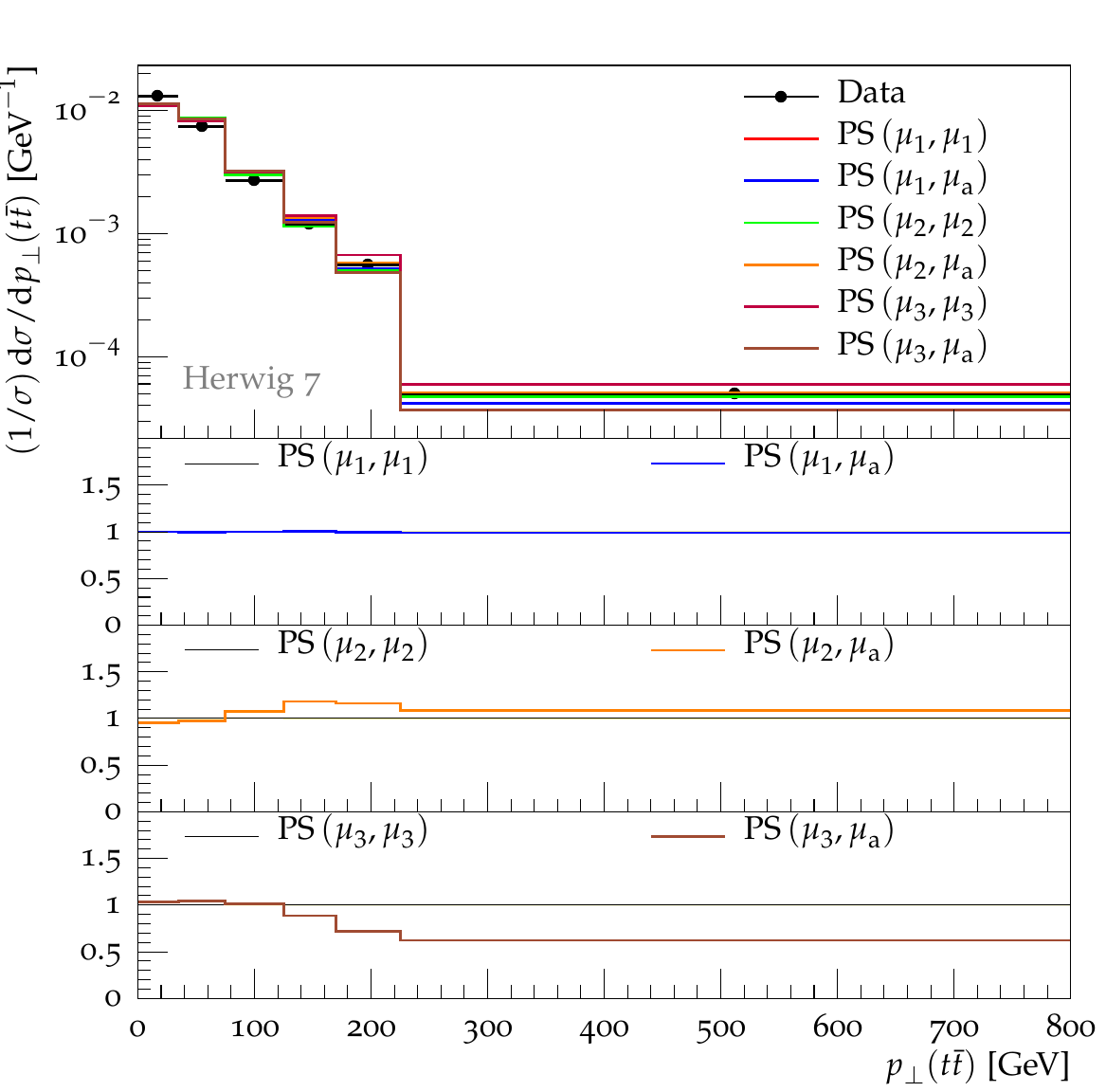}\hspace*{2cm}
  \includegraphics[width=0.4\textwidth]{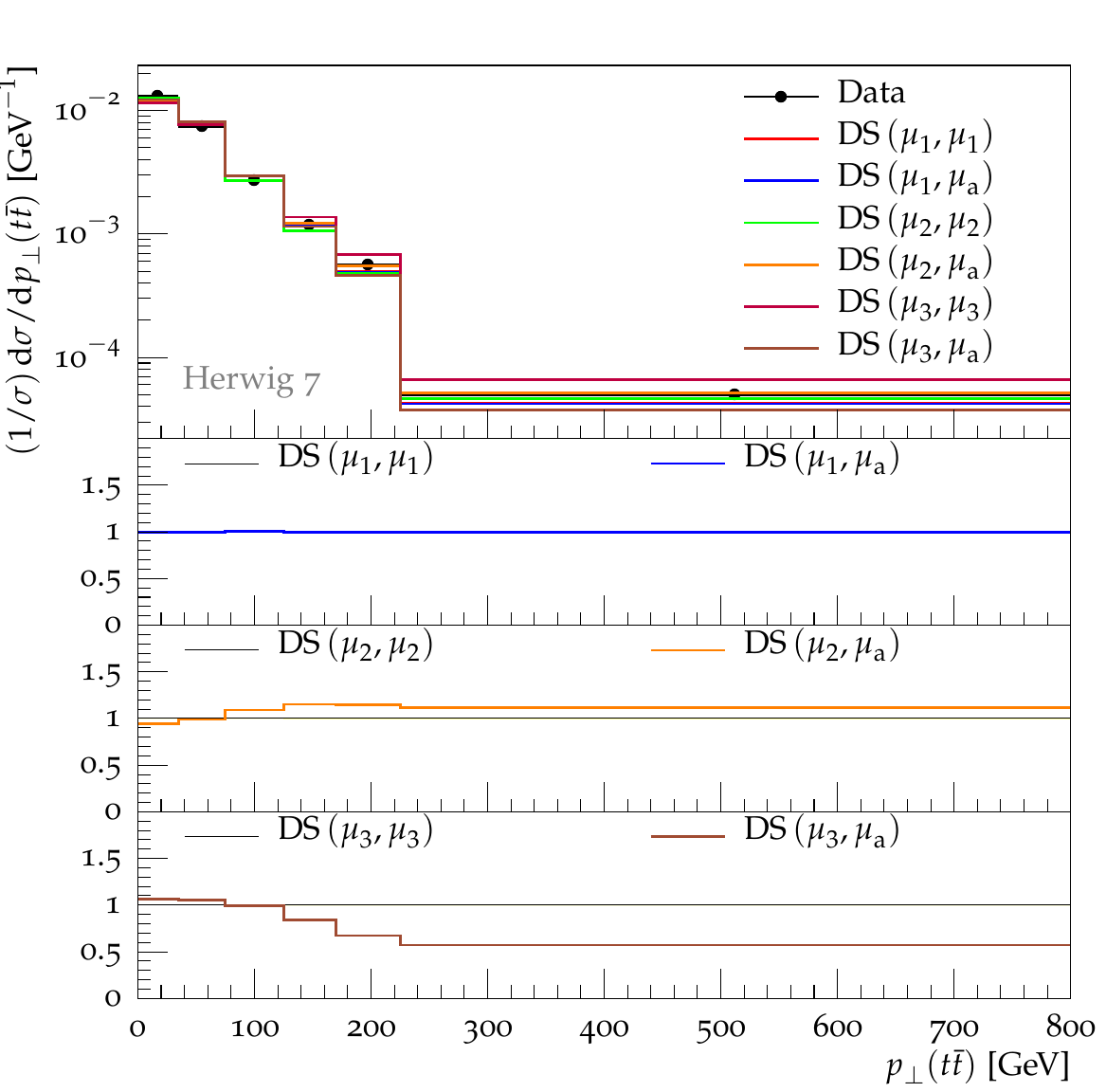}\\
  \includegraphics[width=0.4\textwidth]{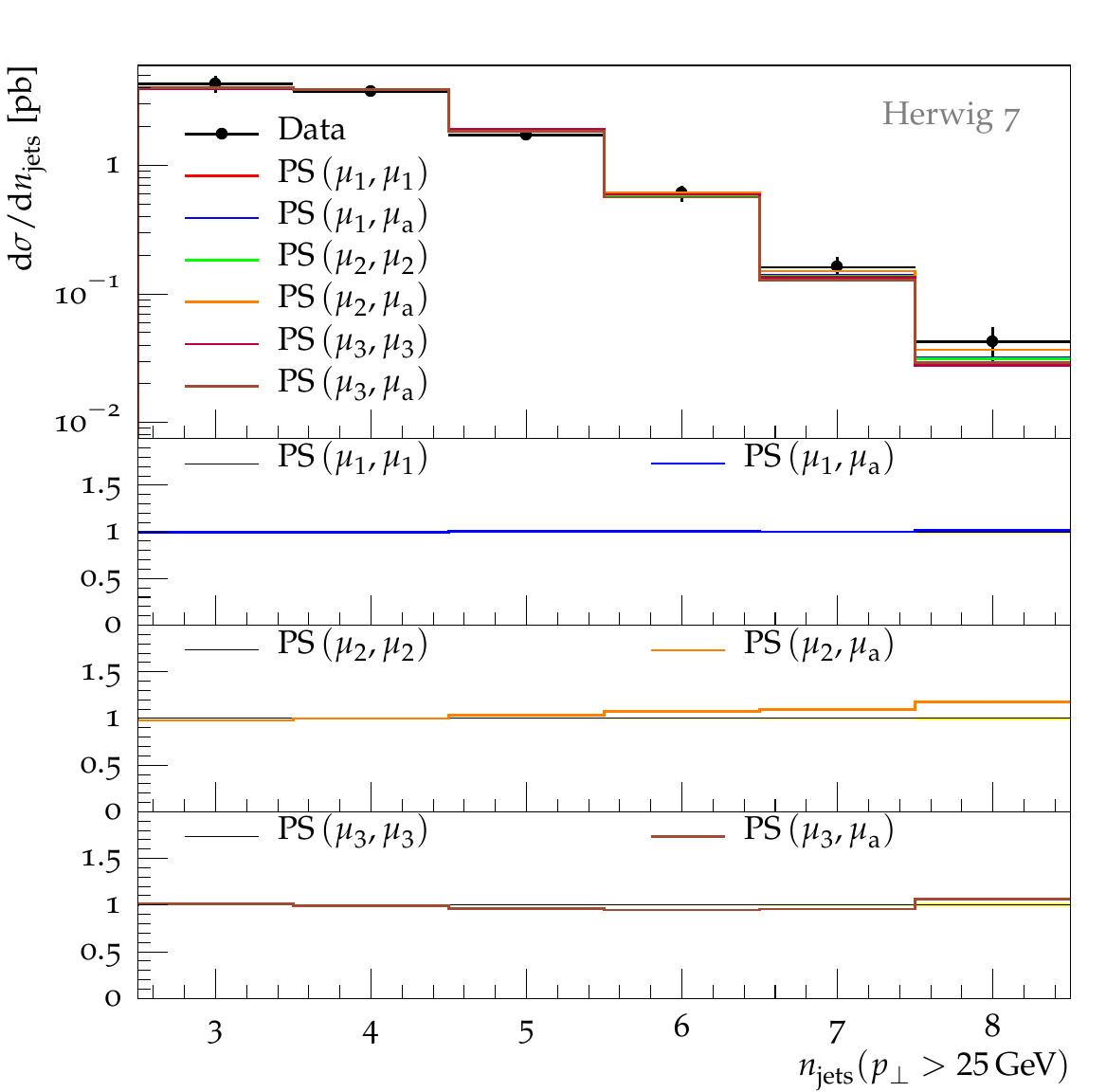}\hspace*{2cm}
  \includegraphics[width=0.4\textwidth]{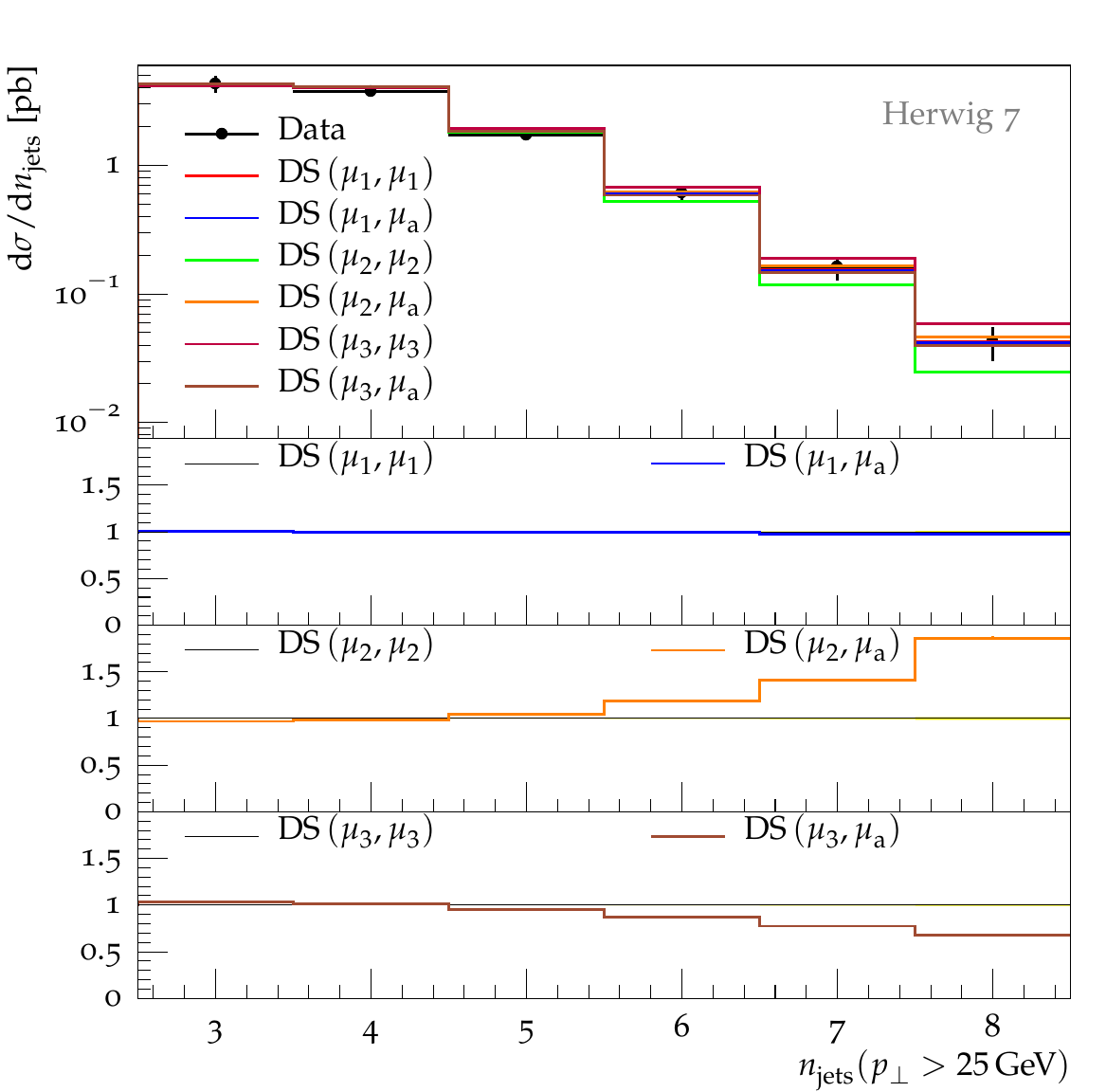}  
  \caption{Top: $t\bar{t}$-pair transverse momentum measured
    by ATLAS in semileptonic 8\,TeV $pp\to t\bar{t}$
events~\cite{Aad:2015mbv} and
    predicted using the angular-ordered~(PS) and dipole~(DS) parton showers, respectively.
    Bottom: The multiplicity distribution of jets with $p_\perp > 25\ {\rm GeV}$
    measured by ATLAS in semileptonic 7\,TeV $pp\to t\bar{t}$
    events~\cite{Aad:2014iaa} and predicted using the angular-ordered and dipole parton
    showers, respectively.
  }
  \label{fig:FullProc-ShowerScale-ttbarPt-nJets25}
\end{figure*}

In order to investigate the effects of the choice of $\vetoScale$ on the
prediction of distributions measured in
experiment we perform the same comparison of scale choices as
in Section~\ref{sec:mcatnlo-veto-scale-prod-level} for the 
full event simulation. We compare
some selected results from experimental analyses that demonstrate the effects
that the choice of $\vetoScale$ can have on data predictions.

We highlight that the hard veto scale is applied in the showering of the
production process only. The veto scale applied in the showering of the decay
process is simply the mass of the decayed particle, {\it i.e.} the top quark,
and is not varied. One should therefore expect the predictions that show the
largest change due to the choice of $\vetoScale$ to be those for observables
that have a direct dependence on radiation from the production process. The
four observables that we present in this section are such observables. First
we discuss the distribution of the transverse momentum of the $t\bar{t}$-pair,
sensitive to the hardest jet emitted in the production process, followed by
three observables that measure the jet activity in each event.

In Section~\ref{sec:mcatnlo-veto-scale-prod-level} we show distributions of
$p_\perp(t\bar{t})$ in the production-level process,
Fig.~\ref{fig:Prod-VetoScale-TopPt-ttbarPt}.  In
Fig.~\ref{fig:FullProc-ShowerScale-ttbarPt-nJets25} we show predictions of the
$p_\perp(t\bar{t})$-distribution measured by\linebreak ATLAS~\cite{Aad:2015mbv} in
semileptonic $t\bar{t}$-events at $\sqrt{s} = 8\text{ TeV}$, obtained using
the angular-ordered  and dipole showers.  Both of the showers display
very similar behaviour as in the production-level case, this is expected given
that the predictions for this distribution depend on the production process
and its showering.  In order to avoid repetition we refer the reader to our
discussion in Section~\ref{sec:mcatnlo-veto-scale-prod-level} for details.

In Section~\ref{sec:mcatnlo-veto-scale-prod-level} we show distributions of
the jet-{\linebreak}multiplicity with a jet-$p_\perp$ cut of $25\ {\rm GeV}$,
Fig.~\ref{fig:Prod-VetoScale-Jet3Pt-nJets25}.  Predictions of the
jet-multiplicity distribution measured by ATLAS~\cite{Aad:2014iaa} in
semileptonic $t\bar{t}$ events at $\sqrt{s} = 7\text{ TeV}$, obtained using
the angular-ordered and dipole showers, are shown in
Fig.~\ref{fig:FullProc-ShowerScale-ttbarPt-nJets25}. As in our
production-level analysis this distribution also implements a
minimum-$p_\perp$ requirement of $25\ {\rm GeV}$ on the jets.  We note that the full process includes additional jets from the
top-quark decays and the hadronic W-boson decay which are not present in the production-level process. Taking this in to account,
the behaviour of these predictions is consistent with the
production-level results and we refer the reader to the discussion in
Section~\ref{sec:mcatnlo-veto-scale-prod-level} for details.

\begin{figure*}
  \centering
  \includegraphics[width=0.4\textwidth]{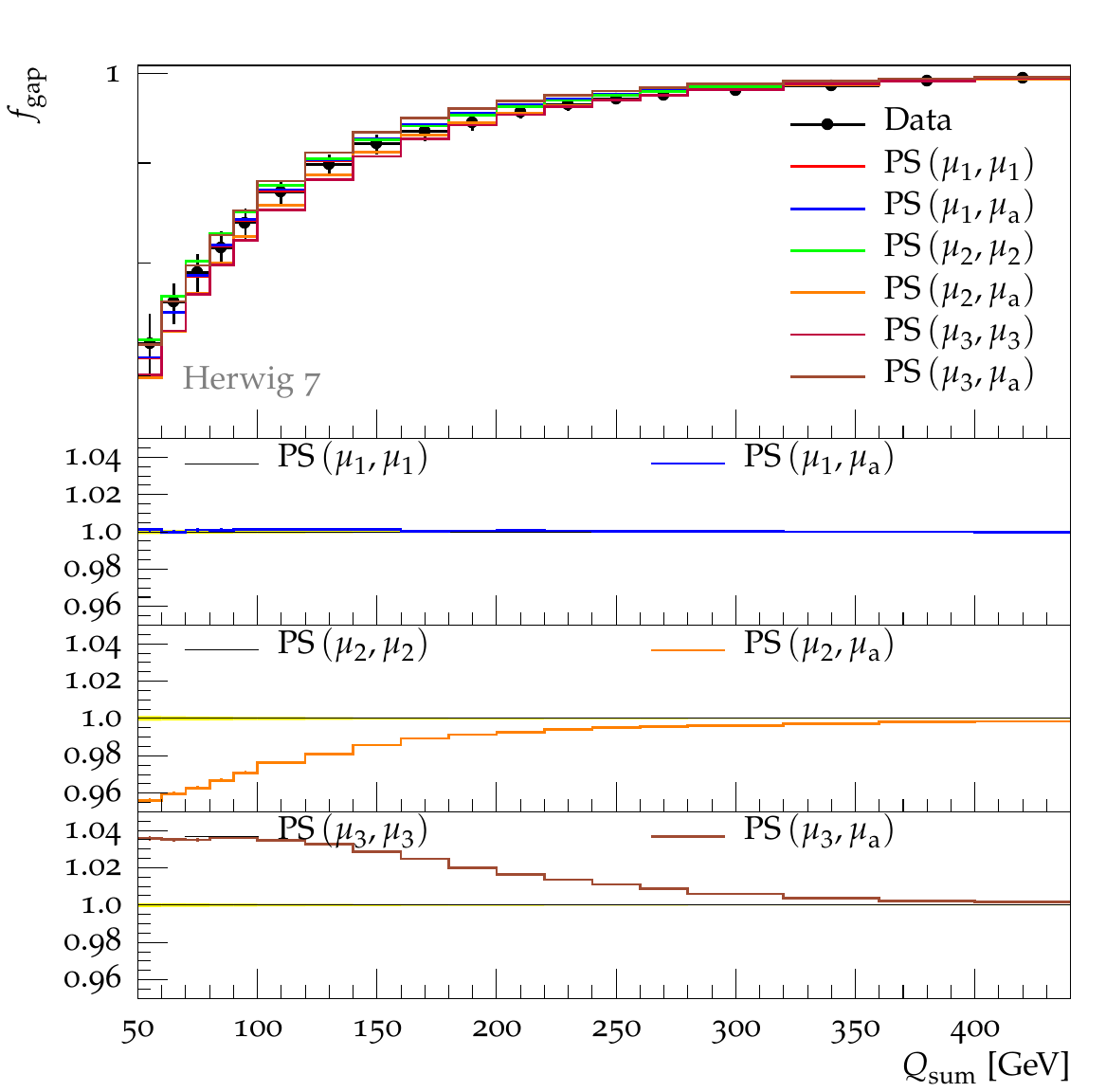}\hspace*{2cm}
  \includegraphics[width=0.4\textwidth]{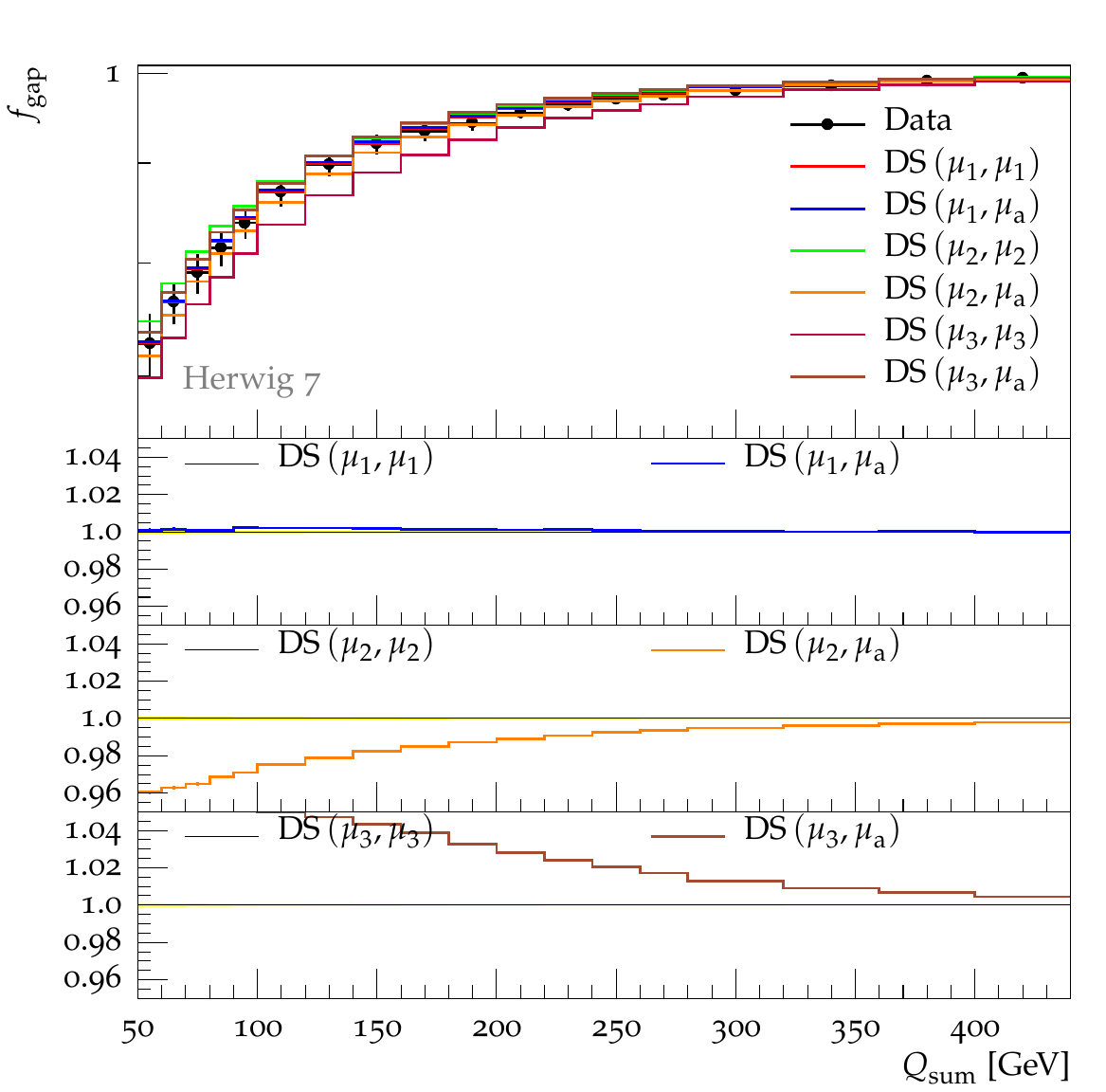}\\
  \includegraphics[width=0.4\textwidth]{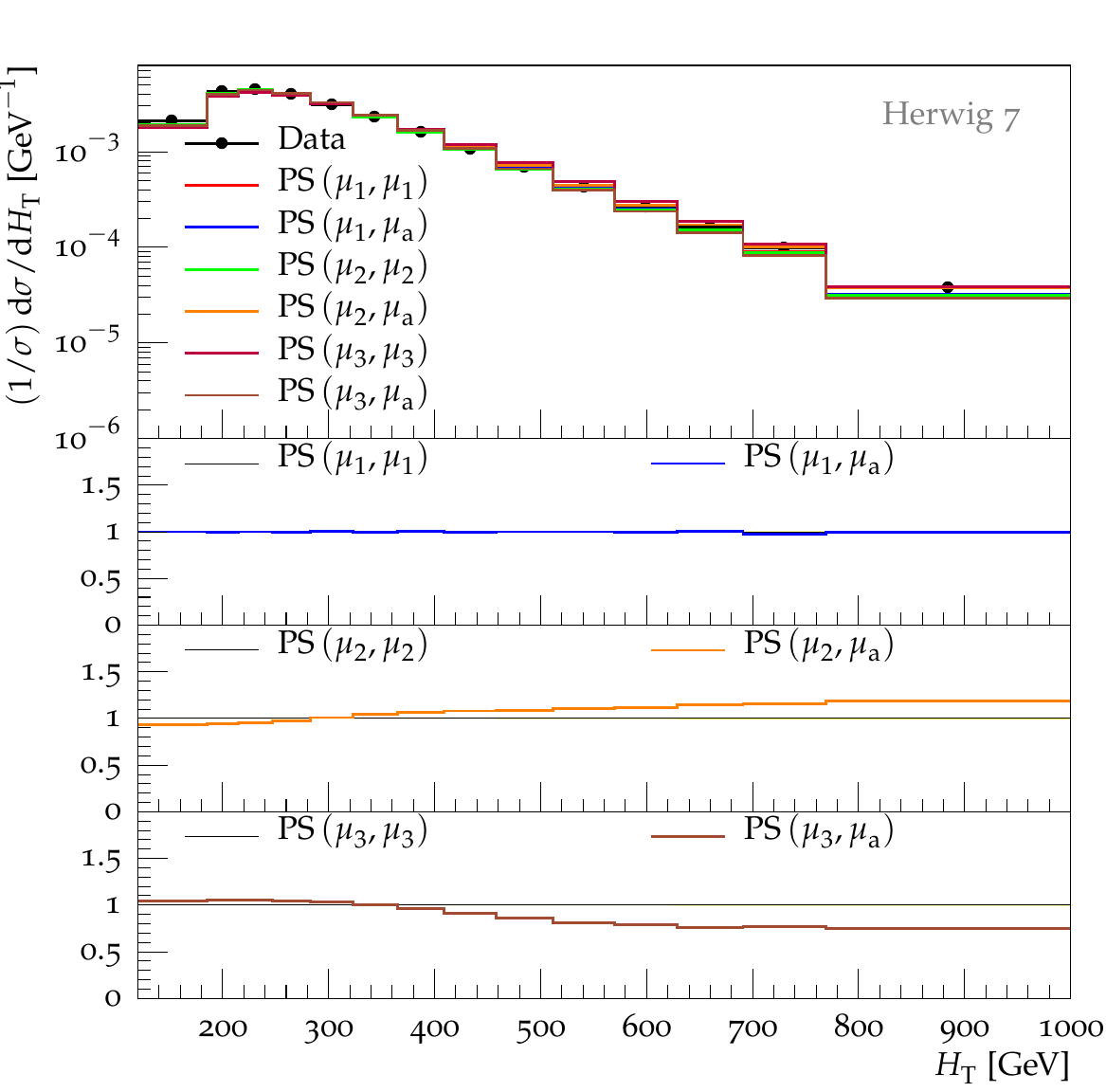}\hspace*{2cm}
  \includegraphics[width=0.4\textwidth]{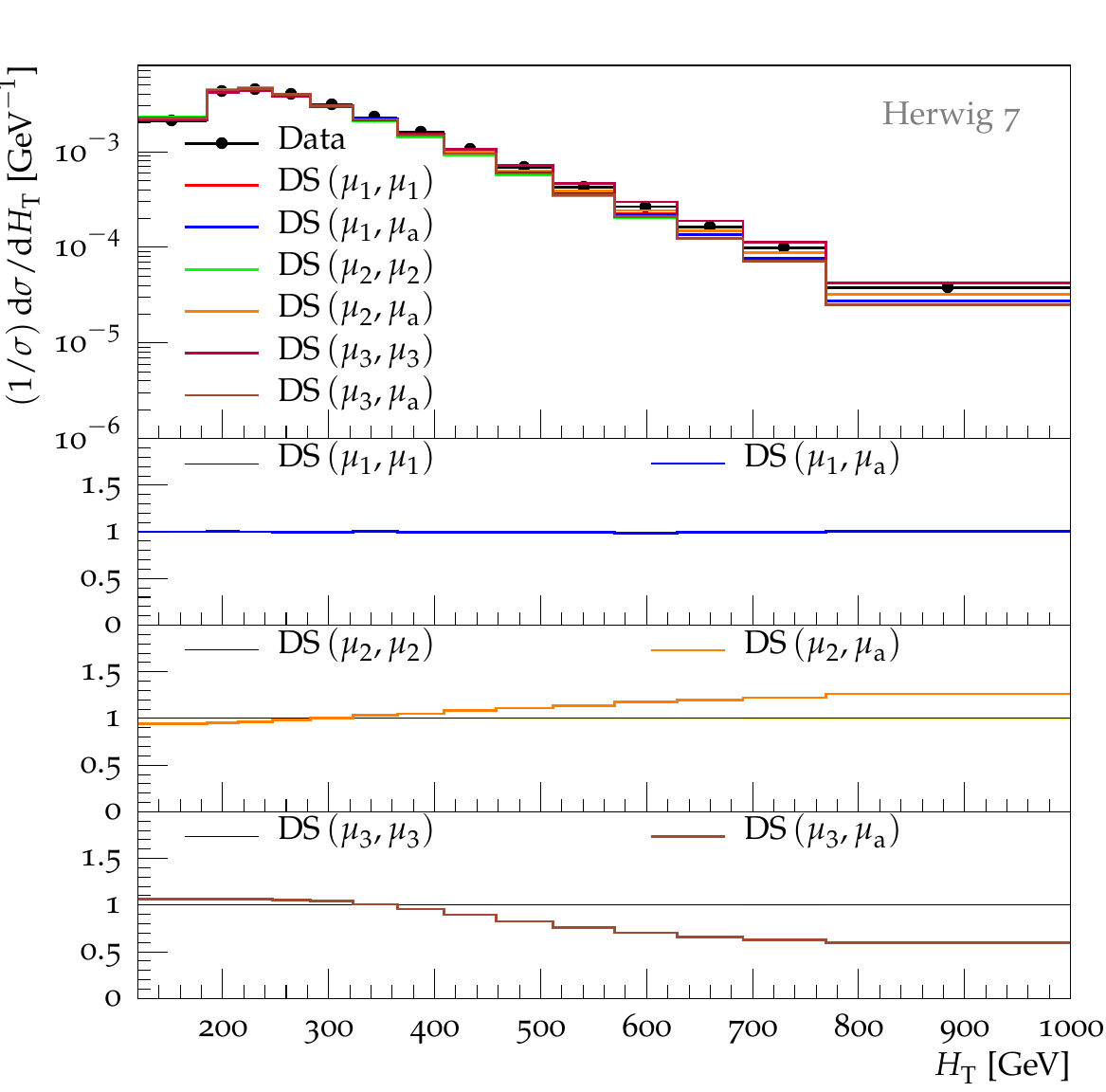}  
  \caption{Top: The gap fraction measured by ATLAS in dileptonic 7\,TeV $pp\to t\bar{t}$
    events~\cite{ATLAS:2012al}, in veto region $\lvert y \rvert < 2.1$, and predicted
    using the angular-ordered~(PS) and dipole~(DS) parton showers, respectively.
    Bottom: Combined lepton channel measurement of the $H_\mathrm{T}$ distribution by
    CMS in semileptonic
    $8\ {\rm TeV}$ $pp\to t\bar{t}$ events~\cite{Khachatryan:2016oou} and predicted using the
    angular-ordered and dipole parton showers, respectively.
  }
  \label{fig:FullProc-ShowerScale-GapFrac-HT}
\end{figure*}

Fig.~\ref{fig:FullProc-ShowerScale-GapFrac-HT} shows predictions of the gap
fraction, $f(Q_\mathrm{sum})$, measured by ATLAS~\cite{ATLAS:2012al} in
dileptonic $t\bar{t}$-events at\linebreak \mbox{$\sqrt{s} = 7\text{ TeV}$}, obtained using the
angular-ordered and\linebreak dipole showers.  The gap fraction is a measure of
additional jet activity in $t\bar{t}$-events, {\it i.e.} jets which originate
from quark and gluon radiation in the event as opposed to the decay products
themselves.  The analysis used selects only events in the dilepton decay
channel so that additional jets can be easily distinguished from the decay
products, {\it i.e.} two leptons and two bottom-tagged jets.  The gap fraction
is defined as
\begin{equation}
  f(Q_\mathrm{sum}) = \frac{n(Q_\mathrm{sum})}{N} \ ,
\end{equation}
where $N$ is the number of $t\bar{t}$ events that pass the analysis cuts and
$n(Q_\mathrm{sum})$ is the number of these events in which the sum of the
scalar transverse momenta of the additional jets in a given rapidity range
is less than the veto scale $Q_\mathrm{sum}$. In particular we present
results for additional jets in the rapidity range $\lvert y \rvert < 2.1$.

As we found for most observables in the production-level results, for the choice
$\hardProcScale = \mu_1$ we see very little difference in the predictions due to the choice
of $\vetoScale$ for both showers. For the scale choice $\hardProcScale = \mu_2$ the
predictions for both showers using $\vetoScale = \mu_\mathrm{a}$ display a decrease
in the gap fraction with decreasing $Q_\mathrm{sum}$ compared to using
$\vetoScale = \hardProcScale$. Conversely for the scale choice $\hardProcScale = \mu_3$ the predictions
for both showers using $\vetoScale = \mu_\mathrm{a}$ display an increase in the gap
fraction with decreasing $Q_\mathrm{sum}$ compared to using $\vetoScale = \hardProcScale$.
This corresponds to an increase in jet activity for $\hardProcScale = \mu_2$ and a
decrease for $\hardProcScale = \mu_3$, using $\vetoScale = \mu_\mathrm{a}$ compared to using
$\vetoScale = \hardProcScale$, as we would expect following our
discussions in Section~\ref{sec:mcatnlo-veto-scale-prod-level}.

Moreover, Fig.~\ref{fig:FullProc-ShowerScale-GapFrac-HT} shows predictions of the
$H_\mathrm{T}$ distribution, as defined in
Eq.~\eqref{eq:htcms}, measured by CMS~\cite{Khachatryan:2016oou} in
semileptonic $t\bar{t}$-events at \mbox{$\sqrt{s} = 8\text{ TeV}$} obtained using the
angular-ordered and dipole showers.  $H_\mathrm{T}$ is another probe of
jet activity in an event.  Given our previous discussion of the effects of the
choice of $\vetoScale$ the predictions obtained using both showers behave as
we would expect.  Using $\hardProcScale = \mu_1$ we see that the choice $\vetoScale =
\mu_\mathrm{a}$ has virtually no effect on the results compared to using
$\vetoScale = \hardProcScale$.  For the scale choice $\hardProcScale = \mu_1$ we see an increase in
the predicted distribution at high-$H_\mathrm{T}$ using $\vetoScale =
\mu_\mathrm{a}$ compared to using $\vetoScale = \hardProcScale$, while for the scale
choice $\hardProcScale = \mu_3$ we see a decrease in the predicted distribution at
high-$H_\mathrm{T}$ using $\vetoScale = \mu_\mathrm{a}$ compared to using
$\vetoScale = \hardProcScale$.

To conclude we re-iterate that we are not suggesting one choice for
$\vetoScale$ over another. Through the discussion in this section we have
demonstrated that the choice of the hard veto scale used in MC@NLO-type matching
can have a significant effect on the prediction of observables of interest in
$t\bar{t}$ production at the LHC. In particular we have presented predictions
using three choices for $\hardProcScale$ and two choices for $\vetoScale$,
the default choice in Herwig and a
new option.  The hard veto scale in MC@NLO-type matching directly effects the
showering of the production process, accordingly the predictions most affected
by the choice of $\vetoScale$ are those sensitive to jet activity in the
production process.

\section{Summary and Outlook}
\label{sec:Conclusions}

In this work we have presented a detailed study of NLO plus parton
shower matched predictions for top pair production at the LHC in the
\Herwig\ 7 event generator. We have considered various sources of
uncertainty, including the matching algorithms themselves for which
two options, a subtractive (MC@NLO-type) and multiplicative
(Powheg-type) paradigm can be used within \Herwig\ 7, as well as all
scale choices involved. We have not only considered NLO corrections to
the production process, but also in the decay process. Both shower
modules in \Herwig\ 7 are now able to handle radiation in both the
production and the decay of top quarks, and we have used this paper as
an opportunity to present a new treatment for radiation from heavy
quarks in the dipole shower.

We have found that no single scale variation encompasses the
entire set of independent variations, therefore all sources need to be
considered to obtain a reliable estimate of the uncertainty on predictions.
We have explicitly shown that NLO matching provides
improvements over a LO plus parton shower simulation where expected.
Higher jet multiplicities, however, do suffer from large uncertainties, even
using NLO matching, a fact which should be considered when using
tuned predictions. In the course of this work we have also considered
boosted topologies, focusing on N-subjettiness ratios which highlight
the internal structure of the jets.

Particular attention has been paid to the choice of the hard veto
scale. This is an ambiguity in matching algorithms which has not been
addressed extensively in the literature but plays an important role in
the handling of real-emission corrections present in the NLO
matching. Inappropriate choices can lead to artificially suppressed or
enhanced radiation, and we have found that scales which identify the
hard objects in the process provide the most reliable results.

The main purpose of this work was to highlight the uncertainties and
ambiguities associated with NLO matching, which need to be compared
between different shower and matching algorithms. The \Herwig~7 event
generator provides unique capabilities to quantify the differences
between predictions obtained using different setups and to benchmark
variations against each other. These sources of uncertainty should be taken into account
when comparing predictions against data, also in view of an improved
simulation based on multi-jet merging, which can more reliably predict
higher jet multiplicities.

\section*{Acknowledgements}

SW acknowledges support from a STFC studentship.
KC and SP are grateful to The University of Manchester for
kind hospitality during part of this work.  This work was supported in
part by the European Union as part of the FP7 Marie Curie Initial
Training Network MCnetITN (PITN-GA-2012-315877).  In addition this
work has received funding from the European Union's Horizon 2020
research and innovation programme as part of the Marie
Skłodowska-Curie Innovative Training Network MCnetITN3 (grant
agreement no. 722104). SP acknowledges partial support by the COST
Action CA16201 PARTICLEFACE, and by a FP7 Marie Curie Intra European
Fellowship under Grant Agreement PIEF- GA-2013-628739, when part of
this work was started.
CR acknowledges current support by the European Union's Horizon 2020
research and innovative programme, under grant agreement No. 668679, 
and recent support by the U.S. Department of Energy, under the grant 
agreements DE-SC0010102 and DE-FG02-13ER41942 when part of this work 
was done, as well as partial support by the German Federal Ministry of 
Education and Research (BMBF) when part of this work was started.
We would like to thank the other members of the \Herwig
collaboration for their continuous support. In particular we would
like to thank Johannes Bellm and Stefan Gieseke for valuable 
discussions, as well as Martin Stoll for his initial contribution to
massive dipoles.

\appendix

\section{Dipole Shower Kinematics}

\subsection{Alternative Formulation for the Final-Final Dipole Kinematics}
\label{app:FFKin:altForm}

The physical momenta of the partons following a splitting from a massive final-final dipole,
written in terms of the splitting variables $\FFz$ and $\FFy$, are
\begin{subequations}
\begin{IEEEeqnarray}{rCl}
      q_i & = & A_i Q + k_\perp  + B_i v_{\parallel},
      \\
      q_j & = & A_j Q - k_\perp + B_j v_{\parallel},
      \\
      q_k & = & A_k Q + B_k v_{\parallel},
      \label{eq:FFKin:pk}
\end{IEEEeqnarray}
\end{subequations}
where
\begin{subequations}
  \begin{IEEEeqnarray}{rCl}
    A_i &=& \frac{1}{s} \left[ m_i^2 + \frac{\bar{s}}{2} \left(\FFy + \FFz(1-\FFy)\right) \right],\\
    A_j &=& \frac{1}{s} \left[ m_j^2 + \frac{\bar{s}}{2} \left(1-\FFz(1-\FFy) \right) \right],\\
    A_k &=& \frac{1}{s} \left[ m_k^2 + \frac{\bar{s}}{2}(1-\FFy) \right],\\
   B_i &=& \frac{1}{B_k}\left( s A_i A_k - \frac{\bar{s}}{2}\FFz(1-\FFy) \right),\\
   B_j &=& \frac{1}{B_k}\left( s A_j A_k- \frac{\bar{s}}{2}(1-\FFz)(1-\FFy) \right),\\
   B_k &=& \sqrt{ \frac{1}{s} \left( m_k^2 + \frac{\bar{s}}{2}(1-\FFy) \right)^2 - m_k^2 }\,,
\end{IEEEeqnarray}
\end{subequations}
and the 4-vector
\begin{IEEEeqnarray}{C}
   v_\parallel = \sqrt{ \frac{4 s}{ \lambda \left(s, m_k^2, m_{ij}^2 \right) } } \left( \tilde{p}_k - \frac{Q \cdot \tilde{p}_k}{s} Q \right),
\end{IEEEeqnarray}
 is expressed using the Kallen function
\begin{IEEEeqnarray}{C}
   \lambda(x,y,z) =  x^2 + y^2 + z^2 - 2xy -2xz - 2yz\,.
\end{IEEEeqnarray}
Note that while it is trivial to write an expression $p_\perp = p_\perp(\FFz,\FFy)$, this expression is cubic in $\FFy$ which leads to a complicated analytic equation for $\FFy = \FFy(p_\perp,\FFz)$. 

\subsection{Final-Final Dipole Kinematics}
\label{app:FFKin}

\subsubsection{Completing the Formulation}

In order to complete our formulation of the splitting kinematics in Section~\ref{sec:DSFFKin} we require expressions for
the scaling parameters $x_k$ and $x_{ij}$ in terms of the variables $p_\perp$ and $z$.
We first write an expression for the virtuality of the emitter and emission produced
in the splitting
\begin{IEEEeqnarray}{rCl}
  Q_{ij}^2 & = & \left( q_i + q_j \right)^2
  \nonumber
  \\
  && = \frac{1}{z(1-z)}\left[ p_\perp^2 + (1-z) m_i^2 + z m_j^2 \right] \ .
\end{IEEEeqnarray}
Defining the variables $\lambda_{ij}$ and $\lambda_k$ as
\begin{equation}
 \lambda_k = 1 + \frac{m_k^2}{\FFinv}, \qquad \lambda_{ij} = 1 + \frac{m_{ij}^2}{\FFinv} \ ,
\end{equation}
we derive the following expressions for the scaling parameters
\begin{IEEEeqnarray}{rCl}
  x_{ij} &=& 1-\frac{m_k^2}{\FFinv}\frac{(1-x_k)}{x_k} \ ,
  \\
  x_k &=& \frac{1}{2\lambda_k}
  \left[ \left( \lambda_{ij}\lambda_k + \frac{m_k^2}{\FFinv} - \frac{Q_{ij}^2}{\FFinv} \right)
    \vphantom{\sqrt{ \left( \lambda_{ij}\lambda_k
          + \frac{m_k^2}{\FFinv} - \frac{Q_{ij}^2}{\FFinv} \right)^2
        - 4 \lambda_{ij} \lambda_k \frac{m_k^2}{\FFinv} }}
    \right.
    \\
    &&\qquad{}
    \pm
    \left.
    \sqrt{ \left( \lambda_{ij}\lambda_k + \frac{m_k^2}{\FFinv} - \frac{Q_{ij}^2}{\FFinv} \right)^2
      - 4 \lambda_{ij} \lambda_k \frac{m_k^2}{\FFinv} } \, \right] .
  \nonumber
\end{IEEEeqnarray}

Finally we require expressions for the splitting variables $\FFz$ and $\FFy$ in terms of the
variables $p_\perp$ and $z$. We write $\FFy$ as
\begin{equation}
  \FFy = \frac{1}{\bar{s} z(1-z)} \left[ p_\perp^2 + (1-z)^2 m_i^2 + z^2 m_j^2 \right] \ ,
\end{equation}
where the invariant quantity $\bar{s}$ is
\begin{equation}
  \bar{s} = s - m_i^2 - m_j^2 - m_k^2 \ .
\end{equation}

In order to express $\FFz$ in terms of $p_\perp$ and $z$ we write
\begin{equation}
  \FFz = \frac{2 q_i \cdot q_k}{(1-\FFy) \bar{s}} \ ,
\end{equation}
where the denominator can be written as
\begin{IEEEeqnarray}{l}
  (1-\FFy) \bar{s} =
  \\
  \frac{1}{z(1-z)} \left[ z(1-z) - (1-z)^2 m_i^2 - z^2 m_j^2 - p_\perp^2 \right] \ ,
  \nonumber
\end{IEEEeqnarray}
and the numerator is given by the expression
\begin{equation}
  2 q_i \cdot q_k = z x_{ij} x_k \FFinv +
  \frac{m_k^2}{z x_{ij} x_k \FFinv} \left( p_\perp^2 + m_i^2 \right) \ .
\end{equation}

\subsubsection{Phase-space Limits}
In order to allow us to efficiently generate values for $p_\perp$ and $z$ according to
the splitting kernels we need to express the single-particle emission phase space
and the limits on it in terms of these variables.

The limits on the dipole splitting variables $\FFz$ and $\FFy$ are given
in Ref.~\cite{Catani:2002hc} and we include them here to provide a complete reference,
\begin{IEEEeqnarray}{rCl}
  y_{ij,k,-} &=& \frac{2 m_i m_j}{\bar{s}} \ ,
  \\
  y_{ij,k,+} &=& 1 - \frac{2 m_k (\sqrt{s}-m_k)}{\bar{s}} \ ,
  \\   
  z_{i,\pm}(\FFy) &=&
  \frac{ 2 m_i^2 + \bar{s}\FFy} {2\left[ m_i^2 + m_j^2 + \bar{s}\FFy \right] }
  ( 1 \pm v_{ij,i}v_{ij,k} ) \ ,
\end{IEEEeqnarray}
where the relative velocities $v_{ij,k}$ and $v_{ij,i}$ are expressed as functions of $\FFy$,
\begin{IEEEeqnarray}{rCl}
  v_{ij,k} &=& \frac{ \sqrt{ \left[ 2 m_k^2 + \bar{s}(1-\FFy) \right]^2 - 4 m_k^2 s }}
  {\bar{s}(1-\FFy)} \ ,
  \\
  v_{ij,i} &=& \frac{\sqrt{ \bar{s}^2 \FFy^2 - 4 m_i^2 m_j^2 }}
  {\bar{s}\FFy + 2 m_i^2} \ .
\end{IEEEeqnarray}

The limits on $p_\perp$ and $z$ follow from the inequality $\FFy < y_{ij,k,+}$,
\begin{equation}
  p_{\perp,\mathrm{max}} = \frac{1}{2(\sqrt{s}-m_k)}
  \sqrt{  \lambda \left(m_i^2, m_j^2, (\sqrt{s}-m_k)^2 \right) } \ ,
\end{equation}
\begin{IEEEeqnarray}{rCl}
  z_{\pm} & = &\frac{1}{2 \left( \sqrt{s}-m_k \right)^2} 
  \left[ m_i^2 - m_j^2 + (\sqrt{s}-m_k)^2 \vphantom{\sqrt{ 1 - \frac{p_\perp^2}{p_{\perp, \mathrm{max}}^2}}} \right.
\\
&& \left. \pm \sqrt{  \lambda \left(m_i^2, m_j^2, (\sqrt{s}-m_k)^2 \right) }
  \sqrt{ 1 - \frac{p_\perp^2}{p_{\perp, \mathrm{max}}^2}} \right] \ .
\end{IEEEeqnarray}

\subsubsection{The Single-Particle Emission Phase Space}

The single-particle emission phase space required to express the branching probability in
Eq.~\eqref{eq:FFKin:branchProb} is written as
\begin{equation}
  \mathrm{d}q_j = \frac{1}{16\pi^2}
  \frac{ \bar{s}^2}{\sqrt{\lambda\left(s, m_{ij}^2, m_k^2 \right)}}
  \left(1-\FFy \right) \mathrm{d}\FFy \mathrm{d}\FFz \frac{\mathrm{d}\phi}{2\pi} \ .
\end{equation}
As we consider only spin-averaged kernels the azimuthal angle, $\phi$, is averaged over in
the phase-space integration and we do not consider it explicitly in the following discussion.

Using the above expression for the single-particle phase space we express the branching probability
as
\begin{IEEEeqnarray}{rCl}
  \mathrm{d}\mathcal{P}_{\mathrm{branching}} &=& \frac{1}{16\pi^2}
  \langle V_{i j, k} \left( \FFz, \FFy \right) \rangle
  \frac{1}{ \left( 1 + \frac{ m_i^2 + m_j^2 - m_{ij}^2}{\bar{s}\FFy} \right)}
  \nonumber
  \\
  && \times 
  \frac{\bar{s}}{\sqrt{\lambda\left(s, m_{ij}^2, m_k^2 \right)}}
  \left(1-\FFy \right)
  \frac{\mathrm{d}\FFy}{\FFy} \mathrm{d}\FFz \ ,
  \nonumber
  \\
\end{IEEEeqnarray}
where we can express the phase-space integration in terms of the generated variables using the
result
\begin{IEEEeqnarray}{rCl}
  \frac{\mathrm{d}\FFy}{\FFy} \mathrm{d}\FFz & = &
  - \left[ \frac{p_\perp^2}{p_\perp^2 + (1-z)^2 m_i^2 + z^2 m_j^2 } \right]
  \\
  &&
  \times
  \left[ 1  -  2\frac{1}{\bar{s}(1-\FFy)}\frac{m_k^2 Q_{ij}^2}{x_{ij} x_k \FFinv} \right]
  \frac{\mathrm{d} p_\perp^2}{p_\perp^2} \mathrm{d}z \ .
  \nonumber
\end{IEEEeqnarray}

\subsection{Final-Initial Dipole Kinematics}

\label{app:FIKin}

\subsubsection{Phase-space Limits}

The upper limit on $\FIx$ is,
\begin{equation}
  x_{ij,b,+} = \frac{\FIinv}{\FIinv - m_{ij}^2 + (m_i + m_j)^2} \ .
\end{equation}
We can derive a lower limit on $\FIx$. We first write the momentum of the incoming
proton as $P$ and the proton momentum-fraction carried by the spectator prior to the
splitting as $x_s$. We can write
\begin{equation}
  q_b = \frac{1}{\FIx} \tilde{p}_{b} = \frac{1}{\FIx} \left( x_s P \right) < P \ ,
\end{equation}
hence we require
\begin{equation}
  \FIx > x_s \ .
  \label{eq:FI:xineq}
\end{equation}

From the inequality in Eq.~\eqref{eq:FI:xineq} we derive the following limits on the
variables $p_\perp$ and $z$
\begin{equation}
	p_{\perp,\mathrm{max}}^2 = \frac{\FIinvprime}{4} \lambda \left(1, \frac{m_i^2}{\FIinvprime}, \frac{m_j^2}{\FIinvprime} \right) \ ,
\end{equation}
\begin{IEEEeqnarray}{rCCl}
	z_\pm & = & \frac{1}{2} &  \left[ 1 + \frac{m_i^2 - m_j^2}{\FIinvprime} \nonumber \right.
	\\
 	&&& \left. \pm \sqrt{  \lambda \left(1, \frac{m_i^2}{\FIinvprime}, \frac{m_j^2}{\FIinvprime} \right) }
      \sqrt{1-\frac{p_\perp^2}{p_{\perp,\mathrm{max}}^2}} \right] \ ,
\end{IEEEeqnarray}
where $\lambda$ is the standard Kallen function and for convenience we have defined the modified invariant
\begin{equation}
  \FIinvprime = \FIinv \left(\frac{1-x_s}{x_s}\right) + m_{ij}^2 \ .
\end{equation}

\subsubsection{The Single-Particle Emission Phase Space}

The single-particle emission phase space required to express the branching probability in
Eq.~\eqref{eq:FIKin:branchProb} is written as
\begin{equation}
	\mathrm{d}q_j = \frac{1}{16 \pi^2} 2\tilde{p}_{ij} \cdot q_b \mathrm{d}\FIz \mathrm{d}\FIx \frac{\mathrm{d}\phi}{2\pi} \ .
\end{equation}
As we consider only spin-averaged kernels the azimuthal angle is averaged over in the phase space
integration.

Using the above expression for the single-particle phase space we express the branching probability
as
\begin{IEEEeqnarray}{rCl}
  \mathrm{d}\mathcal{P}_{\mathrm{branching}}
  &=& \frac{1}{16\pi^2} \frac{f_b(x_s/\FIx)}{f_b(x_s)}
  \langle V_{i j}^b \left( \FIz, \FIx \right) \rangle
  \nonumber
  \\
  &&
  {} \times \frac{1}{\FIx(1-\FIx)} \mathrm{d}\FIz \mathrm{d}\FIx \ .
\end{IEEEeqnarray}
Noting that $\FIz = z$ we can express the phase-space integration in terms of the generated
variables using the result,
\begin{IEEEeqnarray}{l}
  \frac{1}{\FIx(1-\FIx)} \mathrm{d}\FIz \mathrm{d}\FIx =
  \\
  {}- \left[ \frac{p_\perp^2}{ p_\perp^2 + (1-z) m_i^2 + z m_j^2 
      - z(1-z) m_{ij}^2} \right]\frac{ \mathrm{d}p_\perp^2 }{p_\perp^2} \mathrm{d}z \ .
  \nonumber
\end{IEEEeqnarray}

\subsection{Initial-Final Dipole Kinematics}
\label{app:IFKin}

\subsubsection{Phase-space Limits}

The lower and upper limits on $\IFu$ are
\begin{IEEEeqnarray}{rCl}
  u_{j,+} &=& 0
  \\
  u_{j,+} &=& \frac{1 - \IFx}{1 - \IFx(1 - m_k^2/\IFinv)} \ ,
\end{IEEEeqnarray}
and the upper limit on $\IFx$ is
\begin{equation}
  x_{jk,a,+} = 1 \ .
\end{equation}
Following an analogous argument to that used to derive the inequality in Eq.~\eqref{eq:FI:xineq}
we derive a lower limit for $\IFx$
\begin{equation}
  \IFx > x_e \ ,
  \label{eq:IF:xineq}
\end{equation}
where $x_e$ is the proton momentum-fraction carried by the emitter prior to the splitting.

From the inequality in Eq.~\eqref{eq:IF:xineq} we derive the following limits on the
variables $p_\perp$ and $z$
\begin{equation}
  p_{\perp,\mathrm{max}}^2 = \frac{\IFinvprime^2}{4} \left[ \frac{1}{m_k^2 + \IFinvprime} \right] \ ,
\end{equation}
\begin{equation}
  z_{\pm} = \frac{1}{2} \left[ (1+x_e) \pm (1-x_e) \sqrt{1-\frac{p_\perp^2}{p_{\perp,\mathrm{max}}^2}}
  \right] \ ,
\end{equation}
where for convenience we have defined the rescaled invariant
\begin{equation}
	\IFinvprime = \IFinv \left( \frac{1-x_e}{x_e} \right) \ .
\end{equation}

\subsubsection{The Single-Particle Emission Phase Space}

The single-particle emission phase space required to express the branching probability in
Eq.~\eqref{eq:IFKin:branchProb} is written as
\begin{equation}
  \mathrm{d}q_j = \frac{1}{16\pi^2} 2  q_a\cdot\tilde{p}_k \mathrm{d}\IFu \mathrm{d}\IFx 
  \frac{\mathrm{d}\phi}{2\pi} \ .
\end{equation}
As we consider only spin-averaged kernels the azimuthal angle is averaged over in the phase-space
integration.

Using the above expression for the single-particle phase space we express the branching probability
as
\begin{IEEEeqnarray}{rCl}
  \mathrm{d}\mathcal{P}_{\mathrm{branching}} &=&
  \frac{1}{16\pi^2} \frac{f_a(x_e/\IFx)}{\tilde{f}_{aj}(x_e)}
  \langle V^{a j}_k \left( \IFu, \IFx \right) \rangle
  \nonumber
  \\
  &&   {}\times \frac{1}{\IFu} \frac{1}{\IFx} \mathrm{d}\IFu \mathrm{d}\IFx \ ,
\end{IEEEeqnarray}
where we can express the phase-space integration in terms of the generated variables using
the result
\begin{IEEEeqnarray}{l}
  \frac{1}{\IFu}\frac{1}{\IFx}\mathrm{d}\IFu \mathrm{d}\IFx =
  \\
  \left[ \IFu + \IFx - \IFu\IFx\left(1-\frac{m_k^2}{\IFinv}\right) \right]^{-1}
  \frac{\mathrm{d} p_\perp^2}{p_\perp^2} \mathrm{d}z \ .
  \nonumber
\end{IEEEeqnarray}

\section{Generation Cut for Boosted Top Analyses}
\label{app:boostedCut}

We can use the existing cut infrastructure in \Herwig7 to implement a
generation cut to enhance the production of events that include a boosted
top quark or antiquark. We create a `MatchboxFactoryMatcher' that identifies
top quarks and antiquarks and associate this with the existing `JetFinder'
and `JetCuts' objects. Then we assign the existing `FirstJet' and `SecondJet'
objects to the `JetRegions' of the `JetCuts' object. As we have assigned the
matcher of the `JetCuts' object to be our new top quark matcher,
the `FirstJet' and `SecondJet' actually identify top quarks. Therefore we
can set transverse momentum and rapidity cuts on the top quarks by setting
the cuts on the `FirstJet' and `SecondJet'. The code snippet required to do
this is included below.

\begin{verbatim}
################################################
## Cut for boosted top analyses
################################################
cd /Herwig/MatrixElements/Matchbox

# Create a new particle group consisting of top 
# quarks and antiquarks
do Factory:StartParticleGroup ttbar
insert Factory:ParticleGroup 0 /Herwig/
Particles/t
insert Factory:ParticleGroup 0 /Herwig/
Particles/tbar
do Factory:EndParticleGroup

# Create a new matcher and associate it with the 
# top quark particle group
create Herwig::MatchboxFactoryMatcher 
TopAntiTopMatcher
set TopAntiTopMatcher:Factory /Herwig/
MatrixElements/Matchbox/Factory
set TopAntiTopMatcher:Group ttbar

# Set the matcher of the JetFinder and JetCuts
# to the new top quark matcher
set /Herwig/Cuts/JetFinder:UnresolvedMatcher 
TopAntiTopMatcher
set /Herwig/Cuts/JetCuts:UnresolvedMatcher 
TopAntiTopMatcher

cd /Herwig/Cuts

# This snippet sets up JetFinder and JetCuts
read Matchbox/DefaultPPJets.in

# Set up the JetRegions and cuts
# Note: FirstJet and SecondJet are actually top 
# quarks/antiquarks
insert JetCuts:JetRegions 0 FirstJet
insert JetCuts:JetRegions 1 SecondJet

set FirstJet:PtMin 0.*GeV
do FirstJet:YRange -5.0 5.0

set SecondJet:PtMin 0.*GeV
do SecondJet:YRange -5.0 5.0
\end{verbatim}

\bibliography{ttbar-matching}

\end{document}